\documentclass[%
 aip,
 amsmath,amssymb,
 reprint,%
]{revtex4-1}

\usepackage{graphicx}% Include figure files
\usepackage{dcolumn}% Align table columns on decimal point
\usepackage{bm}% bold math
\usepackage{braket}
\usepackage{multirow}
\usepackage{comment}
\usepackage{hyperref}
\usepackage[labelformat=empty]{subfig}
\newcommand{\parz}[2]{ \frac{\partial{#1}}{\partial{#2}}}            %derivate parziali
\newcommand*{\diff}{\mathop{}\!\mathrm{d}}             % per fare i differenziali
\newcommand{\abs}[1]{\left| {#1} \right|}              % abb. sbarrette valore assoluto

\newcommand{\gr}[1]{\textbf{#1}}

\usepackage[utf8]{inputenc}
\usepackage[T1]{fontenc}
\usepackage{mathptmx}

\begin{document}

\preprint{AIP/123-QED}

\graphicspath{{img/}} % Directory in which figures are stored

\title[]{Electronic Transitions for a Fully Polarizable QM/MM Approach Based on Fluctuating Charges and Fluctuating Dipoles: Linear and Corrected Linear Response Regimes}
% Force line breaks with \\

\author{Tommaso Giovannini}
 \affiliation{Department of Chemistry, Norwegian University of Science and Technology, 7491 Trondheim, Norway}
\author{Rosario Roberto Riso}
\affiliation{Scuola Normale Superiore,
             Piazza dei Cavalieri 7, 56126 Pisa, Italy.}  
\author{Matteo Ambrosetti}
\affiliation{Scuola Normale Superiore,
             Piazza dei Cavalieri 7, 56126 Pisa, Italy.}          
\author{Alessandra Puglisi}
\affiliation{Scuola Normale Superiore,
             Piazza dei Cavalieri 7, 56126 Pisa, Italy.}
\author{Chiara Cappelli}
\email{chiara.cappelli@sns.it}
\affiliation{Scuola Normale Superiore,
             Piazza dei Cavalieri 7, 56126 Pisa, Italy.}

\date{\today}% It is always \today, today,

\begin{abstract}
Fully polarizable QM/MM approach based on fluctuating charges and fluctuating dipoles, named QM/FQF$\mu$ (J. Chem. Theory Comput., \textbf{15}, 2233 (2019)), is extended to the calculation of vertical excitation energies of solvated molecular systems. Excitation energies are defined within two different solvation regimes, i.e. linear response (LR), where the response of the MM portion is adjusted to the QM transition density, and corrected-Linear Response (cLR) in which the MM response is adjusted to the relaxed QM density, thus being able to account for charge equilibration in the excited state. The model, which is specified in terms of three physical parameters (electronegativity, chemical hardness, and polarizability)  is applied to vacuo-to-water solvatochromic shifts of aqueous solutions of para-nitroaniline, pyridine and pyrimidine. The results show a good agreement with their experimental counterparts, thus highlighting the potentialities of this approach.
\end{abstract}

\maketitle

%tommaso: il testa di cazzo del due non capisce quale e' la novita', e la stessa cosa dice Gao. Nonostante sia nel titolo (cLR) secondo me va spiegato meglio e piu' diretto nell'introduzione, nella teoria e nelle conclusioni

\section{Introduction}

Excited-state phenomena play a crucial role in many application fields, as for instance photocatalysis, optical information storage and solar cells. In the past decades, theoretical modeling of excited-state properties of molecules in the gas-phase has become a widespread strategy of investigation,\cite{helgaker2012recent} giving precious information on, for instance, the nature of the electronic excitation,\cite{le2011qualitative,guido2013metric,
savarese2017metrics} nuclei-electron coupling effects\cite{baiardi2016general,bloino2016aiming,improta2009vibronic} and excited state electron dynamics.\cite{barbatti2011nonadiabatic,barbatti2010relaxation,gonzalez2012progress}

However, for electronic phenomena taking place in the condensed phase,\cite{cannelli2017understanding,carlotti2018evaluation,
barone2012implementation,
reichardt1994solvatochromic,
mennucci1998excited,jacquemin2011excited,corni2005electronic,
cupellini2019electronic,mennucci2019multiscale} the interplay between the molecule and its environment can substantially alter the electronic response to external electromagnetic fields. Therefore, any accurate modeling of excited states of solvated systems asks for reliable theoretical approaches to include the effects of the environment at all levels of the excitation phenomenon. 

Most of the currently available approaches to describe the effects of the external environment on molecular properties belong to the class of the so-called focused models;\cite{warshel1976theoretical,warshel1972calculation,
miertuvs1981electrostatic,tomasi1994molecular,orozco2000theoretical,tomasi2005} the attention is focused on the molecule and the environment is treated a lower level of sophistication as it modifies, but not determines, the molecular response to the external radiation. In order to keep the atomistic description of the environment, thus substantially overcoming well-known and amply used continuum solvent descriptions,\cite{tomasi2005,tomasi1994molecular,mennucci2007continuum,mennucci2010continuum,mennucci2013modeling,lipparini2016perspective} multiscale QM/Molecular Mechanics (MM) approaches have been developed.\cite{senn2009qm,lin2007qm} In such models, the molecule (solute) is treated at the QM level, whereas the environment (solvent) is modelled by means of classical MM force fields (FF). The interaction between the QM and MM portions is usually described in terms of electrostatic forces, although approaches to include non-electrostatic QM/MM interactions have been proposed.\cite{gordon1996approximate,jensen1998approximate,
ben1998direct,giovannini2017disrep,curutchet2018density,
giovannini2019eprdisrep,naabo2016embedding,reinholdt2017polarizable} In order to fully capure the physics of solute-solvent interactions,  mutual polarization effects between the QM and MM moieties need to be accounted for. Therefore, several polarizable QM/MM approaches have been proposed, based on distributed multipoles,\cite{day1996effective,kairys2000qm,mao2016assessing,loco2016qm,loco2018modelingijqc} induced dipoles,\cite{thole1981molecular,curutchet2009electronic,olsen2011molecular,
steindal2011excitation,jurinovich2014fenna} Drude oscillators\cite{boulanger2012solvent}, Fluctuating Charges (FQ)\cite{rick1994dynamical,rick1996dynamical,cappelli2016integrated} and the recently developed approach based on both Fluctuating Charges and Fluctuating Dipoles (FQF$\mu$).\cite{giovannini2019fqfmu,giovannini2019fqfmuder2}

The latter can be seen as a refinement of QM/FQ, previously developed by some of us, \cite{giovannini2018hyper,cappelli2016integrated,
lipparini2012linear,lipparini2012analytical,lipparini2013gauge,
giovannini2017polarizable,giovannini2016effective,
giovannini2018effective,egidi2019combined,puglisi2019interplay,giovannini2019tpa} where the MM portion is described by means of electric charges, which vary as a function of differences in MM atomic electronegativities and as a response to the electric potential generated by the QM density. Therefore, in QM/FQ only monopoles, i.e. zeroth order of the electrostatic Taylor expansion, are taken into consideration. This means that the intrinsic anisotropy and out-of-plane contributions of  molecule-environment interactions are not explicitly taken into account. To overcome this problem, the electrostatic description given by  FQ has been refined by including  fluctuating dipoles (F$\mu$), thus resulting in the QM/FQF$\mu$ approach. Notice that similar approaches have been developed in other contexts,\cite{stern1999fluctuating,naserifar2017polarizable,oppenheim2018extension,
mayer2007formulation,jensen2009atomistic,rinkevicius2014hybrid} however they are not specifically intended to model molecular systems/properties in solution. 

The potentialities of QM/FQF$\mu$ at predicting electrostatic interaction energies of organic molecules in solution have been previously highlighted.\cite{giovannini2019fqfmu} In this work, this model is further extended to the calculation of vertical excitation energies, by resorting to linear response (LR) theory for Self Consistent Field (SCF) methods. Excitation energies are determined by solving the so-called Casida equations,\cite{Casida95_155} for the QM portion, appropriately modified so to include extra terms due to QM/MM mutual polarization effects.\cite{lipparini2012linear,loco2016qm,rinkevicius2014hybrid,steindal2011excitation} The LR approach may reliably describe electronic transitions involving excited states associated with large transition dipole moments. However, it lacks any accounting of the relaxation of the classical portion as a response to charge equilibration following the excitation of the QM density. For this reason, several state specific (SS) approaches have been proposed in the literature both in case of continuum models\cite{caricato2006formation,marenich2011practical,marenich2014electronic,budzak2014unveiling,improta2006state,
corni2005electronic,caricato2014corrected,guido2018excited,duchemin2018bethe,schroder2018corrected,guido2019description,guido2017excited,
guido2015electronic} and polarizable QM/MM approaches.\cite{zeng2015analytic,loco2016qm}
%tommaso: aggiunto SS dopo perturbative come richiesto dal ref 2
In this work, the perturbative SS corrected linear response (cLR) scheme\cite{caricato2006formation} originally developed for the Polarizable Continuum Model (PCM) is exploited and extended to both QM/FQ and QM/FQF$\mu$ transition energies for the first time. In this way, the classical MM portion is adjusted to the relaxed density matrix of a given QM excited state. Notice that a similar approach has been applied to polarizable QM/AMOEBA calculations\cite{loco2016qm}.
 
To show both the potentialities of QM/FQF$\mu$ at describing excitation energies and the differences arising from resorting to LR and cLR regimes, the method is applied to the calculation of vacuo-to-water solvatochromic shifts of para-nitroaniline (pNA), pyridine and pyrimidine. %, three molecules between those recently studied by Marenich \emph{et al.},\cite{marenich2014electronic} i.e. . 
From such an analysis, it will be possible to (i)highlight differences in the description of the excitation phenomenon  moving from continuum PCM to QM/FQ and QM/FQF$\mu$ fully atomistic approaches; (ii) directly quantify the separate role fluctuating dipoles and fluctuating charges, and their coupling; (iii) evaluate effects arising from the specification of LR or cLR regimes.

The manuscript is organized as follows. In the next section, the FQF$\mu$ force field is presented and its coupling with a QM description at the SCF level (QM/FQF$\mu$) is detailed. LR and cLR working equations are then presented and discussed. After a brief section discussing on the computational protocol which is adopted, numerical results are presented. Some drawn conclusions and a discussion on the future perspectives of the approach end the manuscript.

\section{Theory}

%tommaso: modifico questa sezione come richiesto dal ref1 : va scorciato
\subsection{QM/FQF$\mu$ Approach}

In the FQF$\mu$ force field each MM atom is endowed with both a charge ($q$) and an atomic dipole ($\bm{\mu}$), that are not fixed but can vary according to the external electric potential/ electric field. 

%Both charges and dipoles are described as s-type gaussian distribution functions, $\rho_{q}$ and $\rho_{\bm{\mu}}$, defined as:

%\begin{align}
%\rho_{q_i}(\gr{r}) & = \frac{q_i}{\pi^{\frac{3}{2}} R_{q_i}^3} \exp \left( -\frac{\abs{\gr{r}-\gr{r}_i}^2}{R_q^2}\right) \nonumber \\
%\rho_{\bm{\mu}_i}(\gr{r}) & =\frac{\abs{\bm{\mu}_{i}}}{\pi^{\frac{3}{2}}R_{\bm{\mu}_i}^{3}} \hat{\gr{n}}_i\cdot\boldsymbol{\nabla} \left[\exp \left(-\frac{\left|\gr{r}-\gr{r}_{i}\right|^{2}}{R_{\bm{\mu}_i}^{2}} \right)\right]
%\label{eq:distributioni}
%\end{align}

%where $R_{q_i}$ and $R_{\bm{\mu}_i}$ represent the width of the distributions $\rho_{q_i}$ and $\rho_{\bm{\mu}_i}$, respectively.  $\hat{\gr{n}}_{i}$ is a unit vector pointing to the dipole direction $\bm{\mu}_{i}$.

The total energy $\mathcal{E}$ associated with a gaussian distribution of charges and dipoles reads: \cite{mayer2007formulation}

\begin{align}
\mathcal{E}(\mathbf{q},\bm{\mu}) & = \sum_{i}q_{i}\chi_{i} +  \frac{1}{2}
\sum_{i}q_{i}\eta_{i}q_{i} + \frac{1}{2}\sum_{i}\sum_{j\neq i} q_{i}\text{T}_{ij}^{qq} q_{j} + \nonumber \\ 
& + \sum_{i}\sum_{j\neq i}q_{i}\gr{T}_{ij}^{q\mu}\bm{\mu}_{j} 
 + \frac{1}{2}\sum_{i}\sum_{j\neq i}\bm{\mu}_{i}^{\dagger}\mathbf{T}_{ij}^{\mu\mu}\bm{\mu}_{j} - \frac{1}{2}\sum_{i}\bm{\mu}_{i}^{\dagger}\alpha_{i}^{-1}\bm{\mu}_{i} 
\label{eq:energy-iniziale}
\end{align}

where $\chi$ is the atomic electronegativity, $\eta$ the chemical hardness and $\alpha$ the atomic polarizability. $\text{T}_{ij}^{qq}$, $\mathbf{T}_{ij}^{q\mu}$ and $\mathbf{T}_{ij}^{\mu\mu}$ are the charge-charge, charge-dipole and dipole-dipole interaction kernels, respectively. Their functional form can be found in Refs.\citenum{mayer2007formulation,giovannini2019fqfmu}. 

In order to collect all the charge quadratic terms, the diagonal elements of $\gr{T}^{qq}$ and $\gr{T}^{\mu\mu}$ can be imposed to be related to atomic chemical hardness $\eta$ and atomic polarizability $\alpha$, respectively.\cite{giovannini2019fqfmu} Therefore, the parameters entering the definition of FQF$\mu$ are limited to electronegativity, chemical hardness and polarizability of each atom type. As a result, Eq. \ref{eq:energy-iniziale} can be formally rewritten as:

\begin{align}
\mathcal{E}(\mathbf{q},\bm{\mu}) & = \frac{1}{2}\sum_{i}\sum_{j} q_{i}{\text{T}_{ij}^{qq}}q_{j} +\frac{1}{2}\sum_{i}\sum_{j}\bm{\mu}_{i}^{\dagger}\mathbf{T}_{ij}^{\mu\mu}\bm{\mu}_{j}  + \nonumber \\
& + \sum_{i}\sum_{j}q_{i}\textbf{T}_{ij}^{q\mu}\bm{\mu}_{j}^{\dagger}  + \sum_{i}q_{i}\chi_{i} = \nonumber \\
& = \frac{1}{2}\mathbf{q}^{\dagger}\mathbf{T}^{qq}\mathbf{q} + \frac{1}{2}\bm{\mu}^{\dagger}\mathbf{T}^{\mu\mu}\bm{\mu} + \mathbf{q}^{\dagger}\mathbf{T}^{q\mu}\bm{\mu}+\bm{\chi}^{\dagger}\mathbf{q}
\label{eq:energiaMM_mat}
\end{align}

where a vector notation has been adopted. Similarly to QM/FQ, the equilibrium condition of Eq. \ref{eq:energiaMM_mat} is reached when the Electronegativity Equalization Principle (EEP) is satisfied, i.e. when each atom has the same electronegativity.\cite{sanderson1951} If each MM moiety is constrained to assume a fixed, total charge
Q$_\alpha$, Eq. \ref{eq:energiaMM_mat} can be written by exploiting a set of Lagrangian multipliers ($\lambda_\alpha$), whose number is equal to the total number of moieties in the MM portion:

\begin{align}
\mathcal{E}\left(\mathbf{q},\bm{\mu},\bm{\lambda}\right) & = E\left(\mathbf{q},\bm{\mu}\right) + \sum_{\alpha}\left[\lambda_{\alpha} \sum_{i} \left(q_{\alpha i}\right)-Q_{\alpha}\right] = \nonumber \\
& =\frac{1}{2}\sum_{i\alpha}\sum_{j\beta} q_{i\alpha}{\text{T}_{i\alpha,j\beta}^{qq}}q_{j\beta} +\frac{1}{2}\sum_{i}\sum_{j}\bm{\mu}_{i\alpha}^{\dagger}\mathbf{T}_{i\alpha,j\beta}^{\mu\mu}\bm{\mu}_{j\beta} 
 + \nonumber \\
& + \sum_{i}\sum_{j}q_{i\alpha}\textbf{T}_{i\alpha,j\beta}^{q\mu}\bm{\mu}_{j\beta}^{\dagger} + \sum_{i\alpha}q_{i\alpha}\chi_{i\alpha} + \nonumber \\ 
& + \sum_{\alpha}\lambda_{\alpha}\left[\sum_{i} q_{\alpha i}-Q_{\alpha}\right] = \nonumber \\
& = \frac{1}{2}\mathbf{q}^{\dagger}\mathbf{T}^{qq}\mathbf{q} + \frac{1}{2}\bm{\mu}^{\dagger}\mathbf{T}^{\mu\mu}\bm{\mu} + \mathbf{q}^{\dagger}\mathbf{T}^{q\mu}\bm{\mu}+\bm{\chi}^{\dagger}\mathbf{q} + \bm{\lambda}^\dagger \gr{q}  
\label{eq:energia_finale_MM}
\end{align}

where $\alpha$ and $\beta$ run over MM moieties. % and the constraints $\lambda_{\alpha}$ are meant to preserve the total charge $Q_\alpha$ of every molecule. 
By minimizing $\mathcal{E}$ with respect all variables, the following linear system is defined:\cite{giovannini2019fqfmu}

\begin{equation}
\left(
\begin{array}{ccc}
\mathbf{T}^{qq} & \mathbf{1}_{\bm{\lambda}} & \mathbf{T}^{q\mu} \\ 
\mathbf{1}^{\dagger}_{\bm{\lambda}} & \mathbf{0} & \mathbf{0}   \\
-\mathbf{T}^{q\mu^{\dagger}} & \mathbf{0} & \mathbf{T}^{\mu\mu}
\end{array}
\right)
\left({\begin{array}{c} 
\mathbf{q}\\
\bm{\lambda}\\ 
\bm{\mu} 
\end{array}}\right)
=
\left(\begin{array}{c} -\bm{\chi} \\ 
\mathbf{Q} \\
\mathbf{0}
\end{array}\right)
 \Rightarrow 
\mathbf{D}\mathbf{L}_{\lambda} = -\mathbf{C}_Q
\label{eq:MMlinearsys}
\end{equation}

where $\gr{1}_{\bm{\lambda}}$ is a rectangular matrix, which accounts for the Lagrangians. $\mathbf{C}_Q$ is a vector containing atomic electronegativities and total charge constraints, whereas $\mathbf{L}_\lambda$ is a vector containing charges, dipoles and Lagrange multipliers.

FQF$\mu$ FF can be effectively coupled to a QM SCF description, within a QM/MM framework:
%The interaction between the two portions is defined as:

%\begin{equation}
%E_{QM/MM}= \sum_i V[\rho_{QM}](\gr{r}_i){q_i} - \bm{\mu}_i^{\dagger}\mathbf{E}[\rho_{QM}](\gr{r}_i)
%\end{equation}

%where $V[\rho_{QM}](\gr{r}_i)$ and $\mathbf{E}[\rho_{QM}](\gr{r}_i)$ are the electric potential and electric field, respectively, calculated at the $i$-th charge and $i$-th dipole placed at $\gr{r}_i$.
%Therefore, the global QM/MM energy functional reads:
 
\begin{align}
\mathcal{E}(\mathbf{P},\mathbf{q},\bm{\mu},\bm{\lambda}) & = \text{tr} \bf{hP}+\frac{1}{2}\text{tr}\mathbf{PG(P)} + \frac{1}{2}\mathbf{q}^{\dagger}\mathbf{T}^{qq}\mathbf{q} + \frac{1}{2}\bm{\mu}^{\dagger}\mathbf{T}^{\mu\mu}\bm{\mu} + \nonumber \\ 
& + \mathbf{q}^{\dagger}\mathbf{T}^{q\mu}\bm{\mu}+\bm{\chi}^{\dagger}\mathbf{q} + \bm{\lambda}^\dagger \gr{q} + \mathbf{q}^{\dagger}\mathbf{V}(\mathbf{P}) - \bm{\mu}^{\dagger}\mathbf{E}(\mathbf{P})
\label{eq:funz_qmmm_fqfmu}
\end{align}

where $\mathbf{h}$ and $\mathbf{G}$ are the usual one- and two-electron matrices, and $\gr{P}$ is the density matrix. $\mathbf{q}^{\dagger}\mathbf{V}(\mathbf{P})$ and $\bm{\mu}^{\dagger}\mathbf{E}(\mathbf{P})$ are the electrostatic couplings between the QM density and the FQs and F$\mu$s, respectively.
The effective Fock matrix, i.e. the derivative of the energy with respect to the density matrix, reads:

\begin{equation}
 \label{eq:Fock_fqfmu}
 \tilde{F}_{\mu\nu} = \parz{\mathcal{E}}{P_{\mu\nu}} = h_{\mu\nu} + G_{\mu\nu}(\mathbf{P}) + \mathbf{V}^{\dagger}_{\mu\nu}\mathbf{q} - \mathbf{E}_{\mu\nu}^{\dagger}\bm{\mu}
\end{equation}

Charges and dipoles are obtained by imposing the global functional to be stationary with respect to charges, dipoles and Lagrangian multipliers:
\begin{align}
 \label{eq:QMlinearsystem}
\left(
\begin{array}{ccc}
\mathbf{T}^{qq} & \mathbf{1}_{\bm{\lambda}} & \mathbf{T}^{q\mu} \\ 
\mathbf{1}^{\dagger}_{\bm{\lambda}} & \mathbf{0} & \mathbf{0}   \\
-\mathbf{T}^{q\mu^{\dagger}} & \mathbf{0} & \mathbf{T}^{\mu\mu}
\end{array}
\right)
\left({\begin{array}{c} 
\mathbf{q}\\
\bm{\lambda}\\ 
\bm{\mu} 
\end{array}}\right)
& =
\left(\begin{array}{c} 
-\bm{\chi} \\ 
\textbf{Q}_{\text{tot}} \\
\mathbf{0}
\end{array}\right)
+
\left(\begin{array}{c}
-\gr{V}(\gr{P}) \\
\textbf{0} \\
\gr{E}(\gr{P})
\end{array}
\right) \nonumber \\
\nonumber \\
\mathbf{D}\mathbf{L}_{\lambda} & = -\mathbf{C}_Q - \mathbf{R}(\mathbf{P})
\end{align}

Notice that, with respect to Eq. \ref{eq:MMlinearsys}, a new term, $\mathbf{R}(\mathbf{P}$), appears,  which collects QM polarization sources. $\mathbf{L}_{\lambda}$ is the vector containing charges, dipoles and Lagrangian multipliers.

\subsection{QM/FQF$\mu$  Electronic Transition Energies}

\subsubsection{Linear Response Regime}

LR for polarizable QM/MM approaches has been already discussed in the previous literature.\cite{lipparini2012linear,rinkevicius2014hybrid,steindal2011excitation} 
Electronic excitation energies $\omega$  for SCF Hamiltonians can be obtained by solving Casida's equations:\cite{Casida95_155}

\begin{equation}
 \label{eq-fqfmu:5-FQCasida}
\left (
 \begin{array}{cc}
  \tilde{\mathbf{A}} & \tilde{\mathbf{B}} \\
  \tilde{\mathbf{B}}^* & \tilde{\mathbf{A}}^* \\
 \end{array}
\right )
\left (
 \begin{array}{c}
  \mathbf{X} \\
  \mathbf{Y} \\
 \end{array}
\right )
=
\omega
\left (
 \begin{array}{cc}
  \bm{1}  & \bm{0} \\
  \bm{0}  & -\bm{1} \\
 \end{array}
\right )
\left (
 \begin{array}{c}
  \mathbf{X} \\
  \mathbf{Y} \\
 \end{array}
\right )
\end{equation}

$\tilde{A}$ and $\tilde{B}$ matrices are defined as:

\begin{equation}
 \label{eq-fqfmu:5-Amat}
 \tilde{A}_{ai,bj} = (\epsilon_a - \epsilon_i)\delta_{ab}\delta_{ij} + (ai|bj) - c_x (ab|ij) + c_l f^{xc}_{ai,bj} + C^{pol}_{ai,bj}
\end{equation}
\begin{equation}
 \label{eq-fqfmu:5-Bmat}
 \tilde{B}_{ai,bj} = (ai|bj) - c_x(aj|ib) + C^{pol}_{ai,bj}
\end{equation}

where $(pq|rs)$ are two electron integrals, and $\varepsilon$ are molecular orbital (MO) energies. $c_x$ and $c_l$ are coefficients of which the definition depends on the SCF level adopted ($c_x$ =1, $c_l$ = 0 for HF wavefunctions, $c_x$ =0, $c_l$ = 1 for  DFT). 

In case of polarizable QM/MM approaches, $\tilde{A}$ and $\tilde{B}$ in Eqs. \ref{eq-fqfmu:5-Amat} and \ref{eq-fqfmu:5-Bmat} account for an extra term,  $C^{pol}$, which is specified according to a particular method. In particular, for QM/FQ it reads:

\begin{equation}
\label{eq:Cpol_FQ}
C^{FQ}_{ai,bj} = \sum^{N_q}_{P} \left(\int_{\mathbb{R}^3} \phi_a(\mathbf{r}) \frac{1}{\abs{\mathbf{r}-\mathbf{r}_p}} \phi_i(\mathbf{r}) \diff \mathbf{r} \right) \cdot q^T_{P} (\phi_b,\phi_i)
\end{equation}

where, $q^T$ are the perturbed fluctuating charges adjusted to the transition density $\mathbf{P}^T_K = \mathbf{X}_K + \mathbf{Y}_K$.\cite{lipparini2012linear}
For QM/FQF$\mu$, an extra term appears n the definition of $C^{pol}$, accounting for the presence of fluctuating dipoles:

\begin{align}
\label{eq:Cpol_FQFMu}
C^{FQF\mu}_{ai,bj} & = \sum^{N_q}_{P} \left(\int_{\mathbb{R}^3} \phi_a(\mathbf{r}) \frac{1}{\abs{\mathbf{r}-\mathbf{r}_p}} \phi_i(\mathbf{r}) \diff \mathbf{r} \right) \cdot q^T_{P} (\phi_b,\phi_i) + \nonumber \\ & - \sum^{N_\mu}_{P} \left(\int_{\mathbb{R}^3} \phi_a(\mathbf{r}) \frac{(\mathbf{r}-\mathbf{r}_p)}{\abs{\mathbf{r}-\mathbf{r}_p}^3} \phi_i(\mathbf{r}) \diff \mathbf{r} \right) \cdot \bm{\mu}^T_{P} (\phi_b,\phi_i)
\end{align}

%In QM/FQF$\mu$, $C^{pol}$ depends on both fluctuating charges ($q^T$) and dipoles ($\bm{\mu}^T$) induced by the transition density. 
Perturbed charges ($q^T$) and perturbed dipoles ($\bm{\mu}^T$) are calculated by solving the following system of equations:

\begin{align}
 \label{eq:pert-qmu}
\left(
\begin{array}{ccc}
\mathbf{T}^{qq} & \mathbf{1}_{\bm{\lambda}} & \mathbf{T}^{q\mu} \\ 
\mathbf{1}^{\dagger}_{\bm{\lambda}} & \mathbf{0} & \mathbf{0}   \\
-\mathbf{T}^{q\mu^{\dagger}} & \mathbf{0} & \mathbf{T}^{\mu\mu}
\end{array}
\right)
\left({\begin{array}{c} 
\mathbf{q}\\
\bm{\lambda}\\ 
\bm{\mu} 
\end{array}}\right)
& =
\left(\begin{array}{c}
-\gr{V}(\gr{P}^T_K) \\
\textbf{0} \\
\gr{E}(\gr{P}^T_K)
\end{array}
\right) \nonumber \\
\nonumber \\
\mathbf{D}\mathbf{L}^T_{\lambda} & = - \mathbf{R}(\mathbf{P}^T_K)
\end{align}

where:

\begin{align}
V(\mathbf{P}^T_K) & = -\sum_{ai} \mathbf{P}^T_{K,ai} \int_{R^3} \phi_a(\mathbf{r}) \frac{1}{\abs{\mathbf{r}-\mathbf{r}_p}} \phi_i(\mathbf{r}) \diff \mathbf{r} \\
\textbf{E}(\mathbf{P}^T_K) & = \sum_{ai} \mathbf{P}^T_{K,ai} \int_{R^3} \phi_a(\mathbf{r}) \frac{(\mathbf{r}-\mathbf{r}_p)}{\abs{\mathbf{r}-\mathbf{r}_p}^3} \phi_i(\mathbf{r}) \diff \mathbf{r} 
\end{align}

The right hand side of Eq. \ref{eq:pert-qmu} contains both the electric potential and field due to the perturbed density matrix $\mathbf{P}^T_K$. Notably, the constant term in Eq. \ref{eq:QMlinearsystem} due to atomic electronegativities vanishes. Therefore, polarization arises from the QM density, and not from differences in electronegativity, that similarly to the QM/FQ approach.\cite{lipparini2012linear,giovannini2019simulating}

\subsubsection{Corrected Linear Response Regime}

%tommaso: tolta seconda definizione di SS
LR and SS approaches provide the same result in case of gas-phase systems.\cite{Casida95_155} However, when polarizable QM/classical methods are exploited, the two approaches differ.\cite{corni2005electronic} This is essentially due to the non-linear term introduced in the Hamiltonian (see e.g. Eq. \ref{eq:funz_qmmm_fqfmu}). Different approximations to solve this problem have been proposed.\cite{caricato2006formation,improta2006state,marenich2011practical}%, such as the so-called corrected Linear Response (cLR),\cite{caricato2006formation} state-specific (SS) approaches\cite{improta2006state} and the variational Vertical Excitation Method (VEM) for QM/continuum approaches.\cite{marenich2011practical}
%tommaso:aggiunto SS 
In this work, the SS cLR approach, firstly proposed in for QM/PCM Hamiltoniands and successfully extended to the polarizable QM/AMOEBA method\cite{loco2016qm}, is exploited. %tommaso: la metto in positivo piuttosto che in negativo
%To the best of our knowledge, the extension of cLR to QM/FQ and QM/FQF$\mu$-like methods has never been proposed before.
To the best of our knowledge, this the first time that the extension of cLR to QM/FQ and QM/FQF$\mu$-like methods is proposed.

The common way to discuss the physical differences between LR and cLR regimes in condensed phase is to resort to a two-step protocol, in which the first step is the same in both approaches and consists in the electronic excitation of the solute to the $K$-th excited state by keeping frozen the MM response to the solute ground state. From a computational point of view, such a picture can be achieved by imposing the extra $C^{pol}$ term in Eqs. \ref{eq-fqfmu:5-Amat} and \ref{eq-fqfmu:5-Bmat} to be equal to zero. This means that MM contributions are only included in the specification of MOs through Eq. \ref{eq:Fock_fqfmu}. The excitation energy obtained in this approximation is refereed to as $\omega_0^K$. 
In the second step of the process, MM electronic degrees of freedom adjust to the $K$-th excited state density. Due to the fact that in the LR approach the MM variables are determined as a response of the whole transition densities, LR is unable to catch the energy differences associated to the relaxation of the QM electron density, whereas it only accounts for a correction due to dynamic solute-environment interactions. 
QM/FQ and QM/FQF$\mu$ cLR excitation energies from ground to $K$-th excited state can be written as: 

\begin{equation}
\label{eq:clr-fq}
\omega^{cLR}_{K,FQ} = \omega^0_K + \frac{1}{2} \sum^{Nq}_P q_p (\mathbf{r}_p;\mathbf{P}^{\Delta}_K)V(\mathbf{r}_p,\mathbf{P}^{\Delta}_K)
\end{equation}

%whereas, QM/FQF$\mu$ one as:

\begin{align}
\label{eq:clr-fqfmu}
\omega^{cLR}_{K,FQF\mu} = \omega^0_K & + \frac{1}{2} \sum^{Nq}_P q_p (\mathbf{r}_p;\mathbf{P}^{\Delta}_K)V(\mathbf{r}_p,\mathbf{P}^{\Delta}_K) + \nonumber \\
& - \frac{1}{2} \sum^{N_\mu}_P \bm{\mu}_p (\mathbf{r}_p;\mathbf{P}^{\Delta}_K)\mathbf{E}(\mathbf{r}_p,\mathbf{P}^{\Delta}_K)
\end{align}

%As stated before, $\omega^0_K$ is the vertical excitation energy from the ground to $K$-th excited state response in the so-called density frozen approximation. 
$\mathbf{P}^{\Delta}_K$ is the so-called relaxed-density matrix, computed through the so-called Z-vector approach, as:

\begin{equation}
\mathbf{P}^{\Delta}_K = \mathbf{P}^T_K + \mathbf{Z}_K
\end{equation}

where $\mathbf{P}_K^T$ is the unrelaxed density matrix adopted in the LR approach, whereas $\mathbf{Z}_K$ is the Z-vector contribution which accounts for orbital relaxations.
Fluctuating Charges and Fluctuating Dipole in Eq. \ref{eq:clr-fqfmu} are calculated as in Eq. \ref{eq:pert-qmu}, by substituting $\mathbf{P}^T_K \rightarrow \mathbf{P}^{\Delta}_K$.

\section{Computational Details}

The following computational protocol was exploited:\cite{giovannini2018effective,giovannini2019simulating,giovannini2018hyper,giovannini2019eprdisrep}

\begin{enumerate}
\item \textit{Definition of the system:} The solute (pNA, pyridine and pyrimidine) was put at the center of a cubic box containing a number of water molecules large enough to account for solute-solvent interactions (vide infra). pNA, pyridine and pyrimidine geometries were optimized at the B3LYP/aug-cc-pVDZ level of theory, and RESP charges were calculated by including aqueous solvent effects by means of the PCM approach.\cite{tomasi2005}.
\item \textit{Classical Molecular Dynamics (MD) runs:} For each system, MD simulations were performed by imposing periodic boundary conditions (PBC) on the cubic box. Each MD run was preceded by an equilibration step. From each MD run, a set of 100 uncorrelated snapshots was extracted. 
\item \textit{Definition of the the two-layer scheme:} For each snapshot, a solute-centered sphere was cut. The radius of the sphere was chosen so to describe
all specific water-solute interactions (vide infra).
\item \textit{QM/MM calculations:} QM/FQ and QM/FQF$\mu$ vertical excitation energies were calculated on the spherical frames obtained at the previous step. The results obtained for each frame were convoluted with a gaussian band-shape and averaged to produce the final spectrum:
%tommaso: ref1 --> vuole questa equazione? non so proprio cosa voglia senno..
\begin{equation}
\varepsilon_i(\tilde{\nu}) = \frac{\sqrt{\pi}e^2N_A}{1000 \ln(10) c^2 m_e} \frac{f_i}{\sigma} \exp \left[-\left( \frac{\tilde{\nu} - \tilde{\nu}_i}{\sigma}\right)^2 \right]
\end{equation}
where, $\varepsilon$ is the molar absorptivity in L mol$^{-1}$cm$^{-1}$, $\tilde{\nu}_i$ is the excitation energy of the $i-th$ state, $f_i$ is its oscillator strength, whereas $\sigma$ is the standard deviation of the gaussian convolution. $e$, $N_A$, $c$ and $m_e$ are the electron charge, the Avogadro number, the speed of light, and the electron mass, respectively.
\end{enumerate}

\begin{figure}
\centering
\subfloat[][\gr{a) pNA}]{\includegraphics[width=.12\textwidth]{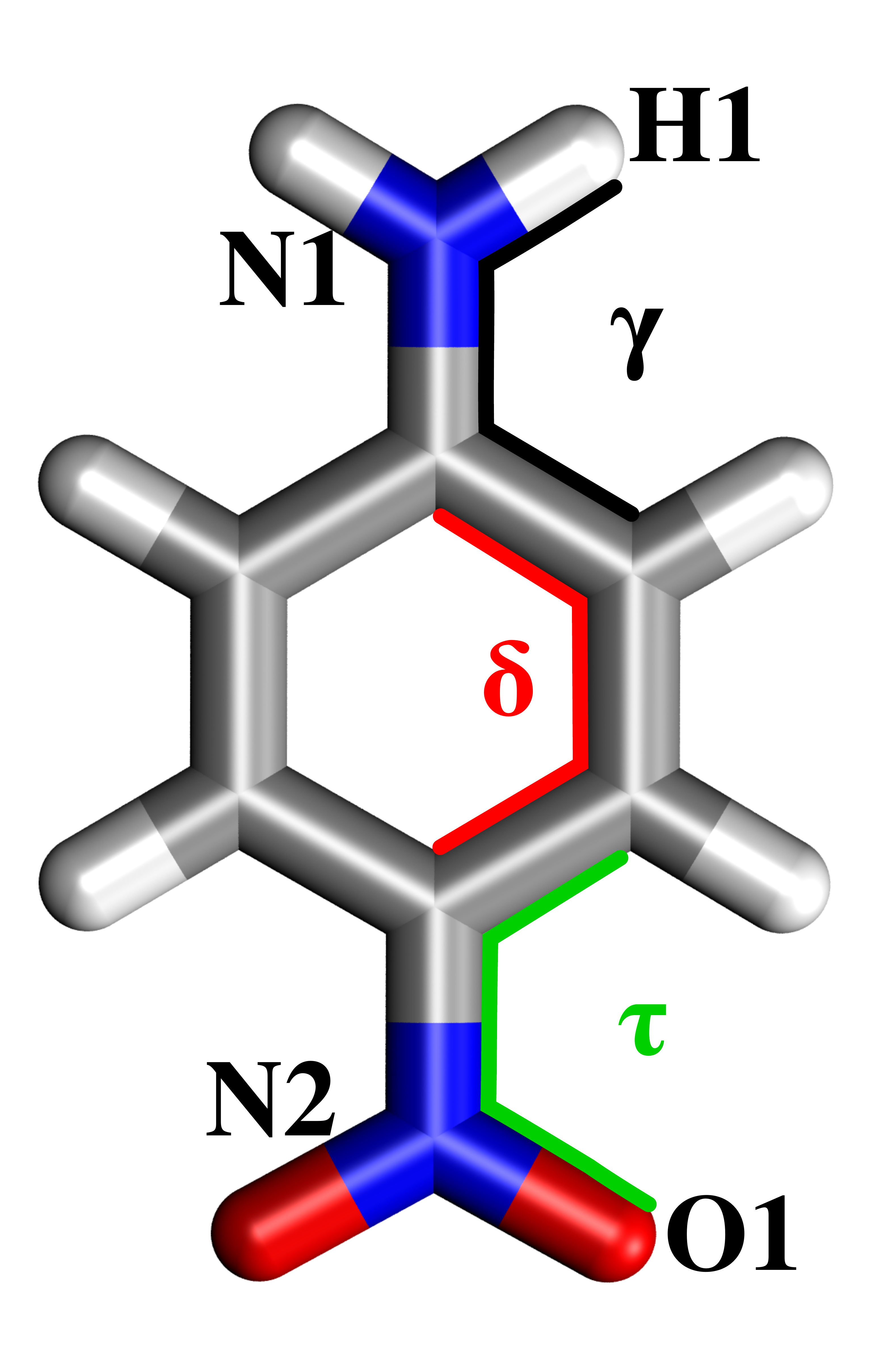}}\qquad
\subfloat[][\gr{b) Pyridine}]{\includegraphics[width=.12\textwidth]{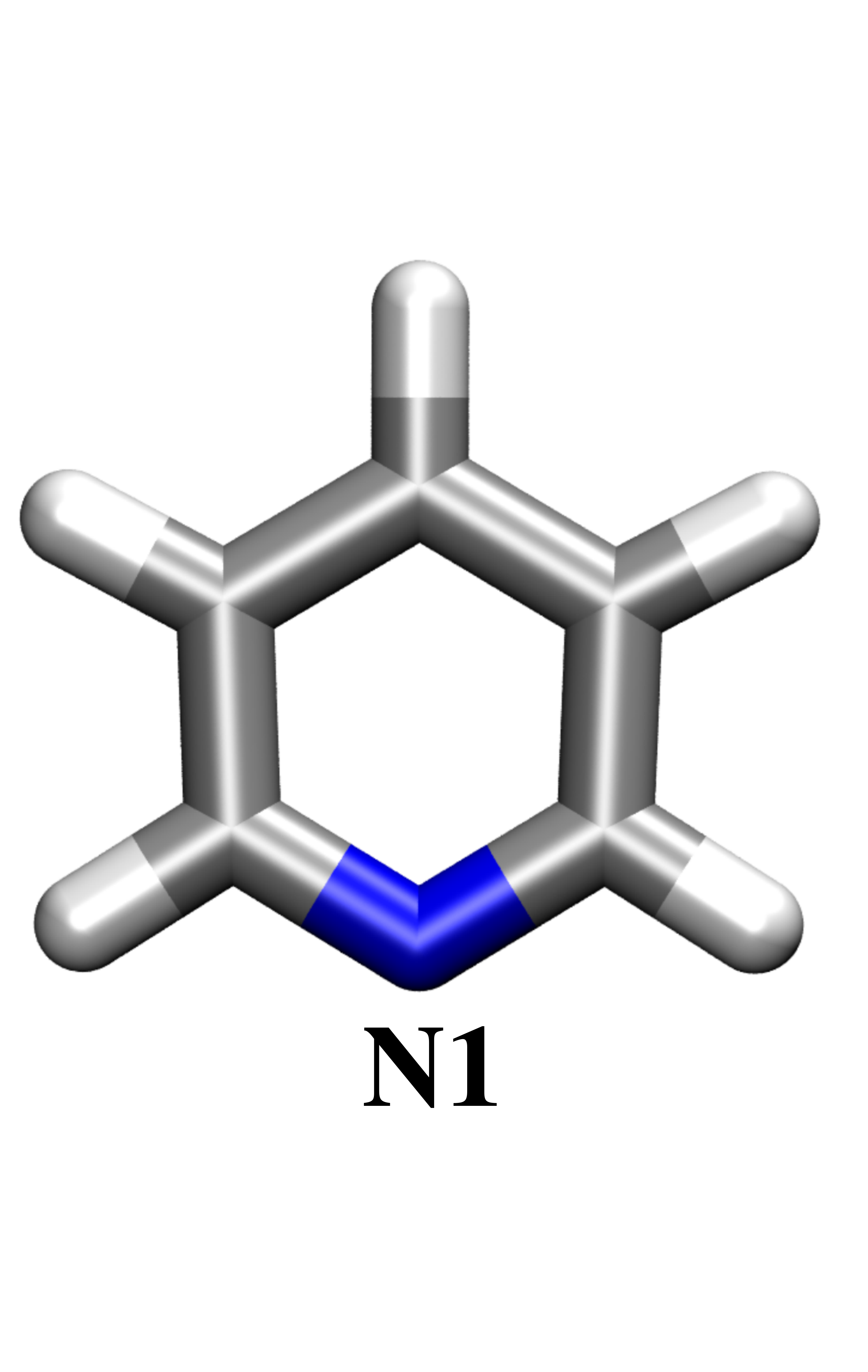}}\qquad
\subfloat[][\gr{c)Pyrimidine}]{\includegraphics[width=.12\textwidth]{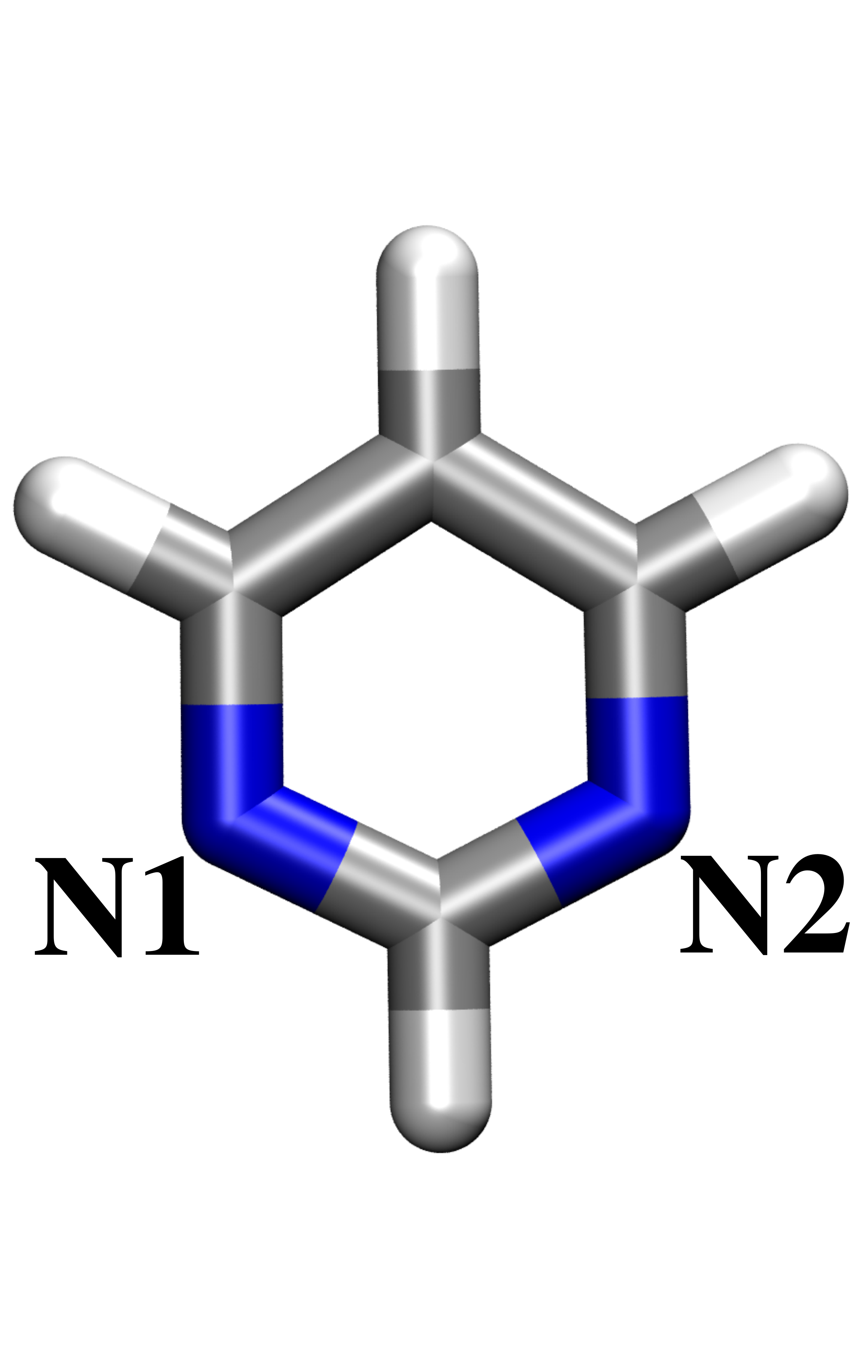}}
\caption{pNA (\gr{a}), Pyridine (\gr{b}) and Pyrimidine (\gr{c}) molecular structures and atom labelling. For pNA, the atoms involved in the definition of $\gamma$, $\delta$ and $\tau$ dihedral angles are highlighted.}
\label{fig:molecole}
\end{figure}

MD simulations were performed using GROMACS\cite{Gromacs5}, by exploiting GAFF \cite{gromos} to describe both intra- and inter-molecular interactions. RESP charges were used to account for electrostatic interactions, and TIP3P FF was used to describe water molecules \cite{mark2001structure}. A single molecule of different solutes was dissolved in a cubic box containing 5781, 5342 and 5324 water molecules in case of pNA, pyridine and pyrimidine, respectively. 
Pyridine and pyrimidine structures were kept fixed during all the steps of the MD run because they exhibit a single conformer in aqueous solution. PNA geometry was kept fixed during the equilibration step only, whereas it was free to move in the production run. The different solutes were brought to 0 K with the steepest descent minimization procedure and  then heated to 298.15 K in an NVT ensemble using the velocity-rescaling\cite{vrescale} method with an integration time step of 1.0 fs and a coupling constant of 0.1 ps for 100 ps. NPT simulations (using the Parrinello-Rahman barostat and a coupling constant of 1.0 ps) for 1 ns were performed to obtain a uniform distribution of molecules in the box. 5 ns production runs in the NVT ensemble were finally carried out, fixing the fastest internal degrees of freedom, i.e. hydrogen atoms, by means of the LINCS algorithm ($\delta$t=2.0 fs)\cite{hess1997lincs}. Electrostatic interactions were treated by using particle-mesh Ewald (PME) \cite{darden1993particle} method with a grid spacing of 1.2 \AA~and a spline interpolation of order 4. Intramolecular interactions between atom pairs separated up to three bonds were excluded. A snapshot every 50 ps was extracted in order to obtain a total of 100 uncorrelated snapshots for each system.

For each snapshot a solute-centered sphere with a radius of 15 \AA~was cut (containing approximately 400 water molecules). On each droplet QM/FQ and QM/FQF$\mu$ vertical excitation energies were calculated by exploiting both LR and cLR regimes. In case of QM/FQ, water molecules were modeled by exploiting polarizable FQ SPC parametrization proposed by \citet{carnimeo2015analytical}, whereas in case of QM/FQF$\mu$ calculations the parametrization proposed by \citet{giovannini2019fqfmu} was adopted.
For the sake of comparison, additional QM calculations on the chromophores treated at the PCM level were performed. PNA was described at the CAM-B3LYP/aug-cc-pVDZ level of theory in agreement with previous studies.\cite{egidi2014stereoelectronic} Pyridine and Pyrimidine were instead treated at the M06/6-311+G(2df,2p) level, according to Ref. \citenum{marenich2014electronic}. All computed absorption spectra were convoluted with a gaussian band-shape with a Full Width at Half Maximum (FWHM) of 0.3 eV. 

All QM/FQ and QM/FQF$\mu$ LR and cLR calculations were performed by using a locally modified version of the Gaussian 16 package\cite{gaussian16}.

\section{Numerical Results}

In this section, computed UV-Vis spectra for  aqueous solutions of pNA, pyridine and pyrimidine\cite{sok2011solvent,kosenkov2010solvent,eriksen2013failures,
sneskov2011scrutinizing,olsen2010excited,eriksen2012importance,frutos2013theoretical,defusco2011modeling,mennucci2002hydrogen,
pagliai2017electronic,biczysko2012integrated,cossi2001time,marenich2011practical,mason1959electronic,millefiori1977electronic,
cai2000low,schreiber2008benchmarks,da2010electronic,prabhumirashi1986solvent,
kovalenko2000femtosecond,de2001monte,silva2010benchmarks} (see Fig. \ref{fig:molecole} for their structures and atom labelling) are discussed. 
First, excitation energies of the three isolated molecules are discussed and compared to experimental values. Then, QM/FQF$\mu$, QM/FQ and QM/PCM LR and cLR solvatochromic shifts are presented
and compared to experimental findings.

\subsection{Gas-Phase Vertical Excitation Energies}

\begin{figure}[htbp!]
\centering
\subfloat[][\gr{a) pNA}]{\includegraphics[width=.16\textwidth]{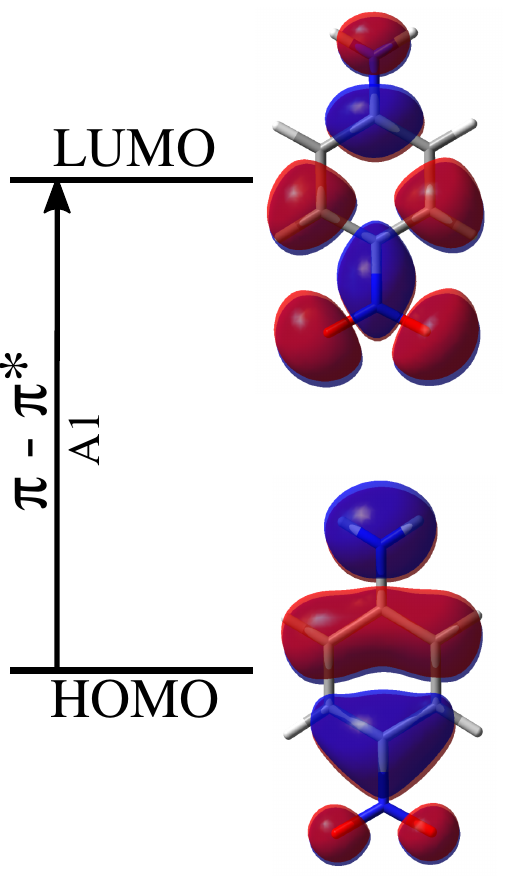}}
\subfloat[][\gr{b) Pyridine}]{\includegraphics[width=.16\textwidth]{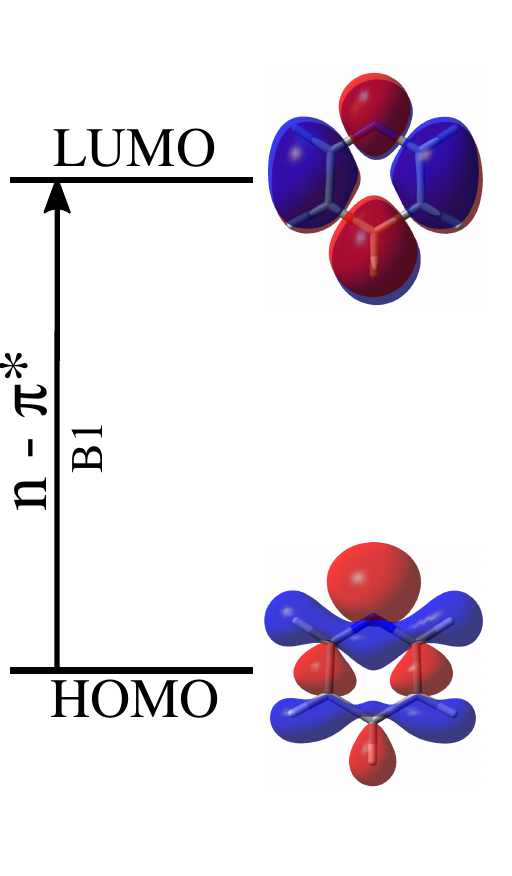}}
\subfloat[][\gr{c)Pyrimidine}]{\includegraphics[width=.16\textwidth]{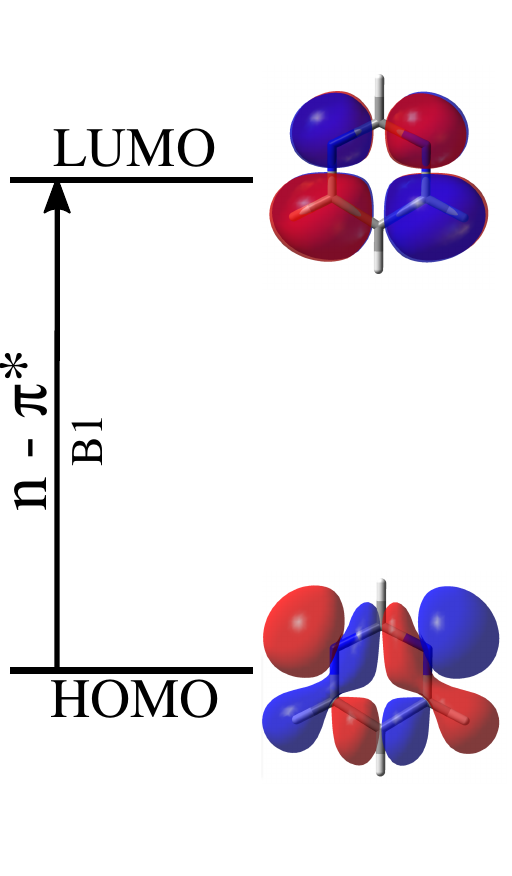}}
\caption{pNA (\gr{a}), Pyridine (\gr{b}) and Pyrimidine (\gr{c}) Molecular Orbitals (MOs) involved in the investigated electronic transitions.}
\label{fig:mos}
\end{figure}

Computed gas phase vertical excitation energies for the three systems are reported in Table \ref{tab:vacuo}, together with experimental values, taken from the literature.
Calculated ground and excited dipole moments are also reported.

\begin{table}[htbp!]
\begin{tabular}{lc|rr|cc}
\hline
Molecule   & Excitation             & $\mu_{g}$    & $\mu_{ex}$   & $E_{vert}$ & Exp. \\
\hline
pNA        & $\pi\rightarrow\pi^*$  & 7.4 & 12.9            &         4.33     &  4.25\text{$\phantom{t}^{\text{a}}$}   \\
pyridine   & $n\rightarrow\pi^*$    & 2.3 & 0.4             &         4.67     &  4.59,\text{$^{\text{b}}$} 4.63,\text{$^{\text{c}}$} 4.74\text{$^{\text{d}}$}    \\
pyrimidine & $n\rightarrow\pi^*$    & 2.4 & 0.6             &         4.17     &  4.18,\text{$^{\text{e}}$} 4.19\text{$^{\text{f}}$}    \\
\hline
\end{tabular}

\text{$\phantom{t}^{\text{a}}$} Ref. \citenum{millefiori1977electronic} \text{$\phantom{t}^{\text{b}}$} Ref. \citenum{schreiber2008benchmarks}  \text{$\phantom{t}^{\text{c}}$} Ref. \citenum{marenich2011practical} \text{$\phantom{t}^{\text{d}}$} Ref. \citenum{cai2000low} \text{$\phantom{t}^{\text{e}}$} Ref. \citenum{da2010electronic} \text{$\phantom{t}^{\text{f}}$} Ref. \citenum{mason1959electronic}.\hfill \\
\caption{pNA (CAM-B3LYP/aug-cc-pVDZ), Pyridine and Pyrimidine (M06/6-311+G(2df,2p) computed vertical excitation energies in gas phase (E$_{vert}$) given in eV. The character of the transition is reported, together with ground ($\mu_g$) and excited ($\mu_{ex}$) molecular dipole moments (Debye). Experimental excitation energies (eV) are also shown.}
\label{tab:vacuo}
\end{table}

The molecular orbitals (MOs) involved in the transitions are graphically depicted in Figure \ref{fig:mos}. The investigated excitation is a pure $\pi \rightarrow \pi^*$ transition in case of pNA, whereas in case of both pyridine and pyrimidine we focus our attention to the dark $n \rightarrow \pi^*$ transition.  pNA $\pi \rightarrow \pi^*$ transition is often considered to have a Charge Transfer (CT) character, due to the fact that it involves a huge variation of the dipole moment between the ground and excited state; this is confirmed by the computed values, see Table \ref{tab:vacuo}. The same also occurs of pyridine and pyrimidine, which differently from pNA show a decrease of the electric dipole moment in their excited states with respect to their ground state.

The comparison between computed and experimental vertical excitation energies in Table \ref{tab:vacuo}, confirms that our choice of DFT functionals/basis sets give computed data that are nicely in agreement with experiments for all systems.

\subsection{MD Analysis}

\begin{figure*}[htbp!]
\centering
\subfloat[][\gr{a) pNA}]{\includegraphics[width=.32\textwidth]{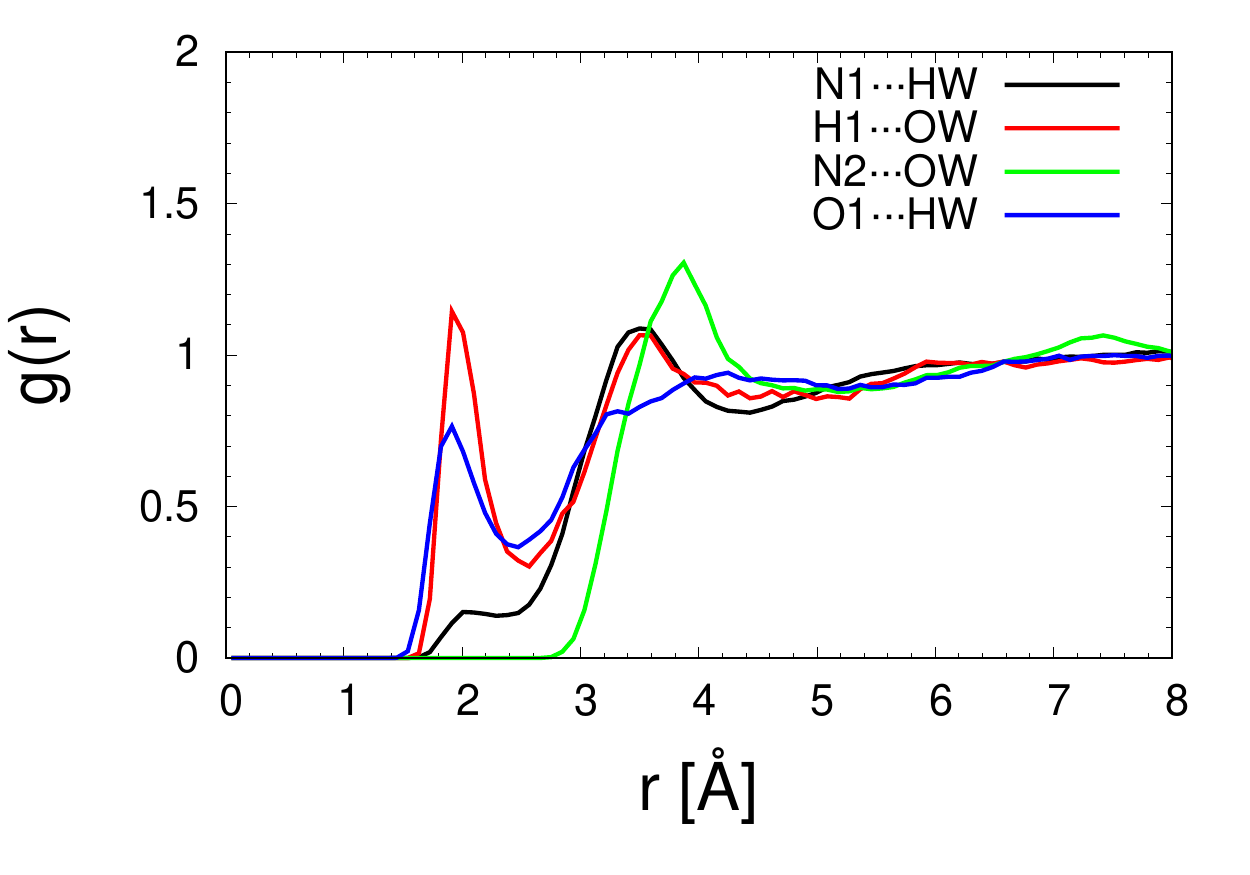}}
\subfloat[][\gr{b) Pyridine}]{\includegraphics[width=.32\textwidth]{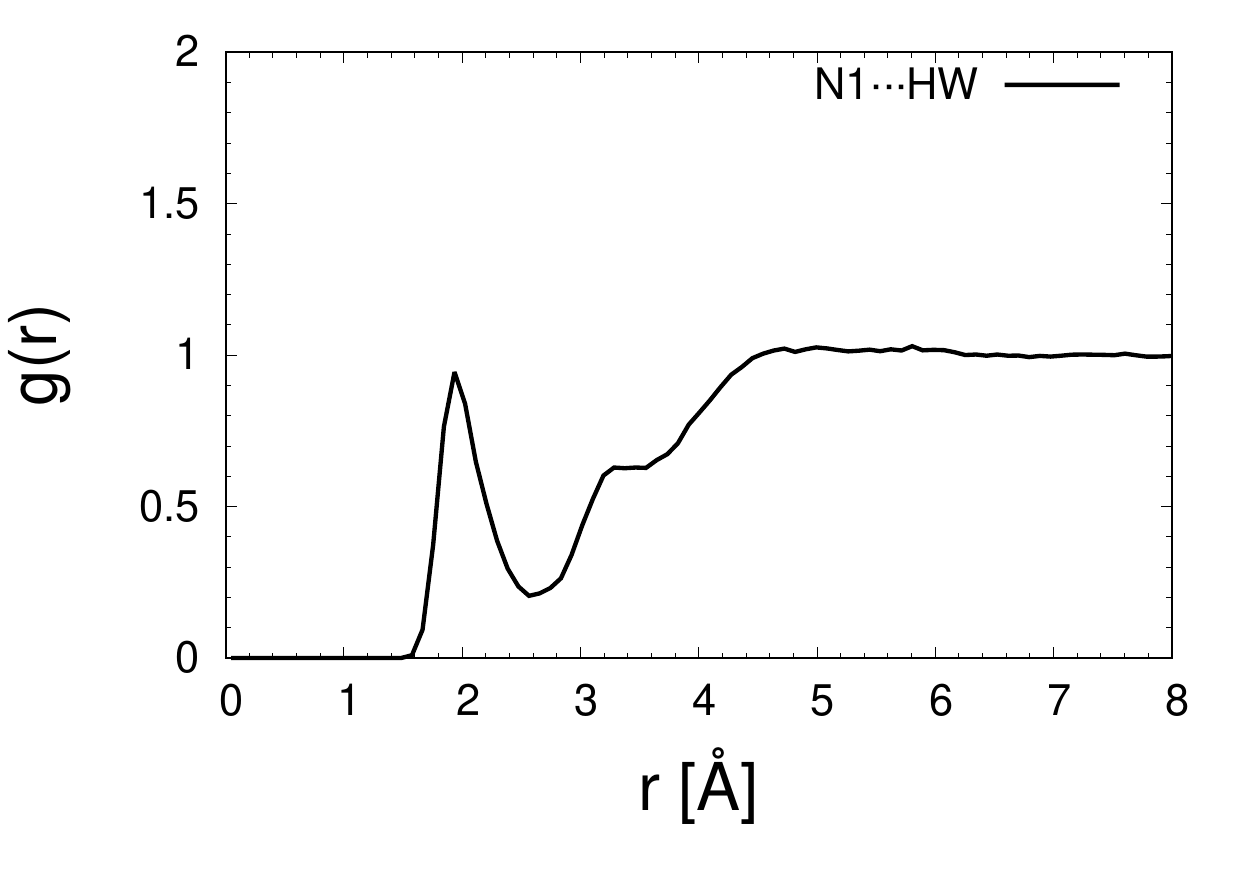}}
\subfloat[][\gr{c) Pyrimidine}]{\includegraphics[width=.32\textwidth]{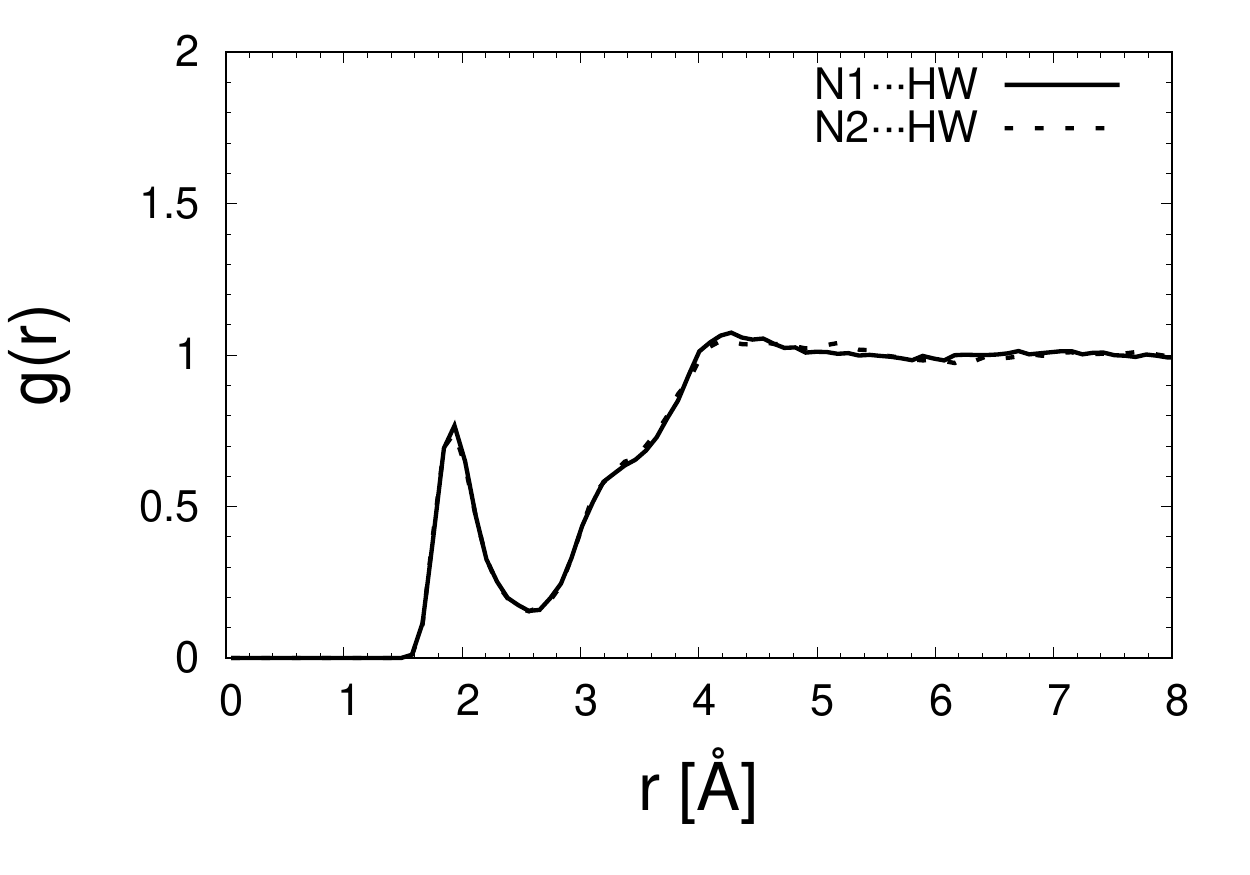}}
\caption{pNA (\gr{a}), Pyridine (\gr{b}) and Pyrimidine (\gr{c}) radial distribution functions for selected atomic sites. See Fig. \ref{fig:molecole} for labeling.}
\label{fig:gdr}
\end{figure*}

\begin{figure*}
\centering
\subfloat[][\gr{a) pNA}]{\includegraphics[width=.32\textwidth]{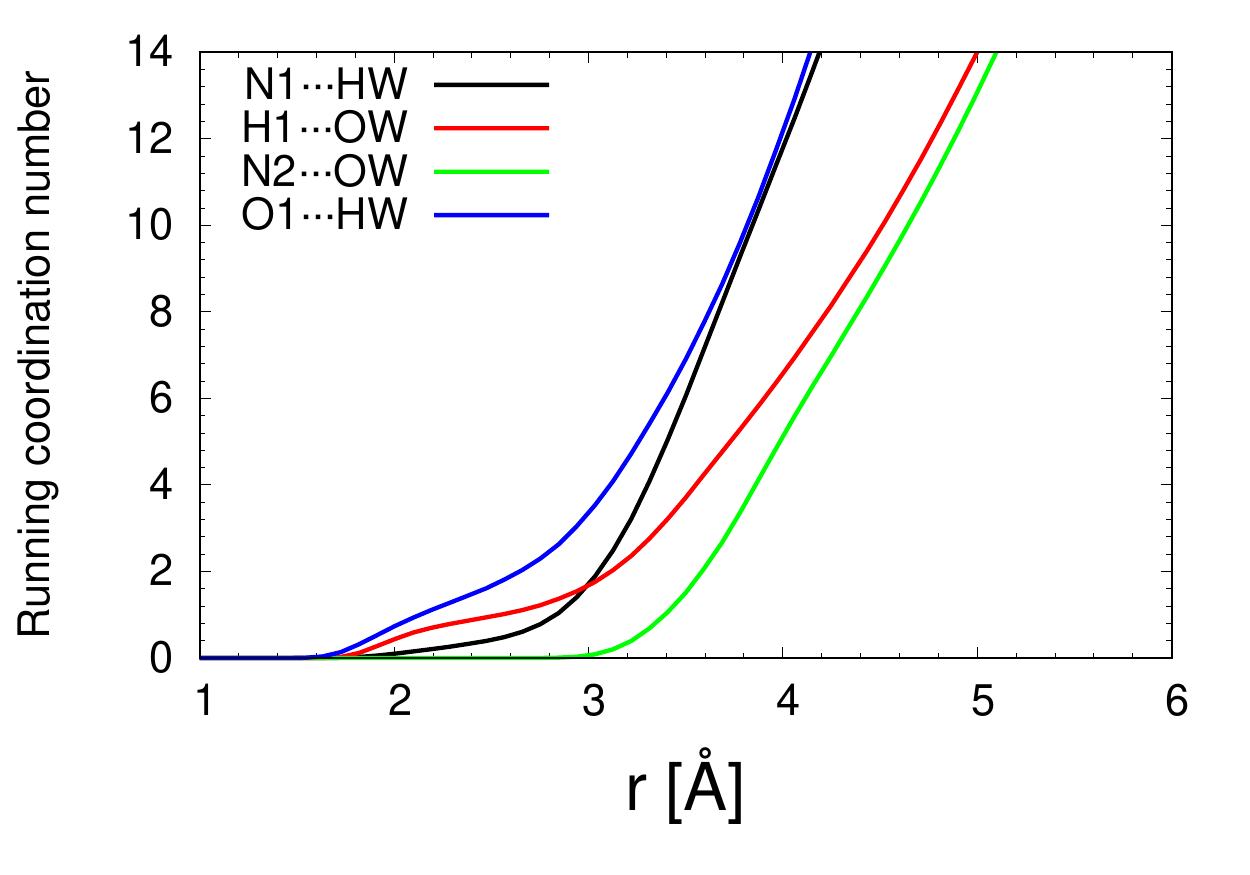}}
\subfloat[][\gr{b) Pyridine}]{\includegraphics[width=.32\textwidth]{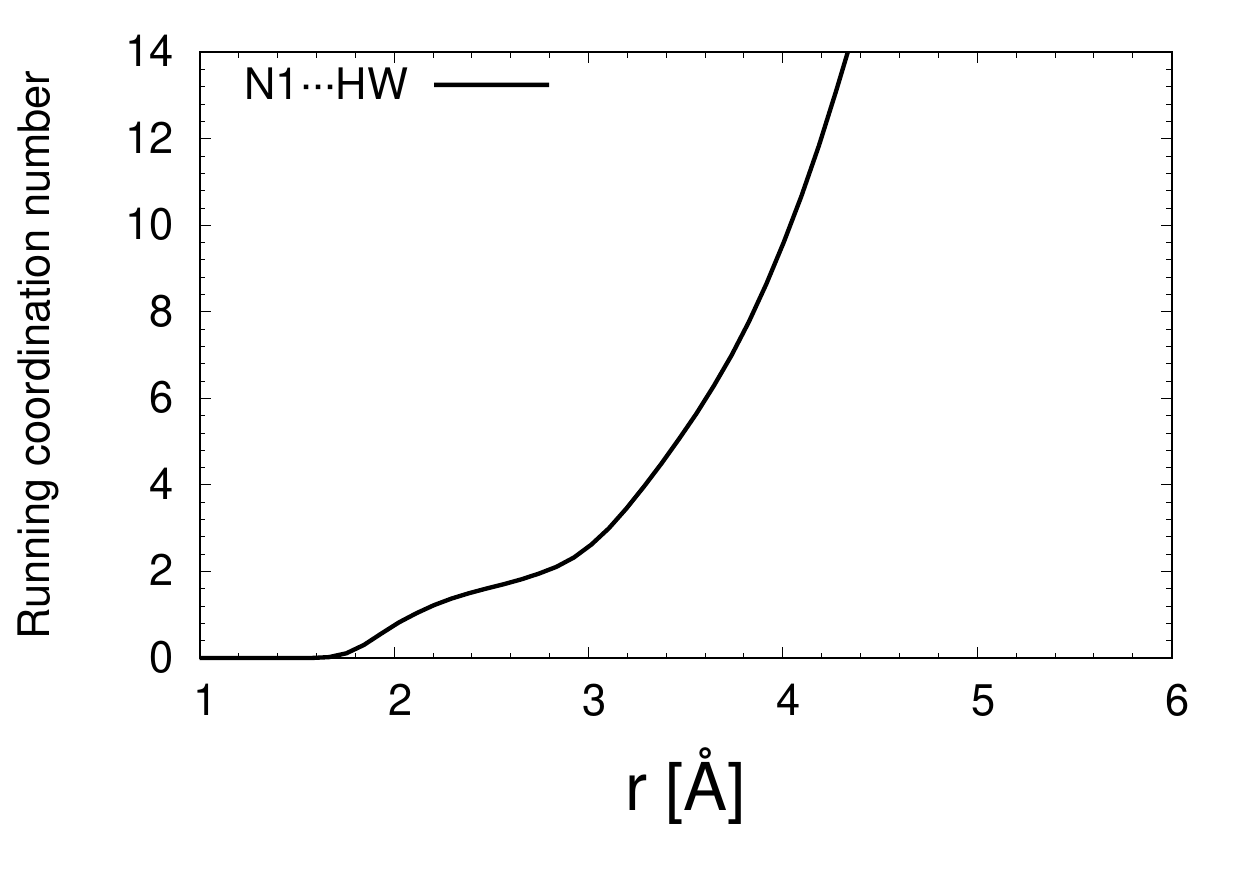}}
\subfloat[][\gr{c) Pyrimidine}]{\includegraphics[width=.32\textwidth]{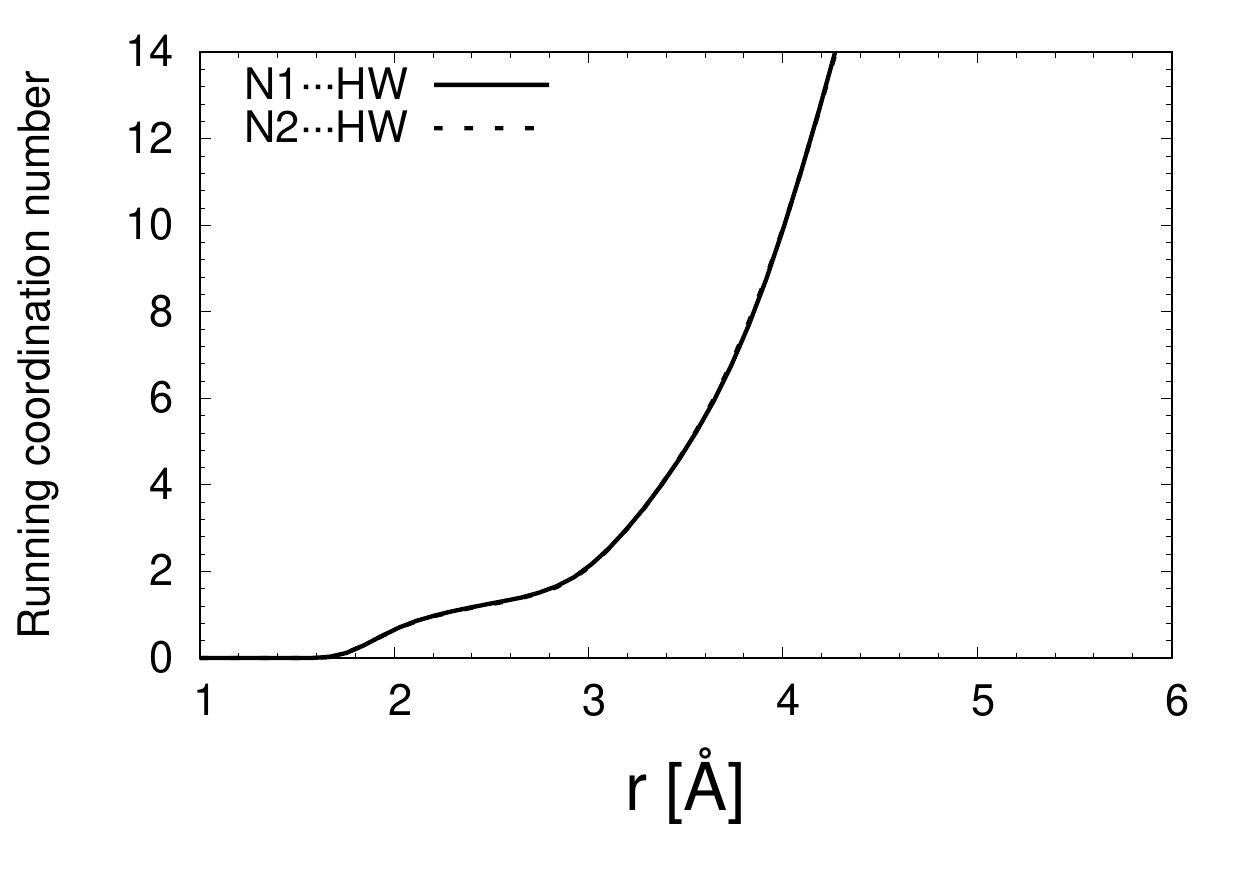}}
\caption{pNA (\gr{a}), Pyridine (\gr{b}) and Pyrimidine (\gr{c}) running coordination number (RCN). See Fig. \ref{fig:molecole} for atom labeling.}
\label{fig:rcn}
\end{figure*}

In this section, MD runs for pNA, pyridine and pyrimidine in aqueous solution are analyzed by using the TRAVIS package, \cite{brehm2011travis} so to extract hydration patterns and analyze the role of hydrogen bonding (HB) interactions, which are crucial to rationalize the trends of vertical excitation energies (vide infra). In particular, the radial distribution function $g(r)$, together with the running coordination number (RCN) between the solute atoms highlighted in Fig. \ref{fig:molecole} and solvent atoms (HW and OW) are studied. the RCN for a specific site is defined as the integral of the $g(r)$ as a function of the distance. The coordination number of a specific hydration shell is the RCN at the corresponding $g(r)$ minima. In addition, in case of pNA, the dihedral distribution function (ddf) is also discussed.

$g(r)$ and RCN for the three studied molecules are reported in Figs. \ref{fig:gdr} and \ref{fig:rcn}. In case of both pyridine and pyrimidine, only the nitrogen atom(s) can be involved in HB interactions with water molecules. For this reason, in Fig. \ref{fig:gdr}, panels b and c, the $g(r)$ for N1$\cdots$OW (and N2$\cdots$OW in case of pyrimidine) are reported. As it can be evinced, the $g(r)$ of both pyridine and pyrimidine  present a sharp peak related to the first solvation shell at about 2 \AA. The coordination numbers associated with the first solvation shell are approximately 2 and 1, for pyridine and pyrimidine, respectively. This indicates that two water molecules are involved in HB interactions with the nitrogen atom(s) of this two solutes. %CHIARA: CONTROLLA!!!!!!

In case of pNA (see Fig. \ref{fig:gdr}, panel a), H1$\cdots$OW and O1$\cdots$HW $g(r)$ are both characterized by a sharp peak at about 2 \AA, whereas N1 and N2 are not involved in HB interactions with the solvent. The coordination numbers of the first shells are around 1, showing that both the push and pull groups are involved in HB interactions with one water molecule, on average. Notice that pNA is symmetric, thus two water molecules undergo HB interactions with both push and pull groups of the solute.

\begin{figure}[htbp!]
\centering
\includegraphics[width=0.48\textwidth]{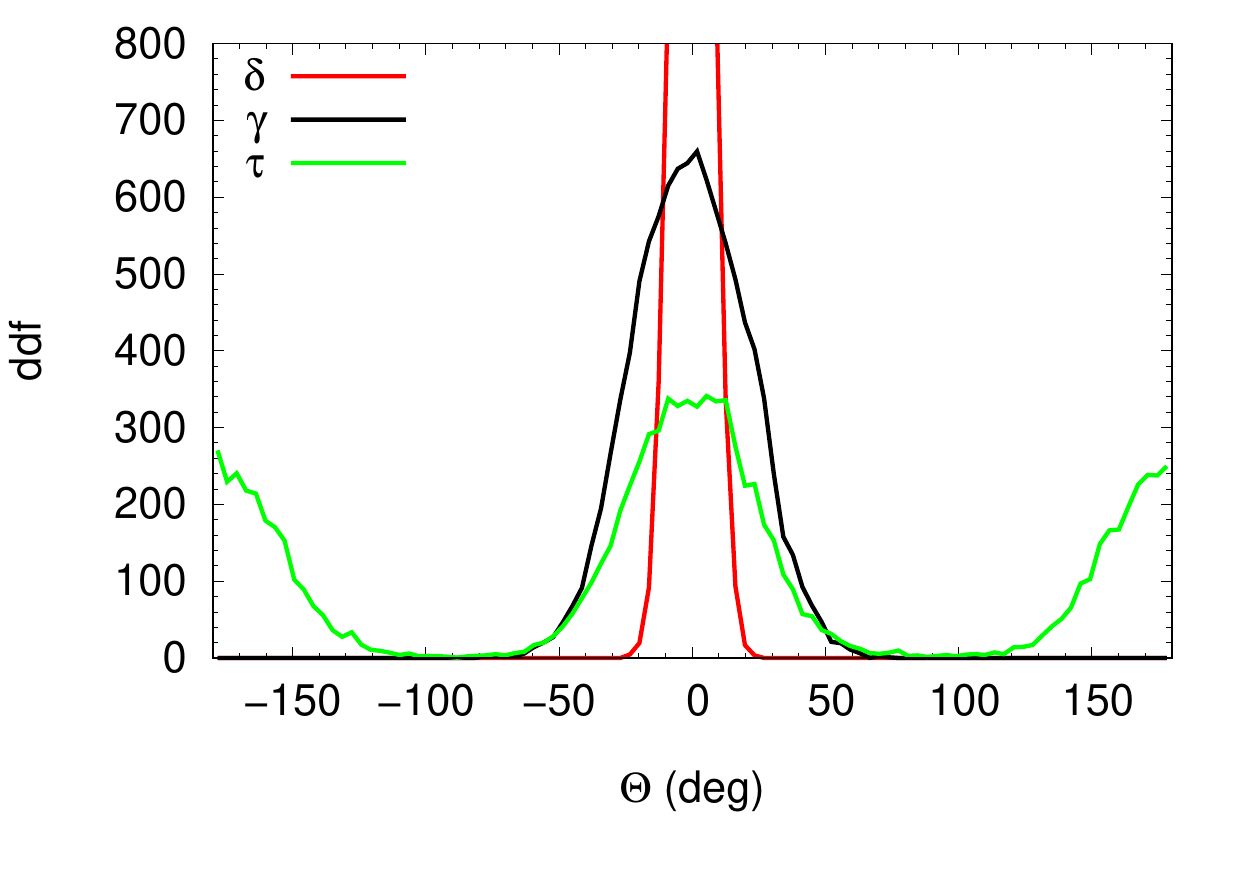}
\caption{pNA dihedral distribution function (ddf). See Fig. \ref{fig:molecole} for atom labelling.}
\label{fig:ddf-pna}
\end{figure}

To end the discussion, the ddf of $\gamma$, $\delta$ and $\tau$ pNA dihedral angles are reported in Fig. \ref{fig:ddf-pna} (see Fig. \ref{fig:molecole} for atoms involved in the definition of the dihedral angles). 
As it can be seen, all the considered dihedral angles fluctuate around 0 degrees (or 180 degrees in case of $\tau$), thus indicating that pNA oscillates around its planar configuration during the MD simulation. 
Specifically, out-of-plane motions of the benzene ring were monitored by analyzing $\delta$ dihedral angle, which only marginally varies during all MD simulations, thus indicating an almost rigid structure of this ring. In addition, the analysis of $\gamma$ and $\theta$ dihedral angles shows that the oxygen atoms of the nitro group can pass the rotational barrier (e.g. they can exchange their positions), whereas the same does not apply to the hydrogen atoms of the amine group. Indeed, they can oscillate in the molecule's plane, but they do not exchange their positions during the MD runs.

\subsection{Vacuo-to-Water Solvatochromism}

QM/FQ and QM/FQF$\mu$ vertical excitation energies of pNA (CAM-B3LYP/aug-cc-pVDZ), pyridine and pyrimidine (M06/6-311+G(2df,2p) in aqueous solution were computed on 100 uncorrelated snapshots extracted from MD runs. 

\begin{figure*}[htbp!]
\centering
\includegraphics[scale=1]{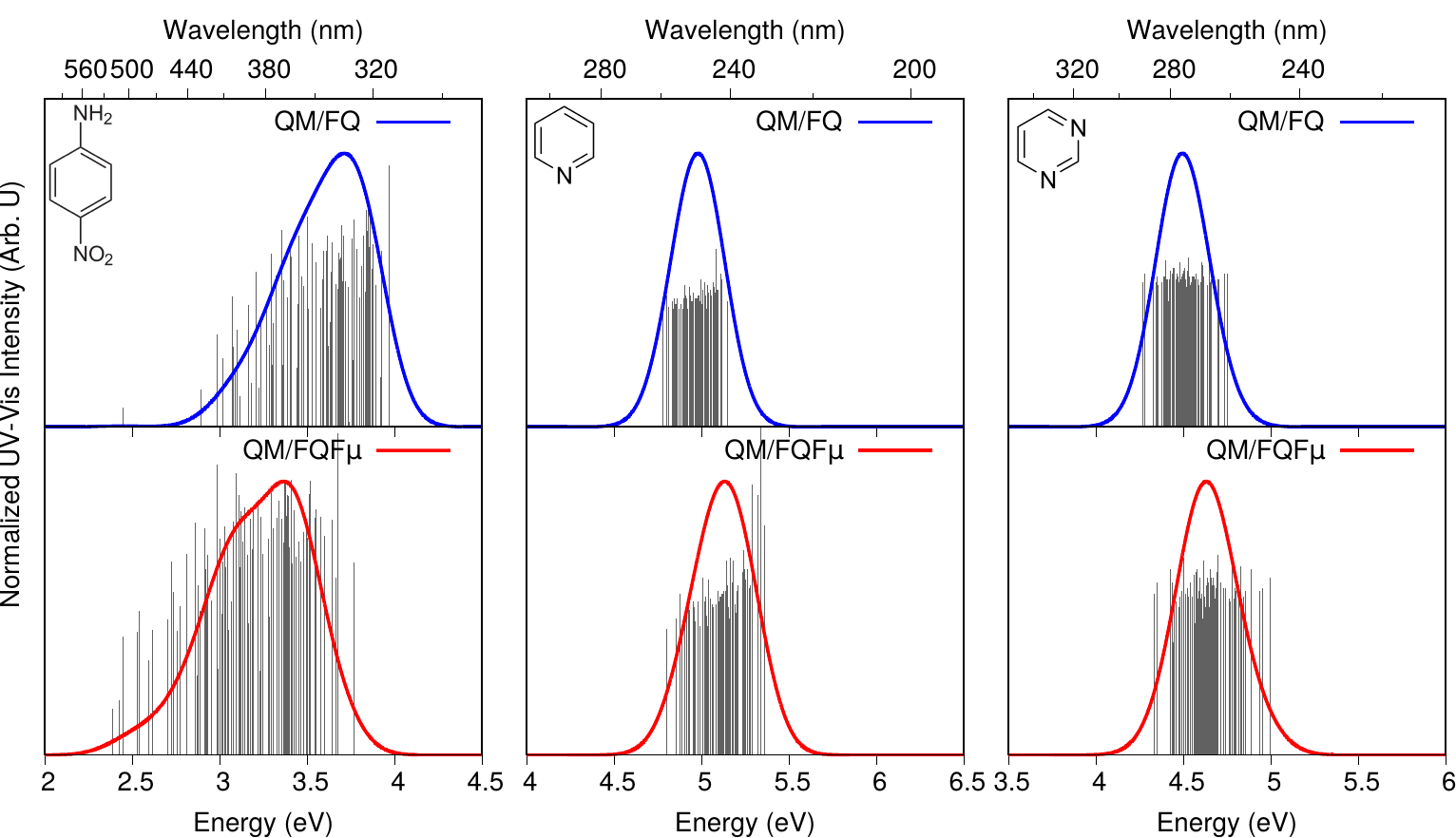}
\caption{pNA (left), Pyridine (middle) and Pyrimidine (right) QM/FQ and QM/FQF$\mu$ UV-Vis cLR raw data (sticks) together with their gaussian convolution (FWHM=0.3 eV)}
\label{fig:stick}
\end{figure*}

%\begin{figure*}[htbp!]
%\centering
%\subfloat[][\gr{a) pNA}]{\includegraphics[width=.3\textwidth]{OPA-pna-stick-cLR}}
%\subfloat[][\gr{b) Pyridine}]{\includegraphics[width=.3\textwidth]{OPA-pyridine-stick-cLR}}
%\subfloat[][\gr{c)Pyrimidine}]{\includegraphics[width=.3\textwidth]{OPA-pyrimidine-stick-cLR}}
%\caption{pNA (\gr{a}), Pyridine (\gr{b}) and Pyrimidine (\gr{c}) QM/FQ and QM/FQF$\mu$ UV-Vis cLR raw data (sticks) together with their gaussian convolution (FWHM=0.3 eV)}
%\label{fig:stick}
%\end{figure*}

%tommaso: cambiate tutte le immagini per il nostro ref 1

In Figure \ref{fig:stick}, pNA, pyridine and pyrimidine cLR absorption intensities as a function of the excitation energy, obtained by exploiting QM/FQ (top) and QM/FQF$\mu$ (bottom), for all 100 snapshots extracted from the MD simulations are plotted. Notice that stick spectra are obtained by plotting the raw data extracted from QM/MM calculations performed on each snapshot. $\omega_0$ and LR stick spectra are reported in Figs. S1 and S2 in the Supplementary Material (SM).

\begin{figure}[htbp!]
\centering
\includegraphics[width=.35\textwidth]{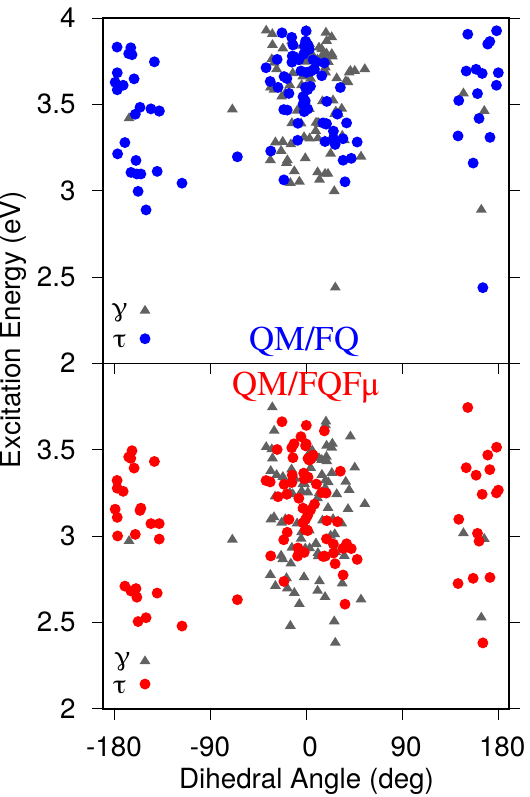}
\caption{QM/FQ (top) and QM/FQF$\mu$ (bottom) cLR excitation energies for pNA as a function of the dihedral angles $\gamma$ and $\tau$ (see Fig. \ref{fig:molecole}, panel a for their definition).}
\label{fig:dihedrals}
\end{figure}

In all cases, the inspection of Fig. \ref{fig:stick} shows that QM/FQF$\mu$ data (sticks) present a larger variability with respect to QM/FQ, which results in a greater broadening of the transition band, and in a different description of vertical excitation energies. To quantify such a variability, the standard deviation of both energies and intensities was computed (see Table S1 given as SM). As expected, QM/FQF$\mu$ energy standard deviations are always larger than QM/FQ ones. Such findings are due to the inclusion of fluctuating dipoles in the MM portion, which increase the molecular dipole moment of MM water molecules. As a consequence, QM/FQF$\mu$ convoluted spectra in Fig. \ref{fig:stick} show a larger inhomogeneous broadening with respect to QM/FQ. 
%tommaso: figura dihedrals
To further investigate the origin of QM/FQF$\mu$ larger inhomogeneous broadening, in Fig. \ref{fig:dihedrals} pNA QM/FQ (top) and QM/FQF$\mu$ (bottom) excitation energies as a function of $\gamma$ and $\tau$ dihedral angles (see Fig. \ref{fig:molecole}) are reported. Such a plot permits to understand whether the observed trend in broadening is related to the solute's geometry fluctuations or to solvent fluctuations, i.e. to the different description of the electrostatic interactions given by QM/FQ and QM/FQF$\mu$. As it is clearly shown by Fig. \ref{fig:dihedrals}, the two approaches give almost identical excitation energies as a function of the dihedral angles. Therefore, it is confirmed that the different broadening is only due to a different description of the QM/MM electrostatic coupling.

\begin{table*}
\centering
\begin{tabular}{cc|cc|cc|cc|cc}
\hline
Molecule & Method & \multicolumn{2}{c}{QM/PCM} & \multicolumn{2}{c}{QM/FQ}   & \multicolumn{2}{c}{QM/FQF$\mu$} & \multicolumn{2}{c}{Exp.} \\
         &        &   E$_{vert}$ & $\Delta$E &   E$_{vert}$ & $\Delta$E &   E$_{vert}$ & $\Delta$E & E$_{vert}$ & $\Delta$E \\
\hline
\multirow{3}{*}{pNA} & $\omega_0$         & 3.92 & 0.41                   & 3.74$\pm$0.03 & 0.59$\pm$0.03 & 3.47$\pm$0.03 & 0.86$\pm$0.03 & \multirow{3}{*}{3.25\text{$\phantom{t}^{\text{a}}$}} & \multirow{3}{*}{0.99} \\
         & LR                             & \textbf{3.81} & \textbf{0.52} & \textbf{3.70$\pm$0.03} & \textbf{0.63$\pm$0.03} & \textbf{3.31$\pm$0.03} & \textbf{1.02$\pm$0.03} &  & \\
         & cLR                            & 3.84 & 0.49                   & 3.71$\pm$0.03 & 0.62$\pm$0.03 & 3.37$\pm$0.03 & 0.96$\pm$0.03 &  & \\
\hline                                    
\multirow{3}{*}{pyridine} & $\omega_0$    & 4.84 & -0.17                   & 5.00$\pm$0.01 & -0.33$\pm$0.01 & 5.21$\pm$0.01 & -0.54$\pm$0.01 & \multirow{3}{*}{4.94\text{$\phantom{t}^{\text{b}}$}} & \multirow{3}{*}{-0.31} \\
         & LR                             & 4.84 & -0.17                   & 5.00$\pm$0.01 & -0.33$\pm$0.01 & 5.20$\pm$0.01 & -0.53$\pm$0.01 & &  \\
         & cLR                            & \textbf{4.79} & \textbf{-0.12} & \textbf{4.97$\pm$0.01} & \textbf{-0.30$\pm$0.01} & \textbf{5.12$\pm$0.01} & \textbf{-0.45$\pm$0.01} & & \\
\hline
\multirow{3}{*}{pyrimidine} & $\omega_0$  & 4.31  & -0.14                   & 4.52$\pm$0.01 & -0.34$\pm$0.01 & 4.67$\pm$0.01   & -0.49$\pm$0.01 & \multirow{3}{*}{4.57\text{$\phantom{t}^{\text{c}}$}} & \multirow{3}{*}{-0.35 $\pm$ 0.06\text{$\phantom{t}^{\text{d}}$}} \\
         & LR                             & 4.30  & -0.13                   & 4.51$\pm$0.01 & -0.34$\pm$0.01 & 4.66$\pm$0.01   & -0.48$\pm$0.01 & & \\
         & cLR                            & \textbf{4.28}  & \textbf{-0.11} & \textbf{4.49$\pm$0.01} & \textbf{-0.31$\pm$0.01} & \textbf{4.63$\pm$0.01}   & \textbf{-0.45$\pm$0.01} & & \\
\hline
\end{tabular}

\text{$\phantom{t}^{\text{a}}$} Ref. \citenum{kovalenko2000femtosecond}. \text{$\phantom{t}^{\text{b}}$} Ref. \citenum{perkampus1996uv}. \text{$\phantom{t}^{\text{c}}$} Ref. \citenum{mason1959electronic}. \text{$\phantom{t}^{\text{d}}$} Ref. \citenum{marenich2014electronic}.\hfill \\
\caption{pNA, Pyridine and Pyrimidine QM/PCM, QM/FQ and QM/FQF$\mu$ $\omega_0$, LR and cLR vertical excitation energies (E$_{vert}$) and vacuo-to-water solvatochromic shifts ($\Delta$E). "Best results" (see text) are in bold. Calculated QM/FQ and QM/FQF$\mu$ standard errors are also reported. They are calculated as $\sigma/\sqrt{N}$, where $\sigma$ is the standard deviation and $N$ is the number of the snapshots used to obtain the average property. Experimental data taken from the literature are reported for comparison's sake. All data are given in eV.}
\label{tab:solv-all}
\end{table*}			         

\begin{figure*}[htbp!]
\centering
\includegraphics[scale=1]{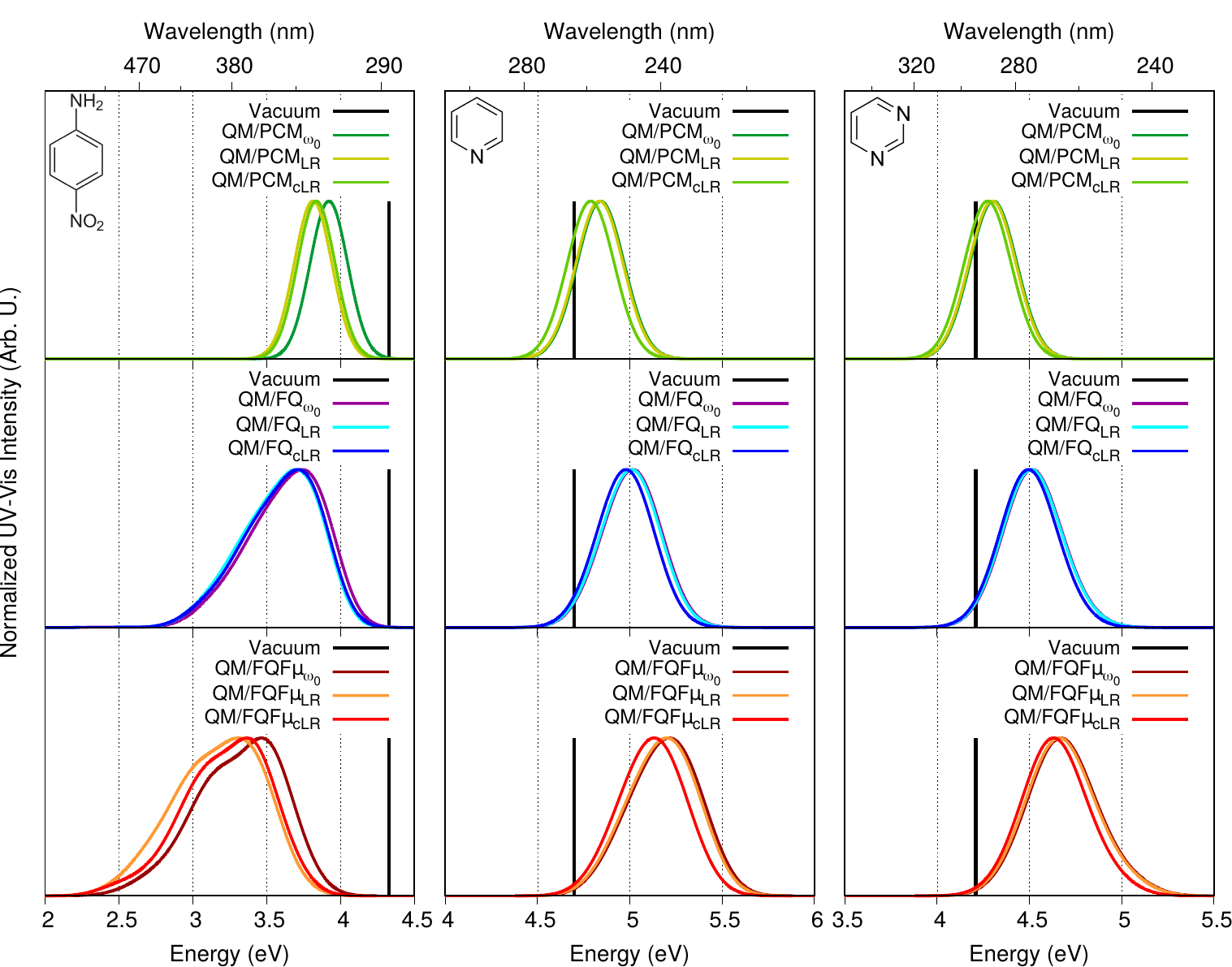}
\caption{pNA (left), Pyridine (middle) and Pyrimidine (right) QM/PCM (top) QM/FQ (middle) and QM/FQF$\mu$ (bottom) $\omega_0$, LR and cLR spectra. (FWHM=0.3 eV). The vertical excitation energy in gas-phase is also reported.}
\label{fig:conv}
\end{figure*}

%CHIARA: NELLE TABELLE VA MESSO a, b, c NELLE NOTE E POI COME APICE NEI NUERI DELLA TABELLA
PNA, pyridine and pyrimidine $\omega_0$, LR and cLR vertical excitation energies in aqueous solution, as calculated by exploiting QM/PCM, QM/FQ and QM/FQF$\mu$, and the corresponding vacuo-to-water solvatochromic shifts are reported in Tab. \ref{tab:solv-all}. The associated spectra are graphically depicted in Fig. \ref{fig:conv}, where gas-phase excitation energies are also given as stick bars. 
By moving from a continuum description (QM/PCM) to an atomistic approach (QM/FQ and QM/FQF$\mu$) a gradual increase of the solvatochromic shift is reported, independently from which vertical excitation energy ($\omega_0$, LR or cLR) is taken in consideration (see Tab. \ref{tab:solv-all} and Fig. \ref{fig:conv}). In particular, in case of pNA, QM/FQ redshifts with respect to QM/PCM of 0.11 eV, whereas QM/FQF$\mu$ of almost 0.50 eV, thus doubling the computed solvatochromism with respect to continuum QM/PCM values. In case of pyridine and pyrimidine, a shift in the opposite direction with respect to pNA is reported by all the solvation models. In particular, QM/FQ and QM/FQF$\mu$ predict an increase of vacuo-to-water solvatochromism of 0.18/0.21 and 0.33/0.35 eV with respect to QM/PCM, respectively. Notice that such an increase corresponds to a relative difference with respect to QM/PCM of almost 300 and 500 \%, in case of QM/FQ and QM/FQF$\mu$, respectively. This huge blueshift can be related to the dynamical atomistic description of the solvent water molecules given by QM/MM approaches coupled to MD simulations. The difference between QM/FQ and QM/FQF$\mu$ results reported by all the selected molecular systems, shows instead the relevance of the inclusion of fluctuating dipoles in addition to fluctuating charges.

%tommaso: cambiate tutte le immagini per il nostro ref

By deepening the analysis of Tab. \ref{tab:solv-all}, the different pictures given by LR and cLR approaches emerge. First, let's focus on pNA $\omega_0$, LR and cLR results. The continuum QM/PCM and atomistic QM/FQ and QM/FQF$\mu$ LR and cLR excitation energies are very similar, and shifted with respect to the frozen density approximation ($\omega_0$) of 0.11, 0.04 and 0.16 eV, respectively. The similarity between LR and cLR energies indicates that in this particular case, the relaxation of the solvent (Which is considered in the cLR approach), is as relevant as the dynamical solute-solvent interactions taken into account by the LR regime. This is not surprising, by considering that the studied excitation is a pure $\pi \rightarrow \pi^*$ transition, for which the LR regime is generally successfull.\cite{guido2015electronic} Furthermore, the major differences with respect to $\omega_0$ are reported for LR energies, which are therefore selected as ``best'' results in case of pNA  (and reported in bold in Tab. \ref{tab:solv-all}). 

Differently from pNA, the $n \rightarrow \pi^*$ transition is studied in case of pyridine and pyrimidine. For such a transition, which is associated to a large variation of the dipole moment together with a transition dipole moment close to zero (i.e. the transition is dark), cLR regime has been previously reported to be essential in order to correctly account for the relaxation of the solvent both in QM/PCM\cite{caricato2006formation} and polarizable QM/MM calculations.\cite{loco2016qm} 
This is confirmed by the data reported in Tab. \ref{tab:solv-all}, where $\omega_0$ and LR excitation energies are predicted to be almost the same by all investigated solvation approaches, whereas a shift in case of cLR is reported. In particular, QM/PCM cLR corrections shift $\omega_0$ (and LR) energies of about 0.05 and 0.03 eV, for pyridine and pyrimidine, respectively; this corresponds to 33\% deviation in both cases. The magnitude of the QM/FQ shift reported for both molecules, and of QM/FQF$\mu$ shift for pyrimidine are very similar to QM/PCM values. This is not surprising, because it has been recently reported to also apply to QM/AMOEBA cLR absorption energies \cite{loco2016qm}.
QM/FQF$\mu$ cLR correction is particularly relevant for pyridine in aqueous solution, for which a shift of 0.09 (20\%) with respect to the frozen density approximation $\omega_0$ (and LR) is obtained.

%cambiato highlighted
Best estimates of QM/PCM, QM/FQ and QM/FQF$\mu$ vacuo-to-water solvatochromic shifts for the three systems are highlighted in Tab. \ref{tab:solv-all}. They are obtained by exploiting the LR regime for pNA and cLR for both pyridine and pyrimidine. In case of QM/FQ and QM/FQF$\mu$, standard errors are also given, which are calculated as $\sigma/\sqrt{N}$, where $\sigma$ is the standard deviation and $N$ is the number of the snapshots used to obtain the average values of the shifts. We note that the continuum QM/PCM is generally unable to recover the experimental vacuo-to-water solvatochromism, that reasonably due to the imporper account of specific solute-solvent interactions and to the "static" description of the solvation phenomenon.\cite{giovannini2017polarizable,loco2018dynamic,loco2016qm} As a consequence, the error on pNA solvatochromism is of 50\%, whereas in case of pyridine and pyrimidine QM/PCM only recovers 35\% of the experimental shift. A much better agreement with the experimental values is achieved by exploiting the polarizable QM/FQ coupled to MD simulations; in fact, this approach is able to recover the experimental vacuo-to-water solvatochromism in case of pyridine and pyrimidine, whereas it fails at reproducing the value for pNA.% However, also in the latter case, QM/FQ vertical excitation energies are closer to the experimental counterpart than QM/PCM ones, reducing the computed error to 33\%. 
pNA is correctly described by QM/FQF$\mu$, which, however, overestimates of about 0.10 eV the values for  pyridine and pyrimidine. Remarkably, in case of QM/FQF$\mu$, cLR shifts $\omega_0$ and LR vertical excitation energies towards their experimental counterparts (see Tab. \ref{tab:solv-all}).

%tommaso a chiara (viene da sopra): aggiungiamo qua la frase? Secondo me potrebbe essere un buon punto.

%In our first paper \cite{giovannini2019fqfmu} we have shown that QM/FQF$\mu$ predicts electrostatic energies in almost perfect agreement with full-QM approaches. Therefore, we speculate that the discrepancies between computed and experimental values are primarly due to the lack of solute-solvent interactions other than electrostatics. Also notice that QM/FQF$\mu$ will show probably the same discrepancy with a full-QM treatment of the whole snapshot for exactly the same reason. However, as stated before, QM/FQF$\mu$ is expected to show a good agreement with a method able to account for the electrostatics-only interaction on the excitation energies, but, to the best of our knoweledge, methods able to decouple the different contributions, such as electrostatics and non-electrostatics, on the excitation energies has not been developed. However, within such a picture, QM/FQF$\mu$ can be taken as a reference for the electrostatic contribution to the vertical excitation energies.

\begin{table}[htbp!]
\centering
\begin{tabular}{llccccc}
\hline
Density & Molecule   & Transition            & Vacuo  &  QM/PCM & QM/FQ   & QM/FQF$\mu$ \\
\hline
\multirow{3}{*}{$\mathbf{P}^T_K$}            & pNA        & $\pi\rightarrow\pi^*$ & 3.14 & 3.11 & 2.60 & 3.35 \\
                                             & pyridine   & $n\rightarrow\pi^*$   & 1.60 & 1.51 & 1.48 & 1.40 \\
                                             & pyrimidine & $n\rightarrow\pi^*$   & 0.99 & 0.99 & 1.00 & 1.00 \\
\hline
\multirow{3}{*}{$\mathbf{P}^{\Delta}_K$}     & pNA        & $\pi\rightarrow\pi^*$ & 2.05 & 2.26 & 1.66 & 2.22 \\ 
                                             & pyridine   & $n\rightarrow\pi^*$   & 0.80 & 1.00 & 0.82 & 0.85 \\
                                             & pyrimidine & $n\rightarrow\pi^*$   & 0.54 & 0.73 & 0.58 & 0.64 \\
\hline
\end{tabular}
\caption{pNA, Pyridine, Pyrimidine vacuo, QM/PCM, QM/FQ and QM/FQF$\mu$ D$_{CT}$ values (\AA) for the studied transitions. The values are calculated by using both the unrelaxed density $\mathbf{P}^T_K$ and relaxed $\mathbf{P}^{\Delta}_K$ densities .}
\label{tab:dct}
\end{table}

To investigate the reasons of the overestimation of vacuo-to-water solvatochromic shifts reported for QM/FQF$\mu$ (which, as stated above, is particularly evident for pyridine and pyrimidine) the charge transfer (CT) nature of the considered transitions was characterized by a quantitative index, denoted as $D_{\mathrm{CT}}$.\cite{le2011qualitative,egidi2018nature} The barycenters of the positive and negative density distributions are calculated as the difference of the ground state (GS) and excited state (ES) densities. The CT length index ($D_{\mathrm{CT}}$ ) is defined as the distance between the two charge barycenters. $D_{\mathrm{CT}}$ indexes, as calculated by considering both the unrelaxed $\mathbf{P}^T_K$ and relaxed densities $\mathbf{P}^{\Delta}_K$, are reported in Tab. \ref{tab:dct}. 

The data in Tab. \ref{tab:dct}, confirm that all the considered transitions are associated with an intramolecular CT. In particular, the largest $D_{\mathrm{CT}}$ values are reported in case of pNA (both in gas phase and in solution), independently from the specific solvation model. Notice that the CT nature of each transition is different (lower) if the calculation is performed by exploiting the relaxed density ($\mathbf{P}^{\Delta}_K$) instead of the unrelaxed one ($\mathbf{P}^{T}_K$), in line with what reported in previous studies.\cite{maschietto2018charge} %CHIARA: CONTROLLA!!!! 
Such an observation supports the necessity of taking into account the relaxation of the solvent polarization, i.e. to adopt a cLR regime rather than the common LR approach for the solvated systems. However, in order to correctly decide the best regime for our systems, molecular dipole moments need to be analyzed. In particular, differences in cLR and LR vertical excitation energies can be related to a different description of the change in the dipole moment moving from GS to ES  ($\Delta\mu = \abs{\mu_{ex}-\mu_g}$) and the transition dipole moment ($\mu_{g,ex}$). %CHIARA: CONTROLLA!!!!! 
In fact, it has been shown that if $\Delta\mu^2$ is smaller than $2\mu^2_{g,ex}$, the LR excitation energy is smaller than cLR one, and vice versa.\cite{caricato2006formation} Therefore, %in addition to $D_{CT}$ indexes, 
the relevant values %GS and ES molecular dipole moments ($\mu_g$ and $\mu_{ex}$) and their squared differences ($\Delta\mu^2 = \abs{\mu_{ex}-\mu_g}^2$) together with GS to the double of the squared ES transition dipole moments ($2 \mu^2_{g,ex}$) 
are reported in Table \ref{tab:dipmom} for the three investigated solvation approaches.

\begin{table*}[htbp!]
\begin{tabular}{l|rrrr|rrrr|rrrr}
\hline
Molecule   & \multicolumn{4}{c}{QM/PCM}                   & \multicolumn{4}{c}{QM/FQ}                & \multicolumn{4}{c}{QM/FQF$\mu$} \\
           & $\mu_{g}$ & $\mu_{ex}$ & $\Delta\mu^2$ & $2\mu^2_{g,ex}$    & $\mu_{g}$ & $\mu_{ex}$ & $\Delta\mu^2$ & $2\mu^2_{g,ex}$  & $\mu_{g}$ & $\mu_{ex}$ & $\Delta\mu^2$ & $2\mu^2_{g,ex}$     \\
\hline
pNA        &  10.5     &  17.6      &  50.4   &   72.5   &  10.3       &  15.0      &   22.1 &   38.1  &   12.9       &     19.0  &    37.2  &   68.0    \\
pyridine   &  3.1      &   0.2      &   8.4   &   0.5    &   3.6        &  0.9      &    7.3 &    0.4  &   4.4        &     1.6   &     7.8  &   0.5      \\
pyrimidine &  3.3      &   0.8      &   6.2   &   0.7    &   3.6        &  1.9      &    2.9 &    0.6  &   4.4        &     2.7   &     2.9  &   0.7     \\
\hline
\end{tabular}
\caption{pNA, Pyridine and Pyrimidine QM/PCM, QM/FQ and QM/FQF$\mu$ GS($\mu_g$) and ES ($\mu_{ex}$) electric dipole moments and their squared difference ($\Delta\mu^2 = \abs{\mu_{ex}-\mu_g}^2$).  ($2 \mu^2_{g,ex}$), with $\mu_{g,ex}$ being the transition dipole moment, is also reported. All dipole data are given in Debye.}
\label{tab:dipmom}
\end{table*}	

We notice that moving from GS to ES the dipole moment decreases for both pyridine and pyrimidine, whereas it increases for pNA.  Remarkably, this is related to the negative solvatochromism reported in case of pyridine and pyrimidine, and positive in case of pNA.\cite{marenich2014electronic} Also, the results obtained in Tab. \ref{tab:solv-all} are coherent with the fact that for pNA $2\mu^2_{g,ex}$ is larger than $\Delta\mu^2$, whereas the opposite applies to both pyridine and pyrimidine. As a consequence, LR excitation energies are smaller than cLR for pNA, whereas cLR ones are smaller than LR for the other two systems (see Table \ref{tab:solv-all}). 
Furthermore, the analysis based on dipole moments (Tab. \ref{tab:dipmom}) suggest that in case of pNA, QM/FQF$\mu$ outperforms QM/FQ because it predicts a larger difference (of almost 25\%) between ES and GS dipole moments, thus resulting in a larger vacuo-to-solvent solvatochromism. The same does not apply to pyridine and pyrimidine, because QM/FQ and QM/FQF$\mu$ predict almost the same difference in dipole moments. The computed larger solvatochromism for QM/FQF$\mu$ is thus probably due to the fact that both GS ans ES dipole moments are increased of almost 50\% with respect to QM/FQ. Howevef, it is worth noticing that molecular dipole moments are highly affected by non-electrostatic interactions, in particular repulsion energy terms, as recently reported by some of the present authors.\cite{giovannini2017disrep} The inclusion of such contributions can in principle reduce the molecular dipole moments of both GS and ES. As a consequence, non-electrostatic interactions should reasonably shift QM/FQF$\mu$ values towards the experimental data. %This topic will be the subject of future communications. 
On the other hand, in light of such considerations, the good performance of QM/FQ for pyridine and pyrimidine might be ascribed to error cancellations between the description of electrostatic effects and the lack of explicit consideration of exchange-repulsion effects.\cite{giovannini2017disrep,curutchet2018density,giovannini2019eprdisrep}%, i.e. to a worst description of the electrostatics balances the lack of exchange-repulsion effects.

\begin{comment}
\begin{table}[htbp!]
\centering
\begin{tabular}{l|rrrr|rrrr|rrrr}
\hline
Molecule   & \multicolumn{4}{c}{QM/PCM}                   & \multicolumn{4}{c}{QM/FQ}                & \multicolumn{4}{c}{QM/FQF$\mu$} \\
           & $\mu_{g}$ & $\mu_{ex}$ & $\Delta(\mu^2)$ & $\mu^{ex}_{g}$    & $\mu_{g}$ & $\mu_{ex}$ & $\Delta(\mu^2)$ & $\mu^{ex}_{g}$  & $\mu_{g}$ & $\mu_{ex}$ & $\Delta(\mu^2)$ & $\mu^{ex}_{g}$     \\
\hline
pNA        &  10.5     &  17.6      &  199.5   &   0.523   &  10.3        &  15.0      &  118.8 &   0.258 &   12.9        &     19.0  &    194.6 &   0.405    \\
pyridine   &  3.1      &   0.2      &   -9.6   &   0.005  &   3.6        &  0.95      &  -12.1 &   0.005 &   4.4         &     1.6   &   -16.8  &   0.006      \\
pyrimidine &  3.3      &   0.8      &  -10.5   &   0.006  &   3.6        &  1.95      &   -9.2 &   0.005 &   4.4         &     2.7   &  -12.1   &   0.006     \\
\hline
\end{tabular}
\caption{pNA, Pyridine and Pyrimidine QM/PCM, QM/FQ and QM/FQF$\mu$ GS and ES electric dipole moments (Debye).}
\label{tab:dipmom}
\end{table}	
\end{comment}

\section{Summary and Conclusions}

In this work, we have extended the fully polarizable QM/FQF$\mu$ approach to the evaluation of excited state energies. The peculiarity of QM/FQF$\mu$ stands in the fact that each MM atom is endowed with both a charge and a dipole, which can vary as a response of the external electric potential and field. QM/FQF$\mu$ is therefore a refinement of QM/FQ, in which only a fluctuating charge is placed on each MM atom. %Therefore, QM/FQF$\mu$ introduces the contribution of anisotropic and out-of-plane polarization modeled in terms of fluctuating dipoles.

%tommaso: aggiunto for the first time dopo introduced 
The extension of QM/FQF$\mu$ to vertical excitation energies, is achieved by first resorting to linear response (LR) theory for SCF methods. Second, a correction, namely corrected Linear Response (cLR), is introduced for the first time for both QM/FQ and QM/FQF$\mu$. The necessity of such a correction is due to the intrinsic limitations of the LR regime when applied to  polarizable embedding approaches. In fact, LR adequately describes environmental effects in case of transitions involving bright states associated with large transition dipole moments, but it is not able to catch the relaxation of the environment as a response to charge equilibration of the QM density on the specific excited state. cLR can instead describe such a relaxation, that is particularly relevant in case of transitions with low (or null) transition dipole moments, and with a large difference between the ground and excited dipole moments.

To test our approach, we selected three molecules, namely para-nitroaniline, pyridine and pyrimidine, in aqueous solution, of which we studied $\pi \rightarrow \pi^*$ transition for pNA, and dark $n \rightarrow \pi^*$ transitions in case of both pyridine and pyrimidine.  Such transitions were selected because vacuo-to-water solvatochromic experimental data are available in the literature.\cite{marenich2014electronic} %tommaso: aggiunta la cosa subito dopo QM/PCM approach nella frase sotto
For all the studied systems, both QM/FQ and QM/FQF$\mu$ over perform the implicit QM/PCM approach, due to the fact that QM/PCM cannot properly describe specific solute-solvent interactions. QM/FQF$\mu$ correctly reproduces the experimental solvachromic shift in case of pNA, whereas it overestimates the experimental value for pyridine and pyrimidine. The analysis of ground and excited state molecular dipole moments shows that such a discrepancy is related to the overestimation of molecular dipole moments, probably related to the electrostatic-only description of the solvation phenomenon which is here adopted. 

%tommaso: qua aggiunta frase per ref1 
It is worth noticing that the comparison between QM/MM results and experiments can be biased by experimental conditions and error bar. As a matter of fact, QM/classical results might be compared with full-QM calculation on clusters (i.e. structures in which few solvent molecules are treated at the QM level). However, full-QM super-molecule approaches would include non-electrostatic interactions, such as Pauli repulsion, solute-solvent and/or solvent-solvent Charge Transfer, and dispersion (in case correlation is taken into account). Therefore, a good agreement between QM/classical approaches and such reference values could be affected by error cancellation.
%tommaso: aggiungiamo magari qua la cosa del vibronico
The inclusion of non-electrostatic interactions, and also vibronic effects,\cite{santoro2016going} which can in principle increase the agreement between computed and experimental data, will be the topic of future communications. 

\section{Supplementary Material}

QM/FQ and QM/FQF$\mu$ standard deviations on vertical excitations and oscillator strengths for pNA, pyridine and pyrimidine in aqueous solution. QM/FQ and QM/FQF$\mu$ $\omega_0$ and cLR stick and convoluted spectra of pNA, pyridine and pyrimidine in aqueous solution.

\section{Acknowledgment}

We are thankful for the computer resources provided by the high performance computer
facilities of the SMART Laboratory (http://smart.sns.it/). CC gratefully
acknowledges the support of H2020-MSCA-ITN-2017 European Training Network
“Computational Spectroscopy In Natural sciences and Engineering” (COSINE),
grant number 765739. TG acknowledges funding from the Research Council of Norway through its grant TheoLight (grant no. 275506)

\section{References}

%\nocite{*}
\bibliography{biblio}% Produces the bibliography via BibTeX.

\begin{thebibliography}{123}%
\makeatletter
\providecommand \@ifxundefined [1]{%
 \@ifx{#1\undefined}
}%
\providecommand \@ifnum [1]{%
 \ifnum #1\expandafter \@firstoftwo
 \else \expandafter \@secondoftwo
 \fi
}%
\providecommand \@ifx [1]{%
 \ifx #1\expandafter \@firstoftwo
 \else \expandafter \@secondoftwo
 \fi
}%
\providecommand \natexlab [1]{#1}%
\providecommand \enquote  [1]{``#1''}%
\providecommand \bibnamefont  [1]{#1}%
\providecommand \bibfnamefont [1]{#1}%
\providecommand \citenamefont [1]{#1}%
\providecommand \href@noop [0]{\@secondoftwo}%
\providecommand \href [0]{\begingroup \@sanitize@url \@href}%
\providecommand \@href[1]{\@@startlink{#1}\@@href}%
\providecommand \@@href[1]{\endgroup#1\@@endlink}%
\providecommand \@sanitize@url [0]{\catcode `\\12\catcode `\$12\catcode
  `\&12\catcode `\#12\catcode `\^12\catcode `\_12\catcode `\%12\relax}%
\providecommand \@@startlink[1]{}%
\providecommand \@@endlink[0]{}%
\providecommand \url  [0]{\begingroup\@sanitize@url \@url }%
\providecommand \@url [1]{\endgroup\@href {#1}{\urlprefix }}%
\providecommand \urlprefix  [0]{URL }%
\providecommand \Eprint [0]{\href }%
\providecommand \doibase [0]{http://dx.doi.org/}%
\providecommand \selectlanguage [0]{\@gobble}%
\providecommand \bibinfo  [0]{\@secondoftwo}%
\providecommand \bibfield  [0]{\@secondoftwo}%
\providecommand \translation [1]{[#1]}%
\providecommand \BibitemOpen [0]{}%
\providecommand \bibitemStop [0]{}%
\providecommand \bibitemNoStop [0]{.\EOS\space}%
\providecommand \EOS [0]{\spacefactor3000\relax}%
\providecommand \BibitemShut  [1]{\csname bibitem#1\endcsname}%
\let\auto@bib@innerbib\@empty
%</preamble>
\bibitem [{\citenamefont {Helgaker}\ \emph {et~al.}(2012)\citenamefont
  {Helgaker}, \citenamefont {Coriani}, \citenamefont {J{\o}rgensen},
  \citenamefont {Kristensen}, \citenamefont {Olsen},\ and\ \citenamefont
  {Ruud}}]{helgaker2012recent}%
  \BibitemOpen
  \bibfield  {author} {\bibinfo {author} {\bibfnamefont {T.}~\bibnamefont
  {Helgaker}}, \bibinfo {author} {\bibfnamefont {S.}~\bibnamefont {Coriani}},
  \bibinfo {author} {\bibfnamefont {P.}~\bibnamefont {J{\o}rgensen}}, \bibinfo
  {author} {\bibfnamefont {K.}~\bibnamefont {Kristensen}}, \bibinfo {author}
  {\bibfnamefont {J.}~\bibnamefont {Olsen}}, \ and\ \bibinfo {author}
  {\bibfnamefont {K.}~\bibnamefont {Ruud}},\ }\bibfield  {title} {\enquote
  {\bibinfo {title} {Recent advances in wave function-based methods of
  molecular-property calculations},}\ }\href@noop {} {\bibfield  {journal}
  {\bibinfo  {journal} {Chem. Rev.}\ }\textbf {\bibinfo {volume} {112}},\
  \bibinfo {pages} {543--631} (\bibinfo {year} {2012})}\BibitemShut {NoStop}%
\bibitem [{\citenamefont {Le~Bahers}, \citenamefont {Adamo},\ and\
  \citenamefont {Ciofini}(2011)}]{le2011qualitative}%
  \BibitemOpen
  \bibfield  {author} {\bibinfo {author} {\bibfnamefont {T.}~\bibnamefont
  {Le~Bahers}}, \bibinfo {author} {\bibfnamefont {C.}~\bibnamefont {Adamo}}, \
  and\ \bibinfo {author} {\bibfnamefont {I.}~\bibnamefont {Ciofini}},\
  }\bibfield  {title} {\enquote {\bibinfo {title} {A qualitative index of
  spatial extent in charge-transfer excitations},}\ }\href@noop {} {\bibfield
  {journal} {\bibinfo  {journal} {J. Chem. Theory Comput.}\ }\textbf {\bibinfo
  {volume} {7}},\ \bibinfo {pages} {2498--2506} (\bibinfo {year}
  {2011})}\BibitemShut {NoStop}%
\bibitem [{\citenamefont {Guido}\ \emph {et~al.}(2013)\citenamefont {Guido},
  \citenamefont {Cortona}, \citenamefont {Mennucci},\ and\ \citenamefont
  {Adamo}}]{guido2013metric}%
  \BibitemOpen
  \bibfield  {author} {\bibinfo {author} {\bibfnamefont {C.~A.}\ \bibnamefont
  {Guido}}, \bibinfo {author} {\bibfnamefont {P.}~\bibnamefont {Cortona}},
  \bibinfo {author} {\bibfnamefont {B.}~\bibnamefont {Mennucci}}, \ and\
  \bibinfo {author} {\bibfnamefont {C.}~\bibnamefont {Adamo}},\ }\bibfield
  {title} {\enquote {\bibinfo {title} {On the metric of charge transfer
  molecular excitations: a simple chemical descriptor},}\ }\href@noop {}
  {\bibfield  {journal} {\bibinfo  {journal} {J. Chem. Theory Comput.}\
  }\textbf {\bibinfo {volume} {9}},\ \bibinfo {pages} {3118--3126} (\bibinfo
  {year} {2013})}\BibitemShut {NoStop}%
\bibitem [{\citenamefont {Savarese}\ \emph {et~al.}(2017)\citenamefont
  {Savarese}, \citenamefont {Guido}, \citenamefont {Bremond}, \citenamefont
  {Ciofini},\ and\ \citenamefont {Adamo}}]{savarese2017metrics}%
  \BibitemOpen
  \bibfield  {author} {\bibinfo {author} {\bibfnamefont {M.}~\bibnamefont
  {Savarese}}, \bibinfo {author} {\bibfnamefont {C.~A.}\ \bibnamefont {Guido}},
  \bibinfo {author} {\bibfnamefont {E.}~\bibnamefont {Bremond}}, \bibinfo
  {author} {\bibfnamefont {I.}~\bibnamefont {Ciofini}}, \ and\ \bibinfo
  {author} {\bibfnamefont {C.}~\bibnamefont {Adamo}},\ }\bibfield  {title}
  {\enquote {\bibinfo {title} {Metrics for molecular electronic excitations: A
  comparison between orbital-and density-based descriptors},}\ }\href@noop {}
  {\bibfield  {journal} {\bibinfo  {journal} {J. Phys. Chem. A}\ }\textbf
  {\bibinfo {volume} {121}},\ \bibinfo {pages} {7543--7549} (\bibinfo {year}
  {2017})}\BibitemShut {NoStop}%
\bibitem [{\citenamefont {Baiardi}, \citenamefont {Bloino},\ and\ \citenamefont
  {Barone}(2016)}]{baiardi2016general}%
  \BibitemOpen
  \bibfield  {author} {\bibinfo {author} {\bibfnamefont {A.}~\bibnamefont
  {Baiardi}}, \bibinfo {author} {\bibfnamefont {J.}~\bibnamefont {Bloino}}, \
  and\ \bibinfo {author} {\bibfnamefont {V.}~\bibnamefont {Barone}},\
  }\bibfield  {title} {\enquote {\bibinfo {title} {General formulation of
  vibronic spectroscopy in internal coordinates},}\ }\href@noop {} {\bibfield
  {journal} {\bibinfo  {journal} {J. Chem. Phys.}\ }\textbf {\bibinfo {volume}
  {144}},\ \bibinfo {pages} {084114} (\bibinfo {year} {2016})}\BibitemShut
  {NoStop}%
\bibitem [{\citenamefont {Bloino}, \citenamefont {Baiardi},\ and\ \citenamefont
  {Biczysko}(2016)}]{bloino2016aiming}%
  \BibitemOpen
  \bibfield  {author} {\bibinfo {author} {\bibfnamefont {J.}~\bibnamefont
  {Bloino}}, \bibinfo {author} {\bibfnamefont {A.}~\bibnamefont {Baiardi}}, \
  and\ \bibinfo {author} {\bibfnamefont {M.}~\bibnamefont {Biczysko}},\
  }\bibfield  {title} {\enquote {\bibinfo {title} {Aiming at an accurate
  prediction of vibrational and electronic spectra for medium-to-large
  molecules: An overview},}\ }\href@noop {} {\bibfield  {journal} {\bibinfo
  {journal} {Int. J. Quantum Chem.}\ }\textbf {\bibinfo {volume} {116}},\
  \bibinfo {pages} {1543--1574} (\bibinfo {year} {2016})}\BibitemShut {NoStop}%
\bibitem [{\citenamefont {Improta}\ \emph {et~al.}(2009)\citenamefont
  {Improta}, \citenamefont {Santoro}, \citenamefont {Barone},\ and\
  \citenamefont {Lami}}]{improta2009vibronic}%
  \BibitemOpen
  \bibfield  {author} {\bibinfo {author} {\bibfnamefont {R.}~\bibnamefont
  {Improta}}, \bibinfo {author} {\bibfnamefont {F.}~\bibnamefont {Santoro}},
  \bibinfo {author} {\bibfnamefont {V.}~\bibnamefont {Barone}}, \ and\ \bibinfo
  {author} {\bibfnamefont {A.}~\bibnamefont {Lami}},\ }\bibfield  {title}
  {\enquote {\bibinfo {title} {Vibronic model for the quantum dynamical study
  of the competition between bright and charge-transfer excited states in
  single-strand polynucleotides: the adenine dimer case},}\ }\href@noop {}
  {\bibfield  {journal} {\bibinfo  {journal} {J. Phys. Chem. A}\ }\textbf
  {\bibinfo {volume} {113}},\ \bibinfo {pages} {15346--15354} (\bibinfo {year}
  {2009})}\BibitemShut {NoStop}%
\bibitem [{\citenamefont {Barbatti}(2011)}]{barbatti2011nonadiabatic}%
  \BibitemOpen
  \bibfield  {author} {\bibinfo {author} {\bibfnamefont {M.}~\bibnamefont
  {Barbatti}},\ }\bibfield  {title} {\enquote {\bibinfo {title} {Nonadiabatic
  dynamics with trajectory surface hopping method},}\ }\href@noop {} {\bibfield
   {journal} {\bibinfo  {journal} {WIREs Comput. Mol. Sci.}\ }\textbf {\bibinfo
  {volume} {1}},\ \bibinfo {pages} {620--633} (\bibinfo {year}
  {2011})}\BibitemShut {NoStop}%
\bibitem [{\citenamefont {Barbatti}\ \emph {et~al.}(2010)\citenamefont
  {Barbatti}, \citenamefont {Aquino}, \citenamefont {Szymczak}, \citenamefont
  {Nachtigallov{\'a}}, \citenamefont {Hobza},\ and\ \citenamefont
  {Lischka}}]{barbatti2010relaxation}%
  \BibitemOpen
  \bibfield  {author} {\bibinfo {author} {\bibfnamefont {M.}~\bibnamefont
  {Barbatti}}, \bibinfo {author} {\bibfnamefont {A.~J.}\ \bibnamefont
  {Aquino}}, \bibinfo {author} {\bibfnamefont {J.~J.}\ \bibnamefont
  {Szymczak}}, \bibinfo {author} {\bibfnamefont {D.}~\bibnamefont
  {Nachtigallov{\'a}}}, \bibinfo {author} {\bibfnamefont {P.}~\bibnamefont
  {Hobza}}, \ and\ \bibinfo {author} {\bibfnamefont {H.}~\bibnamefont
  {Lischka}},\ }\bibfield  {title} {\enquote {\bibinfo {title} {Relaxation
  mechanisms of uv-photoexcited dna and rna nucleobases},}\ }\href@noop {}
  {\bibfield  {journal} {\bibinfo  {journal} {Proc. Natl. Acad. Sci. USA}\
  }\textbf {\bibinfo {volume} {107}},\ \bibinfo {pages} {21453--21458}
  (\bibinfo {year} {2010})}\BibitemShut {NoStop}%
\bibitem [{\citenamefont {Gonz{\'a}lez}, \citenamefont {Escudero},\ and\
  \citenamefont {Serrano-Andr{\'e}s}(2012)}]{gonzalez2012progress}%
  \BibitemOpen
  \bibfield  {author} {\bibinfo {author} {\bibfnamefont {L.}~\bibnamefont
  {Gonz{\'a}lez}}, \bibinfo {author} {\bibfnamefont {D.}~\bibnamefont
  {Escudero}}, \ and\ \bibinfo {author} {\bibfnamefont {L.}~\bibnamefont
  {Serrano-Andr{\'e}s}},\ }\bibfield  {title} {\enquote {\bibinfo {title}
  {Progress and challenges in the calculation of electronic excited states},}\
  }\href@noop {} {\bibfield  {journal} {\bibinfo  {journal} {ChemPhysChem}\
  }\textbf {\bibinfo {volume} {13}},\ \bibinfo {pages} {28--51} (\bibinfo
  {year} {2012})}\BibitemShut {NoStop}%
\bibitem [{\citenamefont {Cannelli}\ \emph {et~al.}(2017)\citenamefont
  {Cannelli}, \citenamefont {Giovannini}, \citenamefont {Baiardi},
  \citenamefont {Carlotti}, \citenamefont {Elisei},\ and\ \citenamefont
  {Cappelli}}]{cannelli2017understanding}%
  \BibitemOpen
  \bibfield  {author} {\bibinfo {author} {\bibfnamefont {O.}~\bibnamefont
  {Cannelli}}, \bibinfo {author} {\bibfnamefont {T.}~\bibnamefont
  {Giovannini}}, \bibinfo {author} {\bibfnamefont {A.}~\bibnamefont {Baiardi}},
  \bibinfo {author} {\bibfnamefont {B.}~\bibnamefont {Carlotti}}, \bibinfo
  {author} {\bibfnamefont {F.}~\bibnamefont {Elisei}}, \ and\ \bibinfo {author}
  {\bibfnamefont {C.}~\bibnamefont {Cappelli}},\ }\bibfield  {title} {\enquote
  {\bibinfo {title} {Understanding the interplay between the solvent and
  nuclear rearrangements in the negative solvatochromism of a push--pull
  flexible quinolinium cation},}\ }\href@noop {} {\bibfield  {journal}
  {\bibinfo  {journal} {Phys. Chem. Chem. Phys.}\ }\textbf {\bibinfo {volume}
  {19}},\ \bibinfo {pages} {32544--32555} (\bibinfo {year} {2017})}\BibitemShut
  {NoStop}%
\bibitem [{\citenamefont {Carlotti}\ \emph {et~al.}(2018)\citenamefont
  {Carlotti}, \citenamefont {Cesaretti}, \citenamefont {Cannelli},
  \citenamefont {Giovannini}, \citenamefont {Cappelli}, \citenamefont
  {Bonaccorso}, \citenamefont {Fortuna}, \citenamefont {Elisei},\ and\
  \citenamefont {Spalletti}}]{carlotti2018evaluation}%
  \BibitemOpen
  \bibfield  {author} {\bibinfo {author} {\bibfnamefont {B.}~\bibnamefont
  {Carlotti}}, \bibinfo {author} {\bibfnamefont {A.}~\bibnamefont {Cesaretti}},
  \bibinfo {author} {\bibfnamefont {O.}~\bibnamefont {Cannelli}}, \bibinfo
  {author} {\bibfnamefont {T.}~\bibnamefont {Giovannini}}, \bibinfo {author}
  {\bibfnamefont {C.}~\bibnamefont {Cappelli}}, \bibinfo {author}
  {\bibfnamefont {C.}~\bibnamefont {Bonaccorso}}, \bibinfo {author}
  {\bibfnamefont {C.~G.}\ \bibnamefont {Fortuna}}, \bibinfo {author}
  {\bibfnamefont {F.}~\bibnamefont {Elisei}}, \ and\ \bibinfo {author}
  {\bibfnamefont {A.}~\bibnamefont {Spalletti}},\ }\bibfield  {title} {\enquote
  {\bibinfo {title} {Evaluation of hyperpolarizability from the solvatochromic
  method: Thiophene containing push- pull cationic dyes as a case study},}\
  }\href@noop {} {\bibfield  {journal} {\bibinfo  {journal} {J. Phys. Chem. C}\
  }\textbf {\bibinfo {volume} {122}},\ \bibinfo {pages} {2285--2296} (\bibinfo
  {year} {2018})}\BibitemShut {NoStop}%
\bibitem [{\citenamefont {Barone}\ \emph {et~al.}(2012)\citenamefont {Barone},
  \citenamefont {Baiardi}, \citenamefont {Biczysko}, \citenamefont {Bloino},
  \citenamefont {Cappelli},\ and\ \citenamefont
  {Lipparini}}]{barone2012implementation}%
  \BibitemOpen
  \bibfield  {author} {\bibinfo {author} {\bibfnamefont {V.}~\bibnamefont
  {Barone}}, \bibinfo {author} {\bibfnamefont {A.}~\bibnamefont {Baiardi}},
  \bibinfo {author} {\bibfnamefont {M.}~\bibnamefont {Biczysko}}, \bibinfo
  {author} {\bibfnamefont {J.}~\bibnamefont {Bloino}}, \bibinfo {author}
  {\bibfnamefont {C.}~\bibnamefont {Cappelli}}, \ and\ \bibinfo {author}
  {\bibfnamefont {F.}~\bibnamefont {Lipparini}},\ }\bibfield  {title} {\enquote
  {\bibinfo {title} {Implementation and validation of a multi-purpose virtual
  spectrometer for large systems in complex environments},}\ }\href@noop {}
  {\bibfield  {journal} {\bibinfo  {journal} {Phys. Chem. Chem. Phys.}\
  }\textbf {\bibinfo {volume} {14}},\ \bibinfo {pages} {12404--12422} (\bibinfo
  {year} {2012})}\BibitemShut {NoStop}%
\bibitem [{\citenamefont {Reichardt}(1994)}]{reichardt1994solvatochromic}%
  \BibitemOpen
  \bibfield  {author} {\bibinfo {author} {\bibfnamefont {C.}~\bibnamefont
  {Reichardt}},\ }\bibfield  {title} {\enquote {\bibinfo {title}
  {Solvatochromic dyes as solvent polarity indicators},}\ }\href@noop {}
  {\bibfield  {journal} {\bibinfo  {journal} {Chem. Rev.}\ }\textbf {\bibinfo
  {volume} {94}},\ \bibinfo {pages} {2319--2358} (\bibinfo {year}
  {1994})}\BibitemShut {NoStop}%
\bibitem [{\citenamefont {Mennucci}, \citenamefont {Cammi},\ and\ \citenamefont
  {Tomasi}(1998)}]{mennucci1998excited}%
  \BibitemOpen
  \bibfield  {author} {\bibinfo {author} {\bibfnamefont {B.}~\bibnamefont
  {Mennucci}}, \bibinfo {author} {\bibfnamefont {R.}~\bibnamefont {Cammi}}, \
  and\ \bibinfo {author} {\bibfnamefont {J.}~\bibnamefont {Tomasi}},\
  }\bibfield  {title} {\enquote {\bibinfo {title} {Excited states and
  solvatochromic shifts within a nonequilibrium solvation approach: A new
  formulation of the integral equation formalism method at the self-consistent
  field, configuration interaction, and multiconfiguration self-consistent
  field level},}\ }\href@noop {} {\bibfield  {journal} {\bibinfo  {journal} {J.
  Chem. Phys.}\ }\textbf {\bibinfo {volume} {109}},\ \bibinfo {pages}
  {2798--2807} (\bibinfo {year} {1998})}\BibitemShut {NoStop}%
\bibitem [{\citenamefont {Jacquemin}, \citenamefont {Mennucci},\ and\
  \citenamefont {Adamo}(2011)}]{jacquemin2011excited}%
  \BibitemOpen
  \bibfield  {author} {\bibinfo {author} {\bibfnamefont {D.}~\bibnamefont
  {Jacquemin}}, \bibinfo {author} {\bibfnamefont {B.}~\bibnamefont {Mennucci}},
  \ and\ \bibinfo {author} {\bibfnamefont {C.}~\bibnamefont {Adamo}},\
  }\bibfield  {title} {\enquote {\bibinfo {title} {Excited-state calculations
  with td-dft: from benchmarks to simulations in complex environments},}\
  }\href@noop {} {\bibfield  {journal} {\bibinfo  {journal} {Phys. Chem. Chem.
  Phys.}\ }\textbf {\bibinfo {volume} {13}},\ \bibinfo {pages} {16987--16998}
  (\bibinfo {year} {2011})}\BibitemShut {NoStop}%
\bibitem [{\citenamefont {Corni}\ \emph {et~al.}(2005)\citenamefont {Corni},
  \citenamefont {Cammi}, \citenamefont {Mennucci},\ and\ \citenamefont
  {Tomasi}}]{corni2005electronic}%
  \BibitemOpen
  \bibfield  {author} {\bibinfo {author} {\bibfnamefont {S.}~\bibnamefont
  {Corni}}, \bibinfo {author} {\bibfnamefont {R.}~\bibnamefont {Cammi}},
  \bibinfo {author} {\bibfnamefont {B.}~\bibnamefont {Mennucci}}, \ and\
  \bibinfo {author} {\bibfnamefont {J.}~\bibnamefont {Tomasi}},\ }\bibfield
  {title} {\enquote {\bibinfo {title} {Electronic excitation energies of
  molecules in solution within continuum solvation models: Investigating the
  discrepancy between state-specific and linear-response methods},}\
  }\href@noop {} {\bibfield  {journal} {\bibinfo  {journal} {J. Chem. Phys.}\
  }\textbf {\bibinfo {volume} {123}},\ \bibinfo {pages} {134512} (\bibinfo
  {year} {2005})}\BibitemShut {NoStop}%
\bibitem [{\citenamefont {Cupellini}\ \emph {et~al.}(2019)\citenamefont
  {Cupellini}, \citenamefont {Corbella}, \citenamefont {Mennucci},\ and\
  \citenamefont {Curutchet}}]{cupellini2019electronic}%
  \BibitemOpen
  \bibfield  {author} {\bibinfo {author} {\bibfnamefont {L.}~\bibnamefont
  {Cupellini}}, \bibinfo {author} {\bibfnamefont {M.}~\bibnamefont {Corbella}},
  \bibinfo {author} {\bibfnamefont {B.}~\bibnamefont {Mennucci}}, \ and\
  \bibinfo {author} {\bibfnamefont {C.}~\bibnamefont {Curutchet}},\ }\bibfield
  {title} {\enquote {\bibinfo {title} {Electronic energy transfer in
  biomacromolecules},}\ }\href@noop {} {\bibfield  {journal} {\bibinfo
  {journal} {WIREs Comput. Mol. Sci.}\ }\textbf {\bibinfo {volume} {9}},\
  \bibinfo {pages} {e1392} (\bibinfo {year} {2019})}\BibitemShut {NoStop}%
\bibitem [{\citenamefont {Mennucci}\ and\ \citenamefont
  {Corni}(2019)}]{mennucci2019multiscale}%
  \BibitemOpen
  \bibfield  {author} {\bibinfo {author} {\bibfnamefont {B.}~\bibnamefont
  {Mennucci}}\ and\ \bibinfo {author} {\bibfnamefont {S.}~\bibnamefont
  {Corni}},\ }\bibfield  {title} {\enquote {\bibinfo {title} {Multiscale
  modelling of photoinduced processes in composite systems},}\ }\href@noop {}
  {\bibfield  {journal} {\bibinfo  {journal} {Nat. Rev. Chem.}\ }\textbf
  {\bibinfo {volume} {3}},\ \bibinfo {pages} {315–330} (\bibinfo {year}
  {2019})}\BibitemShut {NoStop}%
\bibitem [{\citenamefont {Warshel}\ and\ \citenamefont
  {Levitt}(1976)}]{warshel1976theoretical}%
  \BibitemOpen
  \bibfield  {author} {\bibinfo {author} {\bibfnamefont {A.}~\bibnamefont
  {Warshel}}\ and\ \bibinfo {author} {\bibfnamefont {M.}~\bibnamefont
  {Levitt}},\ }\bibfield  {title} {\enquote {\bibinfo {title} {Theoretical
  studies of enzymic reactions: dielectric, electrostatic and steric
  stabilization of the carbonium ion in the reaction of lysozyme},}\
  }\href@noop {} {\bibfield  {journal} {\bibinfo  {journal} {J. Mol. Biol.}\
  }\textbf {\bibinfo {volume} {103}},\ \bibinfo {pages} {227--249} (\bibinfo
  {year} {1976})}\BibitemShut {NoStop}%
\bibitem [{\citenamefont {Warshel}\ and\ \citenamefont
  {Karplus}(1972)}]{warshel1972calculation}%
  \BibitemOpen
  \bibfield  {author} {\bibinfo {author} {\bibfnamefont {A.}~\bibnamefont
  {Warshel}}\ and\ \bibinfo {author} {\bibfnamefont {M.}~\bibnamefont
  {Karplus}},\ }\bibfield  {title} {\enquote {\bibinfo {title} {Calculation of
  ground and excited state potential surfaces of conjugated molecules. i.
  formulation and parametrization},}\ }\href@noop {} {\bibfield  {journal}
  {\bibinfo  {journal} {J. Am. Chem. Soc.}\ }\textbf {\bibinfo {volume} {94}},\
  \bibinfo {pages} {5612--5625} (\bibinfo {year} {1972})}\BibitemShut {NoStop}%
\bibitem [{\citenamefont {Miertu{\v{s}}}, \citenamefont {Scrocco},\ and\
  \citenamefont {Tomasi}(1981)}]{miertuvs1981electrostatic}%
  \BibitemOpen
  \bibfield  {author} {\bibinfo {author} {\bibfnamefont {S.}~\bibnamefont
  {Miertu{\v{s}}}}, \bibinfo {author} {\bibfnamefont {E.}~\bibnamefont
  {Scrocco}}, \ and\ \bibinfo {author} {\bibfnamefont {J.}~\bibnamefont
  {Tomasi}},\ }\bibfield  {title} {\enquote {\bibinfo {title} {Electrostatic
  interaction of a solute with a continuum. a direct utilizaion of ab initio
  molecular potentials for the prevision of solvent effects},}\ }\href@noop {}
  {\bibfield  {journal} {\bibinfo  {journal} {Chem. Phys.}\ }\textbf {\bibinfo
  {volume} {55}},\ \bibinfo {pages} {117--129} (\bibinfo {year}
  {1981})}\BibitemShut {NoStop}%
\bibitem [{\citenamefont {Tomasi}\ and\ \citenamefont
  {Persico}(1994)}]{tomasi1994molecular}%
  \BibitemOpen
  \bibfield  {author} {\bibinfo {author} {\bibfnamefont {J.}~\bibnamefont
  {Tomasi}}\ and\ \bibinfo {author} {\bibfnamefont {M.}~\bibnamefont
  {Persico}},\ }\bibfield  {title} {\enquote {\bibinfo {title} {Molecular
  interactions in solution: an overview of methods based on continuous
  distributions of the solvent},}\ }\href@noop {} {\bibfield  {journal}
  {\bibinfo  {journal} {Chem. Rev.}\ }\textbf {\bibinfo {volume} {94}},\
  \bibinfo {pages} {2027--2094} (\bibinfo {year} {1994})}\BibitemShut {NoStop}%
\bibitem [{\citenamefont {Orozco}\ and\ \citenamefont
  {Luque}(2000)}]{orozco2000theoretical}%
  \BibitemOpen
  \bibfield  {author} {\bibinfo {author} {\bibfnamefont {M.}~\bibnamefont
  {Orozco}}\ and\ \bibinfo {author} {\bibfnamefont {F.~J.}\ \bibnamefont
  {Luque}},\ }\bibfield  {title} {\enquote {\bibinfo {title} {Theoretical
  methods for the description of the solvent effect in biomolecular systems},}\
  }\href@noop {} {\bibfield  {journal} {\bibinfo  {journal} {Chem. Rev.}\
  }\textbf {\bibinfo {volume} {100}},\ \bibinfo {pages} {4187--4226} (\bibinfo
  {year} {2000})}\BibitemShut {NoStop}%
\bibitem [{\citenamefont {Tomasi}, \citenamefont {Mennucci},\ and\
  \citenamefont {Cammi}(2005)}]{tomasi2005}%
  \BibitemOpen
  \bibfield  {author} {\bibinfo {author} {\bibfnamefont {J.}~\bibnamefont
  {Tomasi}}, \bibinfo {author} {\bibfnamefont {B.}~\bibnamefont {Mennucci}}, \
  and\ \bibinfo {author} {\bibfnamefont {R.}~\bibnamefont {Cammi}},\ }\bibfield
   {title} {\enquote {\bibinfo {title} {Quantum mechanical continuum solvation
  models},}\ }\href@noop {} {\bibfield  {journal} {\bibinfo  {journal} {Chem.
  Rev.}\ }\textbf {\bibinfo {volume} {105}},\ \bibinfo {pages} {2999--3094}
  (\bibinfo {year} {2005})}\BibitemShut {NoStop}%
\bibitem [{\citenamefont {Mennucci}\ and\ \citenamefont
  {Cammi}(2007)}]{mennucci2007continuum}%
  \BibitemOpen
  \bibinfo {editor} {\bibfnamefont {B.}~\bibnamefont {Mennucci}}\ and\ \bibinfo
  {editor} {\bibfnamefont {R.}~\bibnamefont {Cammi}},\ eds.,\ \href@noop {}
  {\emph {\bibinfo {title} {Continuum Solvation Models in Chemical Physics}}}\
  (\bibinfo  {publisher} {Wiley, New York},\ \bibinfo {year}
  {2007})\BibitemShut {NoStop}%
\bibitem [{\citenamefont {Mennucci}(2010)}]{mennucci2010continuum}%
  \BibitemOpen
  \bibfield  {author} {\bibinfo {author} {\bibfnamefont {B.}~\bibnamefont
  {Mennucci}},\ }\bibfield  {title} {\enquote {\bibinfo {title} {{Continuum
  Solvation Models: What Else Can We Learn from Them?}}}\ }\href {\doibase
  10.1021/jz100506s} {\bibfield  {journal} {\bibinfo  {journal} {J. Phys. Chem.
  Lett.}\ }\textbf {\bibinfo {volume} {1}},\ \bibinfo {pages} {1666--1674}
  (\bibinfo {year} {2010})},\ \Eprint
  {http://arxiv.org/abs/http://pubs.acs.org/doi/pdf/10.1021/jz100506s}
  {http://pubs.acs.org/doi/pdf/10.1021/jz100506s} \BibitemShut {NoStop}%
\bibitem [{\citenamefont {Mennucci}(2013)}]{mennucci2013modeling}%
  \BibitemOpen
  \bibfield  {author} {\bibinfo {author} {\bibfnamefont {B.}~\bibnamefont
  {Mennucci}},\ }\bibfield  {title} {\enquote {\bibinfo {title} {Modeling
  environment effects on spectroscopies through qm/classical models},}\
  }\href@noop {} {\bibfield  {journal} {\bibinfo  {journal} {Phys. Chem. Chem.
  Phys.}\ }\textbf {\bibinfo {volume} {15}},\ \bibinfo {pages} {6583--6594}
  (\bibinfo {year} {2013})}\BibitemShut {NoStop}%
\bibitem [{\citenamefont {Lipparini}\ and\ \citenamefont
  {Mennucci}(2016)}]{lipparini2016perspective}%
  \BibitemOpen
  \bibfield  {author} {\bibinfo {author} {\bibfnamefont {F.}~\bibnamefont
  {Lipparini}}\ and\ \bibinfo {author} {\bibfnamefont {B.}~\bibnamefont
  {Mennucci}},\ }\bibfield  {title} {\enquote {\bibinfo {title} {Perspective:
  Polarizable continuum models for quantum-mechanical descriptions},}\
  }\href@noop {} {\bibfield  {journal} {\bibinfo  {journal} {J. Chem. Phys.}\
  }\textbf {\bibinfo {volume} {144}},\ \bibinfo {pages} {160901} (\bibinfo
  {year} {2016})}\BibitemShut {NoStop}%
\bibitem [{\citenamefont {Senn}\ and\ \citenamefont
  {Thiel}(2009)}]{senn2009qm}%
  \BibitemOpen
  \bibfield  {author} {\bibinfo {author} {\bibfnamefont {H.~M.}\ \bibnamefont
  {Senn}}\ and\ \bibinfo {author} {\bibfnamefont {W.}~\bibnamefont {Thiel}},\
  }\bibfield  {title} {\enquote {\bibinfo {title} {{QM/MM} methods for
  biomolecular systems},}\ }\href@noop {} {\bibfield  {journal} {\bibinfo
  {journal} {Angew. Chem. Int. Ed.}\ }\textbf {\bibinfo {volume} {48}},\
  \bibinfo {pages} {1198--1229} (\bibinfo {year} {2009})}\BibitemShut {NoStop}%
\bibitem [{\citenamefont {Lin}\ and\ \citenamefont
  {Truhlar}(2007)}]{lin2007qm}%
  \BibitemOpen
  \bibfield  {author} {\bibinfo {author} {\bibfnamefont {H.}~\bibnamefont
  {Lin}}\ and\ \bibinfo {author} {\bibfnamefont {D.~G.}\ \bibnamefont
  {Truhlar}},\ }\bibfield  {title} {\enquote {\bibinfo {title} {{QM/MM}: what
  have we learned, where are we, and where do we go from here?}}\ }\href@noop
  {} {\bibfield  {journal} {\bibinfo  {journal} {Theor. Chem. Acc.}\ }\textbf
  {\bibinfo {volume} {117}},\ \bibinfo {pages} {185--199} (\bibinfo {year}
  {2007})}\BibitemShut {NoStop}%
\bibitem [{\citenamefont {Jensen}\ and\ \citenamefont
  {Gordon}(1996)}]{gordon1996approximate}%
  \BibitemOpen
  \bibfield  {author} {\bibinfo {author} {\bibfnamefont {J.~H.}\ \bibnamefont
  {Jensen}}\ and\ \bibinfo {author} {\bibfnamefont {M.~S.}\ \bibnamefont
  {Gordon}},\ }\bibfield  {title} {\enquote {\bibinfo {title} {An approximate
  formula for the intermolecular pauli repulsion between closed shell
  molecules},}\ }\href@noop {} {\bibfield  {journal} {\bibinfo  {journal} {Mol.
  Phys.}\ }\textbf {\bibinfo {volume} {89}},\ \bibinfo {pages} {1313--1325}
  (\bibinfo {year} {1996})}\BibitemShut {NoStop}%
\bibitem [{\citenamefont {Jensen}\ and\ \citenamefont
  {Gordon}(1998)}]{jensen1998approximate}%
  \BibitemOpen
  \bibfield  {author} {\bibinfo {author} {\bibfnamefont {J.~H.}\ \bibnamefont
  {Jensen}}\ and\ \bibinfo {author} {\bibfnamefont {M.~S.}\ \bibnamefont
  {Gordon}},\ }\bibfield  {title} {\enquote {\bibinfo {title} {An approximate
  formula for the intermolecular pauli repulsion between closed shell
  molecules. ii. application to the effective fragment potential method},}\
  }\href@noop {} {\bibfield  {journal} {\bibinfo  {journal} {J. Chem. Phys.}\
  }\textbf {\bibinfo {volume} {108}},\ \bibinfo {pages} {4772--4782} (\bibinfo
  {year} {1998})}\BibitemShut {NoStop}%
\bibitem [{\citenamefont {Ben-Nun}\ and\ \citenamefont
  {Mart{\'i}nez}(1998)}]{ben1998direct}%
  \BibitemOpen
  \bibfield  {author} {\bibinfo {author} {\bibfnamefont {M.}~\bibnamefont
  {Ben-Nun}}\ and\ \bibinfo {author} {\bibfnamefont {T.~J.}\ \bibnamefont
  {Mart{\'i}nez}},\ }\bibfield  {title} {\enquote {\bibinfo {title} {Direct
  evaluation of the pauli repulsion energy usingclassical'wavefunctions in
  hybrid quantum/classical potential energy surfaces},}\ }\href@noop {}
  {\bibfield  {journal} {\bibinfo  {journal} {Chem. Phys. Lett.}\ }\textbf
  {\bibinfo {volume} {290}},\ \bibinfo {pages} {289--295} (\bibinfo {year}
  {1998})}\BibitemShut {NoStop}%
\bibitem [{\citenamefont {Giovannini}, \citenamefont {Lafiosca},\ and\
  \citenamefont {Cappelli}(2017)}]{giovannini2017disrep}%
  \BibitemOpen
  \bibfield  {author} {\bibinfo {author} {\bibfnamefont {T.}~\bibnamefont
  {Giovannini}}, \bibinfo {author} {\bibfnamefont {P.}~\bibnamefont
  {Lafiosca}}, \ and\ \bibinfo {author} {\bibfnamefont {C.}~\bibnamefont
  {Cappelli}},\ }\bibfield  {title} {\enquote {\bibinfo {title} {A general
  route to include pauli repulsion and quantum dispersion effects in qm/mm
  approaches},}\ }\href@noop {} {\bibfield  {journal} {\bibinfo  {journal} {J.
  Chem. Theory Comput.}\ }\textbf {\bibinfo {volume} {13}},\ \bibinfo {pages}
  {4854--4870} (\bibinfo {year} {2017})}\BibitemShut {NoStop}%
\bibitem [{\citenamefont {Curutchet}\ \emph {et~al.}(2018)\citenamefont
  {Curutchet}, \citenamefont {Cupellini}, \citenamefont {Kongsted},
  \citenamefont {Corni}, \citenamefont {Frediani}, \citenamefont {Steindal},
  \citenamefont {Guido}, \citenamefont {Scalmani},\ and\ \citenamefont
  {Mennucci}}]{curutchet2018density}%
  \BibitemOpen
  \bibfield  {author} {\bibinfo {author} {\bibfnamefont {C.}~\bibnamefont
  {Curutchet}}, \bibinfo {author} {\bibfnamefont {L.}~\bibnamefont
  {Cupellini}}, \bibinfo {author} {\bibfnamefont {J.}~\bibnamefont {Kongsted}},
  \bibinfo {author} {\bibfnamefont {S.}~\bibnamefont {Corni}}, \bibinfo
  {author} {\bibfnamefont {L.}~\bibnamefont {Frediani}}, \bibinfo {author}
  {\bibfnamefont {A.~H.}\ \bibnamefont {Steindal}}, \bibinfo {author}
  {\bibfnamefont {C.~A.}\ \bibnamefont {Guido}}, \bibinfo {author}
  {\bibfnamefont {G.}~\bibnamefont {Scalmani}}, \ and\ \bibinfo {author}
  {\bibfnamefont {B.}~\bibnamefont {Mennucci}},\ }\bibfield  {title} {\enquote
  {\bibinfo {title} {Density-dependent formulation of dispersion--repulsion
  interactions in hybrid multiscale quantum/molecular mechanics (qm/mm)
  models},}\ }\href@noop {} {\bibfield  {journal} {\bibinfo  {journal} {J.
  Chem. Theory Comput.}\ }\textbf {\bibinfo {volume} {14}},\ \bibinfo {pages}
  {1671--1681} (\bibinfo {year} {2018})}\BibitemShut {NoStop}%
\bibitem [{\citenamefont {Giovannini}\ \emph
  {et~al.}(2019{\natexlab{a}})\citenamefont {Giovannini}, \citenamefont
  {Lafiosca}, \citenamefont {Chandramouli}, \citenamefont {Barone},\ and\
  \citenamefont {Cappelli}}]{giovannini2019eprdisrep}%
  \BibitemOpen
  \bibfield  {author} {\bibinfo {author} {\bibfnamefont {T.}~\bibnamefont
  {Giovannini}}, \bibinfo {author} {\bibfnamefont {P.}~\bibnamefont
  {Lafiosca}}, \bibinfo {author} {\bibfnamefont {B.}~\bibnamefont
  {Chandramouli}}, \bibinfo {author} {\bibfnamefont {V.}~\bibnamefont
  {Barone}}, \ and\ \bibinfo {author} {\bibfnamefont {C.}~\bibnamefont
  {Cappelli}},\ }\bibfield  {title} {\enquote {\bibinfo {title} {Effective yet
  reliable computation of hyperfine coupling constants in solution by a qm/mm
  approach: Interplay between electrostatics and non-electrostatic effects},}\
  }\href@noop {} {\bibfield  {journal} {\bibinfo  {journal} {J. Chem. Phys.}\
  }\textbf {\bibinfo {volume} {150}},\ \bibinfo {pages} {124102} (\bibinfo
  {year} {2019}{\natexlab{a}})}\BibitemShut {NoStop}%
\bibitem [{\citenamefont {N{\aa}bo}\ \emph {et~al.}(2016)\citenamefont
  {N{\aa}bo}, \citenamefont {Olsen}, \citenamefont {Holmgaard~List},
  \citenamefont {Solanko}, \citenamefont {W{\"u}stner},\ and\ \citenamefont
  {Kongsted}}]{naabo2016embedding}%
  \BibitemOpen
  \bibfield  {author} {\bibinfo {author} {\bibfnamefont {L.~J.}\ \bibnamefont
  {N{\aa}bo}}, \bibinfo {author} {\bibfnamefont {J.~M.~H.}\ \bibnamefont
  {Olsen}}, \bibinfo {author} {\bibfnamefont {N.}~\bibnamefont
  {Holmgaard~List}}, \bibinfo {author} {\bibfnamefont {L.~M.}\ \bibnamefont
  {Solanko}}, \bibinfo {author} {\bibfnamefont {D.}~\bibnamefont
  {W{\"u}stner}}, \ and\ \bibinfo {author} {\bibfnamefont {J.}~\bibnamefont
  {Kongsted}},\ }\bibfield  {title} {\enquote {\bibinfo {title} {Embedding
  beyond electrostatics—the role of wave function confinement},}\ }\href@noop
  {} {\bibfield  {journal} {\bibinfo  {journal} {J. Chem. Phys.}\ }\textbf
  {\bibinfo {volume} {145}},\ \bibinfo {pages} {104102} (\bibinfo {year}
  {2016})}\BibitemShut {NoStop}%
\bibitem [{\citenamefont {Reinholdt}, \citenamefont {Kongsted},\ and\
  \citenamefont {Olsen}(2017)}]{reinholdt2017polarizable}%
  \BibitemOpen
  \bibfield  {author} {\bibinfo {author} {\bibfnamefont {P.}~\bibnamefont
  {Reinholdt}}, \bibinfo {author} {\bibfnamefont {J.}~\bibnamefont {Kongsted}},
  \ and\ \bibinfo {author} {\bibfnamefont {J.~M.~H.}\ \bibnamefont {Olsen}},\
  }\bibfield  {title} {\enquote {\bibinfo {title} {Polarizable density
  embedding: A solution to the electron spill-out problem in multiscale
  modeling},}\ }\href@noop {} {\bibfield  {journal} {\bibinfo  {journal} {J.
  Phys. Chem. Lett.}\ }\textbf {\bibinfo {volume} {8}},\ \bibinfo {pages}
  {5949--5958} (\bibinfo {year} {2017})}\BibitemShut {NoStop}%
\bibitem [{\citenamefont {Day}\ \emph {et~al.}(1996)\citenamefont {Day},
  \citenamefont {Jensen}, \citenamefont {Gordon}, \citenamefont {Webb},
  \citenamefont {Stevens}, \citenamefont {Krauss}, \citenamefont {Garmer},
  \citenamefont {Basch},\ and\ \citenamefont {Cohen}}]{day1996effective}%
  \BibitemOpen
  \bibfield  {author} {\bibinfo {author} {\bibfnamefont {P.~N.}\ \bibnamefont
  {Day}}, \bibinfo {author} {\bibfnamefont {J.~H.}\ \bibnamefont {Jensen}},
  \bibinfo {author} {\bibfnamefont {M.~S.}\ \bibnamefont {Gordon}}, \bibinfo
  {author} {\bibfnamefont {S.~P.}\ \bibnamefont {Webb}}, \bibinfo {author}
  {\bibfnamefont {W.~J.}\ \bibnamefont {Stevens}}, \bibinfo {author}
  {\bibfnamefont {M.}~\bibnamefont {Krauss}}, \bibinfo {author} {\bibfnamefont
  {D.}~\bibnamefont {Garmer}}, \bibinfo {author} {\bibfnamefont
  {H.}~\bibnamefont {Basch}}, \ and\ \bibinfo {author} {\bibfnamefont
  {D.}~\bibnamefont {Cohen}},\ }\bibfield  {title} {\enquote {\bibinfo {title}
  {An effective fragment method for modeling solvent effects in quantum
  mechanical calculations},}\ }\href@noop {} {\bibfield  {journal} {\bibinfo
  {journal} {J. Chem. Phys.}\ }\textbf {\bibinfo {volume} {105}},\ \bibinfo
  {pages} {1968--1986} (\bibinfo {year} {1996})}\BibitemShut {NoStop}%
\bibitem [{\citenamefont {Kairys}\ and\ \citenamefont
  {Jensen}(2000)}]{kairys2000qm}%
  \BibitemOpen
  \bibfield  {author} {\bibinfo {author} {\bibfnamefont {V.}~\bibnamefont
  {Kairys}}\ and\ \bibinfo {author} {\bibfnamefont {J.~H.}\ \bibnamefont
  {Jensen}},\ }\bibfield  {title} {\enquote {\bibinfo {title} {Qm/mm boundaries
  across covalent bonds: a frozen localized molecular orbital-based approach
  for the effective fragment potential method},}\ }\href@noop {} {\bibfield
  {journal} {\bibinfo  {journal} {J. Phys. Chem. A}\ }\textbf {\bibinfo
  {volume} {104}},\ \bibinfo {pages} {6656--6665} (\bibinfo {year}
  {2000})}\BibitemShut {NoStop}%
\bibitem [{\citenamefont {Mao}\ \emph {et~al.}(2016)\citenamefont {Mao},
  \citenamefont {Demerdash}, \citenamefont {Head-Gordon},\ and\ \citenamefont
  {Head-Gordon}}]{mao2016assessing}%
  \BibitemOpen
  \bibfield  {author} {\bibinfo {author} {\bibfnamefont {Y.}~\bibnamefont
  {Mao}}, \bibinfo {author} {\bibfnamefont {O.}~\bibnamefont {Demerdash}},
  \bibinfo {author} {\bibfnamefont {M.}~\bibnamefont {Head-Gordon}}, \ and\
  \bibinfo {author} {\bibfnamefont {T.}~\bibnamefont {Head-Gordon}},\
  }\bibfield  {title} {\enquote {\bibinfo {title} {Assessing ion--water
  interactions in the amoeba force field using energy decomposition analysis of
  electronic structure calculations},}\ }\href@noop {} {\bibfield  {journal}
  {\bibinfo  {journal} {J. Chem. Theory Comput.}\ }\textbf {\bibinfo {volume}
  {12}},\ \bibinfo {pages} {5422--5437} (\bibinfo {year} {2016})}\BibitemShut
  {NoStop}%
\bibitem [{\citenamefont {Loco}\ \emph {et~al.}(2016)\citenamefont {Loco},
  \citenamefont {Polack}, \citenamefont {Caprasecca}, \citenamefont
  {Lagardere}, \citenamefont {Lipparini}, \citenamefont {Piquemal},\ and\
  \citenamefont {Mennucci}}]{loco2016qm}%
  \BibitemOpen
  \bibfield  {author} {\bibinfo {author} {\bibfnamefont {D.}~\bibnamefont
  {Loco}}, \bibinfo {author} {\bibfnamefont {{\'E}.}~\bibnamefont {Polack}},
  \bibinfo {author} {\bibfnamefont {S.}~\bibnamefont {Caprasecca}}, \bibinfo
  {author} {\bibfnamefont {L.}~\bibnamefont {Lagardere}}, \bibinfo {author}
  {\bibfnamefont {F.}~\bibnamefont {Lipparini}}, \bibinfo {author}
  {\bibfnamefont {J.-P.}\ \bibnamefont {Piquemal}}, \ and\ \bibinfo {author}
  {\bibfnamefont {B.}~\bibnamefont {Mennucci}},\ }\bibfield  {title} {\enquote
  {\bibinfo {title} {A qm/mm approach using the amoeba polarizable embedding:
  from ground state energies to electronic excitations},}\ }\href@noop {}
  {\bibfield  {journal} {\bibinfo  {journal} {J. Chem. Theory Comput.}\
  }\textbf {\bibinfo {volume} {12}},\ \bibinfo {pages} {3654--3661} (\bibinfo
  {year} {2016})}\BibitemShut {NoStop}%
\bibitem [{\citenamefont {Loco}\ and\ \citenamefont
  {Cupellini}(2018)}]{loco2018modelingijqc}%
  \BibitemOpen
  \bibfield  {author} {\bibinfo {author} {\bibfnamefont {D.}~\bibnamefont
  {Loco}}\ and\ \bibinfo {author} {\bibfnamefont {L.}~\bibnamefont
  {Cupellini}},\ }\bibfield  {title} {\enquote {\bibinfo {title} {Modeling the
  absorption lineshape of embedded systems from molecular dynamics: A tutorial
  review},}\ }\href@noop {} {\bibfield  {journal} {\bibinfo  {journal} {Int. J.
  Quantum Chem.}\ ,\ \bibinfo {pages} {DOI: 10.1002/qua.25726}} (\bibinfo
  {year} {2018})}\BibitemShut {NoStop}%
\bibitem [{\citenamefont {Thole}(1981)}]{thole1981molecular}%
  \BibitemOpen
  \bibfield  {author} {\bibinfo {author} {\bibfnamefont {B.~T.}\ \bibnamefont
  {Thole}},\ }\bibfield  {title} {\enquote {\bibinfo {title} {Molecular
  polarizabilities calculated with a modified dipole interaction},}\
  }\href@noop {} {\bibfield  {journal} {\bibinfo  {journal} {Chem. Phys.}\
  }\textbf {\bibinfo {volume} {59}},\ \bibinfo {pages} {341--350} (\bibinfo
  {year} {1981})}\BibitemShut {NoStop}%
\bibitem [{\citenamefont {Curutchet}\ \emph {et~al.}(2009)\citenamefont
  {Curutchet}, \citenamefont {Mu{\~n}oz-Losa}, \citenamefont {Monti},
  \citenamefont {Kongsted}, \citenamefont {Scholes},\ and\ \citenamefont
  {Mennucci}}]{curutchet2009electronic}%
  \BibitemOpen
  \bibfield  {author} {\bibinfo {author} {\bibfnamefont {C.}~\bibnamefont
  {Curutchet}}, \bibinfo {author} {\bibfnamefont {A.}~\bibnamefont
  {Mu{\~n}oz-Losa}}, \bibinfo {author} {\bibfnamefont {S.}~\bibnamefont
  {Monti}}, \bibinfo {author} {\bibfnamefont {J.}~\bibnamefont {Kongsted}},
  \bibinfo {author} {\bibfnamefont {G.~D.}\ \bibnamefont {Scholes}}, \ and\
  \bibinfo {author} {\bibfnamefont {B.}~\bibnamefont {Mennucci}},\ }\bibfield
  {title} {\enquote {\bibinfo {title} {Electronic energy transfer in condensed
  phase studied by a polarizable qm/mm model},}\ }\href@noop {} {\bibfield
  {journal} {\bibinfo  {journal} {J. Chem. Theory Comput.}\ }\textbf {\bibinfo
  {volume} {5}},\ \bibinfo {pages} {1838--1848} (\bibinfo {year}
  {2009})}\BibitemShut {NoStop}%
\bibitem [{\citenamefont {Olsen}\ and\ \citenamefont
  {Kongsted}(2011)}]{olsen2011molecular}%
  \BibitemOpen
  \bibfield  {author} {\bibinfo {author} {\bibfnamefont {J.~M.~H.}\
  \bibnamefont {Olsen}}\ and\ \bibinfo {author} {\bibfnamefont
  {J.}~\bibnamefont {Kongsted}},\ }\bibfield  {title} {\enquote {\bibinfo
  {title} {Molecular properties through polarizable embedding},}\ }in\
  \href@noop {} {\emph {\bibinfo {booktitle} {Adv. Quantum Chem.}}},\
  Vol.~\bibinfo {volume} {61}\ (\bibinfo  {publisher} {Elsevier},\ \bibinfo
  {year} {2011})\ pp.\ \bibinfo {pages} {107--143}\BibitemShut {NoStop}%
\bibitem [{\citenamefont {Steindal}\ \emph {et~al.}(2011)\citenamefont
  {Steindal}, \citenamefont {Ruud}, \citenamefont {Frediani}, \citenamefont
  {Aidas},\ and\ \citenamefont {Kongsted}}]{steindal2011excitation}%
  \BibitemOpen
  \bibfield  {author} {\bibinfo {author} {\bibfnamefont {A.~H.}\ \bibnamefont
  {Steindal}}, \bibinfo {author} {\bibfnamefont {K.}~\bibnamefont {Ruud}},
  \bibinfo {author} {\bibfnamefont {L.}~\bibnamefont {Frediani}}, \bibinfo
  {author} {\bibfnamefont {K.}~\bibnamefont {Aidas}}, \ and\ \bibinfo {author}
  {\bibfnamefont {J.}~\bibnamefont {Kongsted}},\ }\bibfield  {title} {\enquote
  {\bibinfo {title} {Excitation energies in solution: the fully polarizable
  qm/mm/pcm method},}\ }\href@noop {} {\bibfield  {journal} {\bibinfo
  {journal} {J. Phys. Chem. B}\ }\textbf {\bibinfo {volume} {115}},\ \bibinfo
  {pages} {3027--3037} (\bibinfo {year} {2011})}\BibitemShut {NoStop}%
\bibitem [{\citenamefont {Jurinovich}, \citenamefont {Curutchet},\ and\
  \citenamefont {Mennucci}(2014)}]{jurinovich2014fenna}%
  \BibitemOpen
  \bibfield  {author} {\bibinfo {author} {\bibfnamefont {S.}~\bibnamefont
  {Jurinovich}}, \bibinfo {author} {\bibfnamefont {C.}~\bibnamefont
  {Curutchet}}, \ and\ \bibinfo {author} {\bibfnamefont {B.}~\bibnamefont
  {Mennucci}},\ }\bibfield  {title} {\enquote {\bibinfo {title} {The
  fenna--matthews--olson protein revisited: A fully polarizable (td) dft/mm
  description},}\ }\href@noop {} {\bibfield  {journal} {\bibinfo  {journal}
  {ChemPhysChem}\ }\textbf {\bibinfo {volume} {15}},\ \bibinfo {pages}
  {3194--3204} (\bibinfo {year} {2014})}\BibitemShut {NoStop}%
\bibitem [{\citenamefont {Boulanger}\ and\ \citenamefont
  {Thiel}(2012)}]{boulanger2012solvent}%
  \BibitemOpen
  \bibfield  {author} {\bibinfo {author} {\bibfnamefont {E.}~\bibnamefont
  {Boulanger}}\ and\ \bibinfo {author} {\bibfnamefont {W.}~\bibnamefont
  {Thiel}},\ }\bibfield  {title} {\enquote {\bibinfo {title} {Solvent boundary
  potentials for hybrid qm/mm computations using classical drude oscillators: a
  fully polarizable model},}\ }\href@noop {} {\bibfield  {journal} {\bibinfo
  {journal} {J. Chem. Theory Comput.}\ }\textbf {\bibinfo {volume} {8}},\
  \bibinfo {pages} {4527--4538} (\bibinfo {year} {2012})}\BibitemShut {NoStop}%
\bibitem [{\citenamefont {Rick}, \citenamefont {Stuart},\ and\ \citenamefont
  {Berne}(1994)}]{rick1994dynamical}%
  \BibitemOpen
  \bibfield  {author} {\bibinfo {author} {\bibfnamefont {S.~W.}\ \bibnamefont
  {Rick}}, \bibinfo {author} {\bibfnamefont {S.~J.}\ \bibnamefont {Stuart}}, \
  and\ \bibinfo {author} {\bibfnamefont {B.~J.}\ \bibnamefont {Berne}},\
  }\bibfield  {title} {\enquote {\bibinfo {title} {Dynamical fluctuating charge
  force fields: Application to liquid water},}\ }\href@noop {} {\bibfield
  {journal} {\bibinfo  {journal} {J. Chem. Phys.}\ }\textbf {\bibinfo {volume}
  {101}},\ \bibinfo {pages} {6141--6156} (\bibinfo {year} {1994})}\BibitemShut
  {NoStop}%
\bibitem [{\citenamefont {Rick}\ and\ \citenamefont
  {Berne}(1996)}]{rick1996dynamical}%
  \BibitemOpen
  \bibfield  {author} {\bibinfo {author} {\bibfnamefont {S.~W.}\ \bibnamefont
  {Rick}}\ and\ \bibinfo {author} {\bibfnamefont {B.~J.}\ \bibnamefont
  {Berne}},\ }\bibfield  {title} {\enquote {\bibinfo {title} {{Dynamical
  Fluctuating Charge Force Fields: The Aqueous Solvation of Amides}},}\ }\href
  {\doibase 10.1021/ja952535b} {\bibfield  {journal} {\bibinfo  {journal} {J.
  Am. Chem. Soc.}\ }\textbf {\bibinfo {volume} {118}},\ \bibinfo {pages}
  {672--679} (\bibinfo {year} {1996})},\ \Eprint
  {http://arxiv.org/abs/http://pubs.acs.org/doi/pdf/10.1021/ja952535b}
  {http://pubs.acs.org/doi/pdf/10.1021/ja952535b} \BibitemShut {NoStop}%
\bibitem [{\citenamefont {Cappelli}(2016)}]{cappelli2016integrated}%
  \BibitemOpen
  \bibfield  {author} {\bibinfo {author} {\bibfnamefont {C.}~\bibnamefont
  {Cappelli}},\ }\bibfield  {title} {\enquote {\bibinfo {title} {Integrated
  qm/polarizable mm/continuum approaches to model chiroptical properties of
  strongly interacting solute-solvent systems},}\ }\href@noop {} {\bibfield
  {journal} {\bibinfo  {journal} {Int. J. Quantum Chem.}\ }\textbf {\bibinfo
  {volume} {116}},\ \bibinfo {pages} {1532--1542} (\bibinfo {year}
  {2016})}\BibitemShut {NoStop}%
\bibitem [{\citenamefont {Giovannini}\ \emph
  {et~al.}(2019{\natexlab{b}})\citenamefont {Giovannini}, \citenamefont
  {Puglisi}, \citenamefont {Ambrosetti},\ and\ \citenamefont
  {Cappelli}}]{giovannini2019fqfmu}%
  \BibitemOpen
  \bibfield  {author} {\bibinfo {author} {\bibfnamefont {T.}~\bibnamefont
  {Giovannini}}, \bibinfo {author} {\bibfnamefont {A.}~\bibnamefont {Puglisi}},
  \bibinfo {author} {\bibfnamefont {M.}~\bibnamefont {Ambrosetti}}, \ and\
  \bibinfo {author} {\bibfnamefont {C.}~\bibnamefont {Cappelli}},\ }\bibfield
  {title} {\enquote {\bibinfo {title} {Polarizable qm/mm approach with
  fluctuating charges and fluctuating dipoles: the qm/fqf$\mu$ model},}\
  }\href@noop {} {\bibfield  {journal} {\bibinfo  {journal} {J. Chem. Theory
  Comput.}\ }\textbf {\bibinfo {volume} {15}},\ \bibinfo {pages} {2233--2245}
  (\bibinfo {year} {2019}{\natexlab{b}})}\BibitemShut {NoStop}%
\bibitem [{\citenamefont {{Giovannini}}\ \emph {et~al.}(2019)\citenamefont
  {{Giovannini}}, \citenamefont {{Grazioli}}, \citenamefont {{Ambrosetti}},\
  and\ \citenamefont {{Cappelli}}}]{giovannini2019fqfmuder2}%
  \BibitemOpen
  \bibfield  {author} {\bibinfo {author} {\bibfnamefont {T.}~\bibnamefont
  {{Giovannini}}}, \bibinfo {author} {\bibfnamefont {L.}~\bibnamefont
  {{Grazioli}}}, \bibinfo {author} {\bibfnamefont {M.}~\bibnamefont
  {{Ambrosetti}}}, \ and\ \bibinfo {author} {\bibfnamefont {C.}~\bibnamefont
  {{Cappelli}}},\ }\bibfield  {title} {\enquote {\bibinfo {title} {Calculation
  of ir spectra with a fully polarizable qm/mm approach based on fluctuating
  charges and fluctuating dipoles},}\ }\href {\doibase
  10.1021/acs.jctc.9b00574} {\bibfield  {journal} {\bibinfo  {journal} {J.
  Chem. Theory Comput.}\ ,\ \bibinfo {pages} {DOI: 10.1021/acs.jctc.9b00574}}
  (\bibinfo {year} {2019})},\ \bibinfo {note} {pMID: 31436976},\ \Eprint
  {http://arxiv.org/abs/https://doi.org/10.1021/acs.jctc.9b00574}
  {https://doi.org/10.1021/acs.jctc.9b00574} \BibitemShut {NoStop}%
\bibitem [{\citenamefont {Giovannini}, \citenamefont {Ambrosetti},\ and\
  \citenamefont {Cappelli}(2018)}]{giovannini2018hyper}%
  \BibitemOpen
  \bibfield  {author} {\bibinfo {author} {\bibfnamefont {T.}~\bibnamefont
  {Giovannini}}, \bibinfo {author} {\bibfnamefont {M.}~\bibnamefont
  {Ambrosetti}}, \ and\ \bibinfo {author} {\bibfnamefont {C.}~\bibnamefont
  {Cappelli}},\ }\bibfield  {title} {\enquote {\bibinfo {title} {A polarizable
  embedding approach to second harmonic generation (shg) of molecular systems
  in aqueous solutions},}\ }\href@noop {} {\bibfield  {journal} {\bibinfo
  {journal} {Theor. Chem. Acc.}\ }\textbf {\bibinfo {volume} {137}},\ \bibinfo
  {pages} {74} (\bibinfo {year} {2018})}\BibitemShut {NoStop}%
\bibitem [{\citenamefont {Lipparini}, \citenamefont {Cappelli},\ and\
  \citenamefont {Barone}(2012)}]{lipparini2012linear}%
  \BibitemOpen
  \bibfield  {author} {\bibinfo {author} {\bibfnamefont {F.}~\bibnamefont
  {Lipparini}}, \bibinfo {author} {\bibfnamefont {C.}~\bibnamefont {Cappelli}},
  \ and\ \bibinfo {author} {\bibfnamefont {V.}~\bibnamefont {Barone}},\
  }\bibfield  {title} {\enquote {\bibinfo {title} {Linear response theory and
  electronic transition energies for a fully polarizable {QM}/classical
  hamiltonian},}\ }\href@noop {} {\bibfield  {journal} {\bibinfo  {journal} {J.
  Chem. Theory Comput.}\ }\textbf {\bibinfo {volume} {8}},\ \bibinfo {pages}
  {4153--4165} (\bibinfo {year} {2012})}\BibitemShut {NoStop}%
\bibitem [{\citenamefont {Lipparini}\ \emph {et~al.}(2012)\citenamefont
  {Lipparini}, \citenamefont {Cappelli}, \citenamefont {Scalmani},
  \citenamefont {De~Mitri},\ and\ \citenamefont
  {Barone}}]{lipparini2012analytical}%
  \BibitemOpen
  \bibfield  {author} {\bibinfo {author} {\bibfnamefont {F.}~\bibnamefont
  {Lipparini}}, \bibinfo {author} {\bibfnamefont {C.}~\bibnamefont {Cappelli}},
  \bibinfo {author} {\bibfnamefont {G.}~\bibnamefont {Scalmani}}, \bibinfo
  {author} {\bibfnamefont {N.}~\bibnamefont {De~Mitri}}, \ and\ \bibinfo
  {author} {\bibfnamefont {V.}~\bibnamefont {Barone}},\ }\bibfield  {title}
  {\enquote {\bibinfo {title} {Analytical first and second derivatives for a
  fully polarizable {QM}/classical hamiltonian},}\ }\href@noop {} {\bibfield
  {journal} {\bibinfo  {journal} {J. Chem. Theory Comput.}\ }\textbf {\bibinfo
  {volume} {8}},\ \bibinfo {pages} {4270--4278} (\bibinfo {year}
  {2012})}\BibitemShut {NoStop}%
\bibitem [{\citenamefont {Lipparini}, \citenamefont {Cappelli},\ and\
  \citenamefont {Barone}(2013)}]{lipparini2013gauge}%
  \BibitemOpen
  \bibfield  {author} {\bibinfo {author} {\bibfnamefont {F.}~\bibnamefont
  {Lipparini}}, \bibinfo {author} {\bibfnamefont {C.}~\bibnamefont {Cappelli}},
  \ and\ \bibinfo {author} {\bibfnamefont {V.}~\bibnamefont {Barone}},\
  }\bibfield  {title} {\enquote {\bibinfo {title} {A gauge invariant multiscale
  approach to magnetic spectroscopies in condensed phase: General three-layer
  model, computational implementation and pilot applications},}\ }\href@noop {}
  {\bibfield  {journal} {\bibinfo  {journal} {J. Chem. Phys.}\ }\textbf
  {\bibinfo {volume} {138}},\ \bibinfo {pages} {234108} (\bibinfo {year}
  {2013})}\BibitemShut {NoStop}%
\bibitem [{\citenamefont {Giovannini}\ \emph {et~al.}(2017)\citenamefont
  {Giovannini}, \citenamefont {Olsz{\`o}wka}, \citenamefont {Egidi},
  \citenamefont {Cheeseman}, \citenamefont {Scalmani},\ and\ \citenamefont
  {Cappelli}}]{giovannini2017polarizable}%
  \BibitemOpen
  \bibfield  {author} {\bibinfo {author} {\bibfnamefont {T.}~\bibnamefont
  {Giovannini}}, \bibinfo {author} {\bibfnamefont {M.}~\bibnamefont
  {Olsz{\`o}wka}}, \bibinfo {author} {\bibfnamefont {F.}~\bibnamefont {Egidi}},
  \bibinfo {author} {\bibfnamefont {J.~R.}\ \bibnamefont {Cheeseman}}, \bibinfo
  {author} {\bibfnamefont {G.}~\bibnamefont {Scalmani}}, \ and\ \bibinfo
  {author} {\bibfnamefont {C.}~\bibnamefont {Cappelli}},\ }\bibfield  {title}
  {\enquote {\bibinfo {title} {Polarizable embedding approach for the
  analytical calculation of raman and raman optical activity spectra of
  solvated systems},}\ }\href@noop {} {\bibfield  {journal} {\bibinfo
  {journal} {J. Chem. Theory Comput.}\ }\textbf {\bibinfo {volume} {13}},\
  \bibinfo {pages} {4421--4435} (\bibinfo {year} {2017})}\BibitemShut {NoStop}%
\bibitem [{\citenamefont {Giovannini}, \citenamefont {Olszowka},\ and\
  \citenamefont {Cappelli}(2016)}]{giovannini2016effective}%
  \BibitemOpen
  \bibfield  {author} {\bibinfo {author} {\bibfnamefont {T.}~\bibnamefont
  {Giovannini}}, \bibinfo {author} {\bibfnamefont {M.}~\bibnamefont
  {Olszowka}}, \ and\ \bibinfo {author} {\bibfnamefont {C.}~\bibnamefont
  {Cappelli}},\ }\bibfield  {title} {\enquote {\bibinfo {title} {Effective
  fully polarizable qm/mm approach to model vibrational circular dichroism
  spectra of systems in aqueous solution},}\ }\href@noop {} {\bibfield
  {journal} {\bibinfo  {journal} {J. Chem. Theory Comput.}\ }\textbf {\bibinfo
  {volume} {12}},\ \bibinfo {pages} {5483--5492} (\bibinfo {year}
  {2016})}\BibitemShut {NoStop}%
\bibitem [{\citenamefont {Giovannini}\ \emph {et~al.}(2018)\citenamefont
  {Giovannini}, \citenamefont {Del~Frate}, \citenamefont {Lafiosca},\ and\
  \citenamefont {Cappelli}}]{giovannini2018effective}%
  \BibitemOpen
  \bibfield  {author} {\bibinfo {author} {\bibfnamefont {T.}~\bibnamefont
  {Giovannini}}, \bibinfo {author} {\bibfnamefont {G.}~\bibnamefont
  {Del~Frate}}, \bibinfo {author} {\bibfnamefont {P.}~\bibnamefont {Lafiosca}},
  \ and\ \bibinfo {author} {\bibfnamefont {C.}~\bibnamefont {Cappelli}},\
  }\bibfield  {title} {\enquote {\bibinfo {title} {Effective computational
  route towards vibrational optical activity spectra of chiral molecules in
  aqueous solution},}\ }\href@noop {} {\bibfield  {journal} {\bibinfo
  {journal} {Phys. Chem. Chem. Phys.}\ }\textbf {\bibinfo {volume} {20}},\
  \bibinfo {pages} {9181--9197} (\bibinfo {year} {2018})}\BibitemShut {NoStop}%
\bibitem [{\citenamefont {Egidi}\ \emph {et~al.}(2019)\citenamefont {Egidi},
  \citenamefont {Giovannini}, \citenamefont {Del~Frate}, \citenamefont
  {Lemler}, \citenamefont {Vaccaro},\ and\ \citenamefont
  {Cappelli}}]{egidi2019combined}%
  \BibitemOpen
  \bibfield  {author} {\bibinfo {author} {\bibfnamefont {F.}~\bibnamefont
  {Egidi}}, \bibinfo {author} {\bibfnamefont {T.}~\bibnamefont {Giovannini}},
  \bibinfo {author} {\bibfnamefont {G.}~\bibnamefont {Del~Frate}}, \bibinfo
  {author} {\bibfnamefont {P.~M.}\ \bibnamefont {Lemler}}, \bibinfo {author}
  {\bibfnamefont {P.~H.}\ \bibnamefont {Vaccaro}}, \ and\ \bibinfo {author}
  {\bibfnamefont {C.}~\bibnamefont {Cappelli}},\ }\bibfield  {title} {\enquote
  {\bibinfo {title} {A combined experimental and theoretical study of optical
  rotatory dispersion for (r)-glycidyl methyl ether in aqueous solution},}\
  }\href@noop {} {\bibfield  {journal} {\bibinfo  {journal} {Phys. Chem. Chem.
  Phys.}\ }\textbf {\bibinfo {volume} {21}},\ \bibinfo {pages} {3644--3655}
  (\bibinfo {year} {2019})}\BibitemShut {NoStop}%
\bibitem [{\citenamefont {Puglisi}\ \emph {et~al.}(2019)\citenamefont
  {Puglisi}, \citenamefont {Giovannini}, \citenamefont {Antonov},\ and\
  \citenamefont {Cappelli}}]{puglisi2019interplay}%
  \BibitemOpen
  \bibfield  {author} {\bibinfo {author} {\bibfnamefont {A.}~\bibnamefont
  {Puglisi}}, \bibinfo {author} {\bibfnamefont {T.}~\bibnamefont {Giovannini}},
  \bibinfo {author} {\bibfnamefont {L.}~\bibnamefont {Antonov}}, \ and\
  \bibinfo {author} {\bibfnamefont {C.}~\bibnamefont {Cappelli}},\ }\bibfield
  {title} {\enquote {\bibinfo {title} {Interplay between conformational and
  solvent effects in uv-visible absorption spectra: curcumin tautomers as a
  case study},}\ }\href {\doibase 10.1039/C9CP00907H} {\bibfield  {journal}
  {\bibinfo  {journal} {Phys. Chem. Chem. Phys.}\ }\textbf {\bibinfo {volume}
  {21}},\ \bibinfo {pages} {15504--15514} (\bibinfo {year} {2019})}\BibitemShut
  {NoStop}%
\bibitem [{\citenamefont {Di~Remigio}\ \emph {et~al.}(2019)\citenamefont
  {Di~Remigio}, \citenamefont {Giovannini}, \citenamefont {Ambrosetti},
  \citenamefont {Cappelli},\ and\ \citenamefont
  {Frediani}}]{giovannini2019tpa}%
  \BibitemOpen
  \bibfield  {author} {\bibinfo {author} {\bibfnamefont {R.}~\bibnamefont
  {Di~Remigio}}, \bibinfo {author} {\bibfnamefont {T.}~\bibnamefont
  {Giovannini}}, \bibinfo {author} {\bibfnamefont {M.}~\bibnamefont
  {Ambrosetti}}, \bibinfo {author} {\bibfnamefont {C.}~\bibnamefont
  {Cappelli}}, \ and\ \bibinfo {author} {\bibfnamefont {L.}~\bibnamefont
  {Frediani}},\ }\bibfield  {title} {\enquote {\bibinfo {title} {Fully
  polarizable qm/fluctuating charge approach to two-photon absorption of
  aqueous solutions},}\ }\href {\doibase 10.1021/acs.jctc.9b00305} {\bibfield
  {journal} {\bibinfo  {journal} {J. Chem. Theory Comput.}\ }\textbf {\bibinfo
  {volume} {15}},\ \bibinfo {pages} {4056--4068} (\bibinfo {year} {2019})},\
  \bibinfo {note} {pMID: 31244130},\ \Eprint
  {http://arxiv.org/abs/https://doi.org/10.1021/acs.jctc.9b00305}
  {https://doi.org/10.1021/acs.jctc.9b00305} \BibitemShut {NoStop}%
\bibitem [{\citenamefont {Stern}\ \emph {et~al.}(1999)\citenamefont {Stern},
  \citenamefont {Kaminski}, \citenamefont {Banks}, \citenamefont {Zhou},
  \citenamefont {Berne},\ and\ \citenamefont
  {Friesner}}]{stern1999fluctuating}%
  \BibitemOpen
  \bibfield  {author} {\bibinfo {author} {\bibfnamefont {H.~A.}\ \bibnamefont
  {Stern}}, \bibinfo {author} {\bibfnamefont {G.~A.}\ \bibnamefont {Kaminski}},
  \bibinfo {author} {\bibfnamefont {J.~L.}\ \bibnamefont {Banks}}, \bibinfo
  {author} {\bibfnamefont {R.}~\bibnamefont {Zhou}}, \bibinfo {author}
  {\bibfnamefont {B.}~\bibnamefont {Berne}}, \ and\ \bibinfo {author}
  {\bibfnamefont {R.~A.}\ \bibnamefont {Friesner}},\ }\bibfield  {title}
  {\enquote {\bibinfo {title} {Fluctuating charge, polarizable dipole, and
  combined models: parameterization from ab initio quantum chemistry},}\
  }\href@noop {} {\bibfield  {journal} {\bibinfo  {journal} {J. Phys. Chem. B}\
  }\textbf {\bibinfo {volume} {103}},\ \bibinfo {pages} {4730--4737} (\bibinfo
  {year} {1999})}\BibitemShut {NoStop}%
\bibitem [{\citenamefont {Naserifar}\ \emph {et~al.}(2017)\citenamefont
  {Naserifar}, \citenamefont {Brooks}, \citenamefont {Goddard~III},\ and\
  \citenamefont {Cvicek}}]{naserifar2017polarizable}%
  \BibitemOpen
  \bibfield  {author} {\bibinfo {author} {\bibfnamefont {S.}~\bibnamefont
  {Naserifar}}, \bibinfo {author} {\bibfnamefont {D.~J.}\ \bibnamefont
  {Brooks}}, \bibinfo {author} {\bibfnamefont {W.~A.}\ \bibnamefont
  {Goddard~III}}, \ and\ \bibinfo {author} {\bibfnamefont {V.}~\bibnamefont
  {Cvicek}},\ }\bibfield  {title} {\enquote {\bibinfo {title} {Polarizable
  charge equilibration model for predicting accurate electrostatic interactions
  in molecules and solids},}\ }\href@noop {} {\bibfield  {journal} {\bibinfo
  {journal} {J. Chem. Phys.}\ }\textbf {\bibinfo {volume} {146}},\ \bibinfo
  {pages} {124117} (\bibinfo {year} {2017})}\BibitemShut {NoStop}%
\bibitem [{\citenamefont {Oppenheim}, \citenamefont {Naserifar},\ and\
  \citenamefont {Goddard~III}(2018)}]{oppenheim2018extension}%
  \BibitemOpen
  \bibfield  {author} {\bibinfo {author} {\bibfnamefont {J.~J.}\ \bibnamefont
  {Oppenheim}}, \bibinfo {author} {\bibfnamefont {S.}~\bibnamefont
  {Naserifar}}, \ and\ \bibinfo {author} {\bibfnamefont {W.~A.}\ \bibnamefont
  {Goddard~III}},\ }\bibfield  {title} {\enquote {\bibinfo {title} {Extension
  of the polarizable charge equilibration model to higher oxidation states with
  applications to ge, as, se, br, sn, sb, te, i, pb, bi, po, and at
  elements},}\ }\href@noop {} {\bibfield  {journal} {\bibinfo  {journal} {J.
  Phys. Chem. A}\ }\textbf {\bibinfo {volume} {122}},\ \bibinfo {pages}
  {639--645} (\bibinfo {year} {2018})}\BibitemShut {NoStop}%
\bibitem [{\citenamefont {Mayer}(2007)}]{mayer2007formulation}%
  \BibitemOpen
  \bibfield  {author} {\bibinfo {author} {\bibfnamefont {A.}~\bibnamefont
  {Mayer}},\ }\bibfield  {title} {\enquote {\bibinfo {title} {Formulation in
  terms of normalized propagators of a charge-dipole model enabling the
  calculation of the polarization properties of fullerenes and carbon
  nanotubes},}\ }\href@noop {} {\bibfield  {journal} {\bibinfo  {journal}
  {Phys. Rev. B}\ }\textbf {\bibinfo {volume} {75}},\ \bibinfo {pages} {045407}
  (\bibinfo {year} {2007})}\BibitemShut {NoStop}%
\bibitem [{\citenamefont {Jensen}\ and\ \citenamefont
  {Jensen}(2009)}]{jensen2009atomistic}%
  \BibitemOpen
  \bibfield  {author} {\bibinfo {author} {\bibfnamefont {L.~L.}\ \bibnamefont
  {Jensen}}\ and\ \bibinfo {author} {\bibfnamefont {L.}~\bibnamefont
  {Jensen}},\ }\bibfield  {title} {\enquote {\bibinfo {title} {Atomistic
  electrodynamics model for optical properties of silver nanoclusters},}\
  }\href@noop {} {\bibfield  {journal} {\bibinfo  {journal} {J. Phys. Chem. C}\
  }\textbf {\bibinfo {volume} {113}},\ \bibinfo {pages} {15182--15190}
  (\bibinfo {year} {2009})}\BibitemShut {NoStop}%
\bibitem [{\citenamefont {Rinkevicius}\ \emph {et~al.}(2014)\citenamefont
  {Rinkevicius}, \citenamefont {Li}, \citenamefont {Sandberg}, \citenamefont
  {Mikkelsen},\ and\ \citenamefont {{\AA}gren}}]{rinkevicius2014hybrid}%
  \BibitemOpen
  \bibfield  {author} {\bibinfo {author} {\bibfnamefont {Z.}~\bibnamefont
  {Rinkevicius}}, \bibinfo {author} {\bibfnamefont {X.}~\bibnamefont {Li}},
  \bibinfo {author} {\bibfnamefont {J.~A.}\ \bibnamefont {Sandberg}}, \bibinfo
  {author} {\bibfnamefont {K.~V.}\ \bibnamefont {Mikkelsen}}, \ and\ \bibinfo
  {author} {\bibfnamefont {H.}~\bibnamefont {{\AA}gren}},\ }\bibfield  {title}
  {\enquote {\bibinfo {title} {A hybrid density functional theory/molecular
  mechanics approach for linear response properties in heterogeneous
  environments},}\ }\href@noop {} {\bibfield  {journal} {\bibinfo  {journal}
  {J. Chem. Theory Comput.}\ }\textbf {\bibinfo {volume} {10}},\ \bibinfo
  {pages} {989--1003} (\bibinfo {year} {2014})}\BibitemShut {NoStop}%
\bibitem [{\citenamefont {Casida}(1995)}]{Casida95_155}%
  \BibitemOpen
  \bibfield  {author} {\bibinfo {author} {\bibfnamefont {M.~E.}\ \bibnamefont
  {Casida}},\ }\bibfield  {title} {\enquote {\bibinfo {title} {Time-dependent
  density functional response theory for molecules},}\ }in\ \href@noop {}
  {\emph {\bibinfo {booktitle} {Recent Advances in Density Functional Methods
  Part I}}},\ \bibinfo {editor} {edited by\ \bibinfo {editor} {\bibfnamefont
  {D.~P.}\ \bibnamefont {Chong}}}\ (\bibinfo  {publisher} {World Scientific,
  Singapore},\ \bibinfo {year} {1995})\ pp.\ \bibinfo {pages}
  {155--192}\BibitemShut {NoStop}%
\bibitem [{\citenamefont {Caricato}\ \emph {et~al.}(2006)\citenamefont
  {Caricato}, \citenamefont {Mennucci}, \citenamefont {Tomasi}, \citenamefont
  {Ingrosso}, \citenamefont {Cammi}, \citenamefont {Corni},\ and\ \citenamefont
  {Scalmani}}]{caricato2006formation}%
  \BibitemOpen
  \bibfield  {author} {\bibinfo {author} {\bibfnamefont {M.}~\bibnamefont
  {Caricato}}, \bibinfo {author} {\bibfnamefont {B.}~\bibnamefont {Mennucci}},
  \bibinfo {author} {\bibfnamefont {J.}~\bibnamefont {Tomasi}}, \bibinfo
  {author} {\bibfnamefont {F.}~\bibnamefont {Ingrosso}}, \bibinfo {author}
  {\bibfnamefont {R.}~\bibnamefont {Cammi}}, \bibinfo {author} {\bibfnamefont
  {S.}~\bibnamefont {Corni}}, \ and\ \bibinfo {author} {\bibfnamefont
  {G.}~\bibnamefont {Scalmani}},\ }\bibfield  {title} {\enquote {\bibinfo
  {title} {Formation and relaxation of excited states in solution: A new time
  dependent polarizable continuum model based on time dependent density
  functional theory},}\ }\href@noop {} {\bibfield  {journal} {\bibinfo
  {journal} {J. Chem. Phys.}\ }\textbf {\bibinfo {volume} {124}},\ \bibinfo
  {pages} {124520} (\bibinfo {year} {2006})}\BibitemShut {NoStop}%
\bibitem [{\citenamefont {Marenich}\ \emph {et~al.}(2011)\citenamefont
  {Marenich}, \citenamefont {Cramer}, \citenamefont {Truhlar}, \citenamefont
  {Guido}, \citenamefont {Mennucci}, \citenamefont {Scalmani},\ and\
  \citenamefont {Frisch}}]{marenich2011practical}%
  \BibitemOpen
  \bibfield  {author} {\bibinfo {author} {\bibfnamefont {A.~V.}\ \bibnamefont
  {Marenich}}, \bibinfo {author} {\bibfnamefont {C.~J.}\ \bibnamefont
  {Cramer}}, \bibinfo {author} {\bibfnamefont {D.~G.}\ \bibnamefont {Truhlar}},
  \bibinfo {author} {\bibfnamefont {C.~A.}\ \bibnamefont {Guido}}, \bibinfo
  {author} {\bibfnamefont {B.}~\bibnamefont {Mennucci}}, \bibinfo {author}
  {\bibfnamefont {G.}~\bibnamefont {Scalmani}}, \ and\ \bibinfo {author}
  {\bibfnamefont {M.~J.}\ \bibnamefont {Frisch}},\ }\bibfield  {title}
  {\enquote {\bibinfo {title} {Practical computation of electronic excitation
  in solution: vertical excitation model},}\ }\href@noop {} {\bibfield
  {journal} {\bibinfo  {journal} {Chem. Sci.}\ }\textbf {\bibinfo {volume}
  {2}},\ \bibinfo {pages} {2143--2161} (\bibinfo {year} {2011})}\BibitemShut
  {NoStop}%
\bibitem [{\citenamefont {Marenich}, \citenamefont {Cramer},\ and\
  \citenamefont {Truhlar}(2015)}]{marenich2014electronic}%
  \BibitemOpen
  \bibfield  {author} {\bibinfo {author} {\bibfnamefont {A.~V.}\ \bibnamefont
  {Marenich}}, \bibinfo {author} {\bibfnamefont {C.~J.}\ \bibnamefont
  {Cramer}}, \ and\ \bibinfo {author} {\bibfnamefont {D.~G.}\ \bibnamefont
  {Truhlar}},\ }\bibfield  {title} {\enquote {\bibinfo {title} {Electronic
  absorption spectra and solvatochromic shifts by the vertical excitation
  model: solvated clusters and molecular dynamics sampling},}\ }\href@noop {}
  {\bibfield  {journal} {\bibinfo  {journal} {J. Phys. Chem. B}\ }\textbf
  {\bibinfo {volume} {119}},\ \bibinfo {pages} {958--967} (\bibinfo {year}
  {2015})}\BibitemShut {NoStop}%
\bibitem [{\citenamefont {Budzak}\ \emph {et~al.}(2014)\citenamefont {Budzak},
  \citenamefont {Medved}, \citenamefont {Mennucci},\ and\ \citenamefont
  {Jacquemin}}]{budzak2014unveiling}%
  \BibitemOpen
  \bibfield  {author} {\bibinfo {author} {\bibfnamefont {S.}~\bibnamefont
  {Budzak}}, \bibinfo {author} {\bibfnamefont {M.}~\bibnamefont {Medved}},
  \bibinfo {author} {\bibfnamefont {B.}~\bibnamefont {Mennucci}}, \ and\
  \bibinfo {author} {\bibfnamefont {D.}~\bibnamefont {Jacquemin}},\ }\bibfield
  {title} {\enquote {\bibinfo {title} {Unveiling solvents effect on
  excited-state polarizabilities with the corrected linear-response model},}\
  }\href@noop {} {\bibfield  {journal} {\bibinfo  {journal} {J. Phys. Chem. A}\
  }\textbf {\bibinfo {volume} {118}},\ \bibinfo {pages} {5652--5656} (\bibinfo
  {year} {2014})}\BibitemShut {NoStop}%
\bibitem [{\citenamefont {Improta}\ \emph {et~al.}(2006)\citenamefont
  {Improta}, \citenamefont {Barone}, \citenamefont {Scalmani},\ and\
  \citenamefont {Frisch}}]{improta2006state}%
  \BibitemOpen
  \bibfield  {author} {\bibinfo {author} {\bibfnamefont {R.}~\bibnamefont
  {Improta}}, \bibinfo {author} {\bibfnamefont {V.}~\bibnamefont {Barone}},
  \bibinfo {author} {\bibfnamefont {G.}~\bibnamefont {Scalmani}}, \ and\
  \bibinfo {author} {\bibfnamefont {M.~J.}\ \bibnamefont {Frisch}},\ }\bibfield
   {title} {\enquote {\bibinfo {title} {A state-specific polarizable continuum
  model time dependent density functional theory method for excited state
  calculations in solution},}\ }\href@noop {} {\bibfield  {journal} {\bibinfo
  {journal} {J. Chem. Phys.}\ }\textbf {\bibinfo {volume} {125}},\ \bibinfo
  {pages} {054103} (\bibinfo {year} {2006})}\BibitemShut {NoStop}%
\bibitem [{\citenamefont {Caricato}(2014)}]{caricato2014corrected}%
  \BibitemOpen
  \bibfield  {author} {\bibinfo {author} {\bibfnamefont {M.}~\bibnamefont
  {Caricato}},\ }\bibfield  {title} {\enquote {\bibinfo {title} {A
  corrected-linear response formalism for the calculation of electronic
  excitation energies of solvated molecules with the ccsd-pcm method},}\
  }\href@noop {} {\bibfield  {journal} {\bibinfo  {journal} {Comput. Theor.
  Chem.}\ }\textbf {\bibinfo {volume} {1040}},\ \bibinfo {pages} {99--105}
  (\bibinfo {year} {2014})}\BibitemShut {NoStop}%
\bibitem [{\citenamefont {Guido}\ \emph {et~al.}(2018)\citenamefont {Guido},
  \citenamefont {Mennucci}, \citenamefont {Scalmani},\ and\ \citenamefont
  {Jacquemin}}]{guido2018excited}%
  \BibitemOpen
  \bibfield  {author} {\bibinfo {author} {\bibfnamefont {C.~A.}\ \bibnamefont
  {Guido}}, \bibinfo {author} {\bibfnamefont {B.}~\bibnamefont {Mennucci}},
  \bibinfo {author} {\bibfnamefont {G.}~\bibnamefont {Scalmani}}, \ and\
  \bibinfo {author} {\bibfnamefont {D.}~\bibnamefont {Jacquemin}},\ }\bibfield
  {title} {\enquote {\bibinfo {title} {Excited state dipole moments in
  solution: Comparison between state-specific and linear-response td-dft
  values},}\ }\href@noop {} {\bibfield  {journal} {\bibinfo  {journal} {J.
  Chem. Theory Comput.}\ }\textbf {\bibinfo {volume} {14}},\ \bibinfo {pages}
  {1544--1553} (\bibinfo {year} {2018})}\BibitemShut {NoStop}%
\bibitem [{\citenamefont {Duchemin}\ \emph {et~al.}(2018)\citenamefont
  {Duchemin}, \citenamefont {Guido}, \citenamefont {Jacquemin},\ and\
  \citenamefont {Blase}}]{duchemin2018bethe}%
  \BibitemOpen
  \bibfield  {author} {\bibinfo {author} {\bibfnamefont {I.}~\bibnamefont
  {Duchemin}}, \bibinfo {author} {\bibfnamefont {C.~A.}\ \bibnamefont {Guido}},
  \bibinfo {author} {\bibfnamefont {D.}~\bibnamefont {Jacquemin}}, \ and\
  \bibinfo {author} {\bibfnamefont {X.}~\bibnamefont {Blase}},\ }\bibfield
  {title} {\enquote {\bibinfo {title} {The bethe--salpeter formalism with
  polarisable continuum embedding: reconciling linear-response and
  state-specific features},}\ }\href@noop {} {\bibfield  {journal} {\bibinfo
  {journal} {Chem. Sci.}\ }\textbf {\bibinfo {volume} {9}},\ \bibinfo {pages}
  {4430--4443} (\bibinfo {year} {2018})}\BibitemShut {NoStop}%
\bibitem [{\citenamefont {Schr{\"o}der}\ and\ \citenamefont
  {Schwabe}(2018)}]{schroder2018corrected}%
  \BibitemOpen
  \bibfield  {author} {\bibinfo {author} {\bibfnamefont {H.}~\bibnamefont
  {Schr{\"o}der}}\ and\ \bibinfo {author} {\bibfnamefont {T.}~\bibnamefont
  {Schwabe}},\ }\bibfield  {title} {\enquote {\bibinfo {title} {Corrected
  polarizable embedding: Improving the induction contribution to perichromism
  for linear response theory},}\ }\href@noop {} {\bibfield  {journal} {\bibinfo
   {journal} {J. Chem. Theory Comput.}\ }\textbf {\bibinfo {volume} {14}},\
  \bibinfo {pages} {833--842} (\bibinfo {year} {2018})}\BibitemShut {NoStop}%
\bibitem [{\citenamefont {Guido}\ and\ \citenamefont
  {Caprasecca}(2019)}]{guido2019description}%
  \BibitemOpen
  \bibfield  {author} {\bibinfo {author} {\bibfnamefont {C.~A.}\ \bibnamefont
  {Guido}}\ and\ \bibinfo {author} {\bibfnamefont {S.}~\bibnamefont
  {Caprasecca}},\ }\bibfield  {title} {\enquote {\bibinfo {title} {On the
  description of the environment polarization response to electronic
  transitions},}\ }\href@noop {} {\bibfield  {journal} {\bibinfo  {journal}
  {Int. J. Quantum Chem.}\ }\textbf {\bibinfo {volume} {119}},\ \bibinfo
  {pages} {e25711} (\bibinfo {year} {2019})}\BibitemShut {NoStop}%
\bibitem [{\citenamefont {Guido}\ \emph {et~al.}(2017)\citenamefont {Guido},
  \citenamefont {Scalmani}, \citenamefont {Mennucci},\ and\ \citenamefont
  {Jacquemin}}]{guido2017excited}%
  \BibitemOpen
  \bibfield  {author} {\bibinfo {author} {\bibfnamefont {C.~A.}\ \bibnamefont
  {Guido}}, \bibinfo {author} {\bibfnamefont {G.}~\bibnamefont {Scalmani}},
  \bibinfo {author} {\bibfnamefont {B.}~\bibnamefont {Mennucci}}, \ and\
  \bibinfo {author} {\bibfnamefont {D.}~\bibnamefont {Jacquemin}},\ }\bibfield
  {title} {\enquote {\bibinfo {title} {Excited state gradients for a
  state-specific continuum solvation approach: The vertical excitation model
  within a lagrangian tddft formulation},}\ }\href@noop {} {\bibfield
  {journal} {\bibinfo  {journal} {J. Chem. Phys.}\ }\textbf {\bibinfo {volume}
  {146}},\ \bibinfo {pages} {204106} (\bibinfo {year} {2017})}\BibitemShut
  {NoStop}%
\bibitem [{\citenamefont {Guido}\ \emph {et~al.}(2015)\citenamefont {Guido},
  \citenamefont {Jacquemin}, \citenamefont {Adamo},\ and\ \citenamefont
  {Mennucci}}]{guido2015electronic}%
  \BibitemOpen
  \bibfield  {author} {\bibinfo {author} {\bibfnamefont {C.~A.}\ \bibnamefont
  {Guido}}, \bibinfo {author} {\bibfnamefont {D.}~\bibnamefont {Jacquemin}},
  \bibinfo {author} {\bibfnamefont {C.}~\bibnamefont {Adamo}}, \ and\ \bibinfo
  {author} {\bibfnamefont {B.}~\bibnamefont {Mennucci}},\ }\bibfield  {title}
  {\enquote {\bibinfo {title} {Electronic excitations in solution: the
  interplay between state specific approaches and a time-dependent density
  functional theory description},}\ }\href@noop {} {\bibfield  {journal}
  {\bibinfo  {journal} {J. Chem. Theory Comput.}\ }\textbf {\bibinfo {volume}
  {11}},\ \bibinfo {pages} {5782--5790} (\bibinfo {year} {2015})}\BibitemShut
  {NoStop}%
\bibitem [{\citenamefont {Zeng}\ and\ \citenamefont
  {Liang}(2015)}]{zeng2015analytic}%
  \BibitemOpen
  \bibfield  {author} {\bibinfo {author} {\bibfnamefont {Q.}~\bibnamefont
  {Zeng}}\ and\ \bibinfo {author} {\bibfnamefont {W.}~\bibnamefont {Liang}},\
  }\bibfield  {title} {\enquote {\bibinfo {title} {Analytic energy gradient of
  excited electronic state within tddft/mmpol framework: Benchmark tests and
  parallel implementation},}\ }\href@noop {} {\bibfield  {journal} {\bibinfo
  {journal} {J. Chem. Phys.}\ }\textbf {\bibinfo {volume} {143}},\ \bibinfo
  {pages} {134104} (\bibinfo {year} {2015})}\BibitemShut {NoStop}%
\bibitem [{\citenamefont {Sanderson}(1951)}]{sanderson1951}%
  \BibitemOpen
  \bibfield  {author} {\bibinfo {author} {\bibfnamefont {R.}~\bibnamefont
  {Sanderson}},\ }\bibfield  {title} {\enquote {\bibinfo {title} {An
  interpretation of bond lengths and a classification of bonds},}\ }\href@noop
  {} {\bibfield  {journal} {\bibinfo  {journal} {Science}\ }\textbf {\bibinfo
  {volume} {114}},\ \bibinfo {pages} {670--672} (\bibinfo {year}
  {1951})}\BibitemShut {NoStop}%
\bibitem [{\citenamefont {Giovannini}\ \emph {et~al.}(2019)\citenamefont
  {Giovannini}, \citenamefont {Macchiagodena}, \citenamefont {Ambrosetti},
  \citenamefont {Puglisi}, \citenamefont {Lafiosca}, \citenamefont {Lo~Gerfo},
  \citenamefont {Egidi},\ and\ \citenamefont
  {Cappelli}}]{giovannini2019simulating}%
  \BibitemOpen
  \bibfield  {author} {\bibinfo {author} {\bibfnamefont {T.}~\bibnamefont
  {Giovannini}}, \bibinfo {author} {\bibfnamefont {M.}~\bibnamefont
  {Macchiagodena}}, \bibinfo {author} {\bibfnamefont {M.}~\bibnamefont
  {Ambrosetti}}, \bibinfo {author} {\bibfnamefont {A.}~\bibnamefont {Puglisi}},
  \bibinfo {author} {\bibfnamefont {P.}~\bibnamefont {Lafiosca}}, \bibinfo
  {author} {\bibfnamefont {G.}~\bibnamefont {Lo~Gerfo}}, \bibinfo {author}
  {\bibfnamefont {F.}~\bibnamefont {Egidi}}, \ and\ \bibinfo {author}
  {\bibfnamefont {C.}~\bibnamefont {Cappelli}},\ }\bibfield  {title} {\enquote
  {\bibinfo {title} {Simulating vertical excitation energies of solvated dyes:
  From continuum to polarizable discrete modeling},}\ }\href@noop {} {\bibfield
   {journal} {\bibinfo  {journal} {Int. J. Quantum Chem.}\ }\textbf {\bibinfo
  {volume} {119}},\ \bibinfo {pages} {e25684} (\bibinfo {year}
  {2019})}\BibitemShut {NoStop}%
\bibitem [{\citenamefont {Abrahama}\ \emph {et~al.}(2015)\citenamefont
  {Abrahama}, \citenamefont {Murtola}, \citenamefont {Schulz}, \citenamefont
  {P{\'{a}}lla}, \citenamefont {Smith}, \citenamefont {Hess},\ and\
  \citenamefont {Lindahl}}]{Gromacs5}%
  \BibitemOpen
  \bibfield  {author} {\bibinfo {author} {\bibfnamefont {M.~J.}\ \bibnamefont
  {Abrahama}}, \bibinfo {author} {\bibfnamefont {T.}~\bibnamefont {Murtola}},
  \bibinfo {author} {\bibfnamefont {R.}~\bibnamefont {Schulz}}, \bibinfo
  {author} {\bibfnamefont {S.}~\bibnamefont {P{\'{a}}lla}}, \bibinfo {author}
  {\bibfnamefont {J.~C.}\ \bibnamefont {Smith}}, \bibinfo {author}
  {\bibfnamefont {B.}~\bibnamefont {Hess}}, \ and\ \bibinfo {author}
  {\bibfnamefont {E.}~\bibnamefont {Lindahl}},\ }\bibfield  {title} {\enquote
  {\bibinfo {title} {{GROMACS}: High performance molecular simulations through
  multi-level parallelism from laptops to supercomputers},}\ }\href@noop {}
  {\bibfield  {journal} {\bibinfo  {journal} {SoftwareX}\ }\textbf {\bibinfo
  {volume} {1-2}},\ \bibinfo {pages} {19--25} (\bibinfo {year}
  {2015})}\BibitemShut {NoStop}%
\bibitem [{\citenamefont {Oostenbrink}\ \emph {et~al.}(2004)\citenamefont
  {Oostenbrink}, \citenamefont {Villa}, \citenamefont {Mark},\ and\
  \citenamefont {Van~Gunsteren}}]{gromos}%
  \BibitemOpen
  \bibfield  {author} {\bibinfo {author} {\bibfnamefont {C.}~\bibnamefont
  {Oostenbrink}}, \bibinfo {author} {\bibfnamefont {A.}~\bibnamefont {Villa}},
  \bibinfo {author} {\bibfnamefont {A.~E.}\ \bibnamefont {Mark}}, \ and\
  \bibinfo {author} {\bibfnamefont {W.~F.}\ \bibnamefont {Van~Gunsteren}},\
  }\bibfield  {title} {\enquote {\bibinfo {title} {A biomolecular force field
  based on the free enthalpy of hydration and solvation: The gromos force-field
  parameter sets 53a5 and 53a6},}\ }\href {\doibase 10.1002/jcc.20090}
  {\bibfield  {journal} {\bibinfo  {journal} {J. Comput. Chem.}\ }\textbf
  {\bibinfo {volume} {25}},\ \bibinfo {pages} {1656--1676} (\bibinfo {year}
  {2004})}\BibitemShut {NoStop}%
\bibitem [{\citenamefont {Mark}\ and\ \citenamefont
  {Nilsson}(2001)}]{mark2001structure}%
  \BibitemOpen
  \bibfield  {author} {\bibinfo {author} {\bibfnamefont {P.}~\bibnamefont
  {Mark}}\ and\ \bibinfo {author} {\bibfnamefont {L.}~\bibnamefont {Nilsson}},\
  }\bibfield  {title} {\enquote {\bibinfo {title} {Structure and dynamics of
  the tip3p, spc, and spc/e water models at 298 k},}\ }\href@noop {} {\bibfield
   {journal} {\bibinfo  {journal} {J. Phys. Chem. A}\ }\textbf {\bibinfo
  {volume} {105}},\ \bibinfo {pages} {9954--9960} (\bibinfo {year}
  {2001})}\BibitemShut {NoStop}%
\bibitem [{\citenamefont {Bussi}, \citenamefont {Donadio},\ and\ \citenamefont
  {Parrinello}(2007)}]{vrescale}%
  \BibitemOpen
  \bibfield  {author} {\bibinfo {author} {\bibfnamefont {G.}~\bibnamefont
  {Bussi}}, \bibinfo {author} {\bibfnamefont {D.}~\bibnamefont {Donadio}}, \
  and\ \bibinfo {author} {\bibfnamefont {M.}~\bibnamefont {Parrinello}},\
  }\bibfield  {title} {\enquote {\bibinfo {title} {Canonical sampling through
  velocity rescaling},}\ }\href {\doibase http://dx.doi.org/10.1063/1.2408420}
  {\bibfield  {journal} {\bibinfo  {journal} {J. Chem. Phys.}\ }\textbf
  {\bibinfo {volume} {126}},\ \bibinfo {eid} {014101} (\bibinfo {year}
  {2007}),\ http://dx.doi.org/10.1063/1.2408420}\BibitemShut {NoStop}%
\bibitem [{\citenamefont {Hess}\ \emph {et~al.}(1997)\citenamefont {Hess},
  \citenamefont {Bekker}, \citenamefont {Berendsen},\ and\ \citenamefont
  {Fraaije}}]{hess1997lincs}%
  \BibitemOpen
  \bibfield  {author} {\bibinfo {author} {\bibfnamefont {B.}~\bibnamefont
  {Hess}}, \bibinfo {author} {\bibfnamefont {H.}~\bibnamefont {Bekker}},
  \bibinfo {author} {\bibfnamefont {H.~J.}\ \bibnamefont {Berendsen}}, \ and\
  \bibinfo {author} {\bibfnamefont {J.~G.}\ \bibnamefont {Fraaije}},\
  }\bibfield  {title} {\enquote {\bibinfo {title} {Lincs: a linear constraint
  solver for molecular simulations},}\ }\href@noop {} {\bibfield  {journal}
  {\bibinfo  {journal} {J. Comput. Chem.}\ }\textbf {\bibinfo {volume} {18}},\
  \bibinfo {pages} {1463--1472} (\bibinfo {year} {1997})}\BibitemShut {NoStop}%
\bibitem [{\citenamefont {Darden}, \citenamefont {York},\ and\ \citenamefont
  {Pedersen}(1993)}]{darden1993particle}%
  \BibitemOpen
  \bibfield  {author} {\bibinfo {author} {\bibfnamefont {T.}~\bibnamefont
  {Darden}}, \bibinfo {author} {\bibfnamefont {D.}~\bibnamefont {York}}, \ and\
  \bibinfo {author} {\bibfnamefont {L.}~\bibnamefont {Pedersen}},\ }\bibfield
  {title} {\enquote {\bibinfo {title} {Particle mesh ewald: An nlog(n) method
  for ewald sums in large systems},}\ }\href@noop {} {\bibfield  {journal}
  {\bibinfo  {journal} {J. Chem. Phys.}\ }\textbf {\bibinfo {volume} {98}},\
  \bibinfo {pages} {10089--10092} (\bibinfo {year} {1993})}\BibitemShut
  {NoStop}%
\bibitem [{\citenamefont {Carnimeo}, \citenamefont {Cappelli},\ and\
  \citenamefont {Barone}(2015)}]{carnimeo2015analytical}%
  \BibitemOpen
  \bibfield  {author} {\bibinfo {author} {\bibfnamefont {I.}~\bibnamefont
  {Carnimeo}}, \bibinfo {author} {\bibfnamefont {C.}~\bibnamefont {Cappelli}},
  \ and\ \bibinfo {author} {\bibfnamefont {V.}~\bibnamefont {Barone}},\
  }\bibfield  {title} {\enquote {\bibinfo {title} {Analytical gradients for
  mp2, double hybrid functionals, and td-dft with polarizable embedding
  described by fluctuating charges},}\ }\href@noop {} {\bibfield  {journal}
  {\bibinfo  {journal} {J. Comput. Chem.}\ }\textbf {\bibinfo {volume} {36}},\
  \bibinfo {pages} {2271--2290} (\bibinfo {year} {2015})}\BibitemShut {NoStop}%
\bibitem [{\citenamefont {Egidi}\ \emph {et~al.}(2014)\citenamefont {Egidi},
  \citenamefont {Giovannini}, \citenamefont {Piccardo}, \citenamefont {Bloino},
  \citenamefont {Cappelli},\ and\ \citenamefont
  {Barone}}]{egidi2014stereoelectronic}%
  \BibitemOpen
  \bibfield  {author} {\bibinfo {author} {\bibfnamefont {F.}~\bibnamefont
  {Egidi}}, \bibinfo {author} {\bibfnamefont {T.}~\bibnamefont {Giovannini}},
  \bibinfo {author} {\bibfnamefont {M.}~\bibnamefont {Piccardo}}, \bibinfo
  {author} {\bibfnamefont {J.}~\bibnamefont {Bloino}}, \bibinfo {author}
  {\bibfnamefont {C.}~\bibnamefont {Cappelli}}, \ and\ \bibinfo {author}
  {\bibfnamefont {V.}~\bibnamefont {Barone}},\ }\bibfield  {title} {\enquote
  {\bibinfo {title} {Stereoelectronic, vibrational, and environmental
  contributions to polarizabilities of large molecular systems: A feasible
  anharmonic protocol},}\ }\href@noop {} {\bibfield  {journal} {\bibinfo
  {journal} {J. Chem. Theory Comput\.}\ }\textbf {\bibinfo {volume} {10}},\
  \bibinfo {pages} {2456--2464} (\bibinfo {year} {2014})}\BibitemShut {NoStop}%
\bibitem [{\citenamefont {Frisch}\ \emph {et~al.}(2016)\citenamefont {Frisch},
  \citenamefont {Trucks}, \citenamefont {Schlegel}, \citenamefont {Scuseria},
  \citenamefont {Robb}, \citenamefont {Cheeseman}, \citenamefont {Scalmani},
  \citenamefont {Barone}, \citenamefont {Petersson}, \citenamefont {Nakatsuji},
  \citenamefont {Li}, \citenamefont {Caricato}, \citenamefont {Marenich},
  \citenamefont {Bloino}, \citenamefont {Janesko}, \citenamefont {Gomperts},
  \citenamefont {Mennucci}, \citenamefont {Hratchian}, \citenamefont {Ortiz},
  \citenamefont {Izmaylov}, \citenamefont {Sonnenberg}, \citenamefont
  {Williams-Young}, \citenamefont {Ding}, \citenamefont {Lipparini},
  \citenamefont {Egidi}, \citenamefont {Goings}, \citenamefont {Peng},
  \citenamefont {Petrone}, \citenamefont {Henderson}, \citenamefont
  {Ranasinghe}, \citenamefont {Zakrzewski}, \citenamefont {Gao}, \citenamefont
  {Rega}, \citenamefont {Zheng}, \citenamefont {Liang}, \citenamefont {Hada},
  \citenamefont {Ehara}, \citenamefont {Toyota}, \citenamefont {Fukuda},
  \citenamefont {Hasegawa}, \citenamefont {Ishida}, \citenamefont {Nakajima},
  \citenamefont {Honda}, \citenamefont {Kitao}, \citenamefont {Nakai},
  \citenamefont {Vreven}, \citenamefont {Throssell}, \citenamefont
  {Montgomery}, \citenamefont {Peralta}, \citenamefont {Ogliaro}, \citenamefont
  {Bearpark}, \citenamefont {Heyd}, \citenamefont {Brothers}, \citenamefont
  {Kudin}, \citenamefont {Staroverov}, \citenamefont {Keith}, \citenamefont
  {Kobayashi}, \citenamefont {Normand}, \citenamefont {Raghavachari},
  \citenamefont {Rendell}, \citenamefont {Burant}, \citenamefont {Iyengar},
  \citenamefont {Tomasi}, \citenamefont {Cossi}, \citenamefont {Millam},
  \citenamefont {Klene}, \citenamefont {Adamo}, \citenamefont {Cammi},
  \citenamefont {Ochterski}, \citenamefont {Martin}, \citenamefont {Morokuma},
  \citenamefont {Farkas}, \citenamefont {Foresman},\ and\ \citenamefont
  {Fox}}]{gaussian16}%
  \BibitemOpen
  \bibfield  {author} {\bibinfo {author} {\bibfnamefont {M.~J.}\ \bibnamefont
  {Frisch}}, \bibinfo {author} {\bibfnamefont {G.~W.}\ \bibnamefont {Trucks}},
  \bibinfo {author} {\bibfnamefont {H.~B.}\ \bibnamefont {Schlegel}}, \bibinfo
  {author} {\bibfnamefont {G.~E.}\ \bibnamefont {Scuseria}}, \bibinfo {author}
  {\bibfnamefont {M.~A.}\ \bibnamefont {Robb}}, \bibinfo {author}
  {\bibfnamefont {J.~R.}\ \bibnamefont {Cheeseman}}, \bibinfo {author}
  {\bibfnamefont {G.}~\bibnamefont {Scalmani}}, \bibinfo {author}
  {\bibfnamefont {V.}~\bibnamefont {Barone}}, \bibinfo {author} {\bibfnamefont
  {G.~A.}\ \bibnamefont {Petersson}}, \bibinfo {author} {\bibfnamefont
  {H.}~\bibnamefont {Nakatsuji}}, \bibinfo {author} {\bibfnamefont
  {X.}~\bibnamefont {Li}}, \bibinfo {author} {\bibfnamefont {M.}~\bibnamefont
  {Caricato}}, \bibinfo {author} {\bibfnamefont {A.~V.}\ \bibnamefont
  {Marenich}}, \bibinfo {author} {\bibfnamefont {J.}~\bibnamefont {Bloino}},
  \bibinfo {author} {\bibfnamefont {B.~G.}\ \bibnamefont {Janesko}}, \bibinfo
  {author} {\bibfnamefont {R.}~\bibnamefont {Gomperts}}, \bibinfo {author}
  {\bibfnamefont {B.}~\bibnamefont {Mennucci}}, \bibinfo {author}
  {\bibfnamefont {H.~P.}\ \bibnamefont {Hratchian}}, \bibinfo {author}
  {\bibfnamefont {J.~V.}\ \bibnamefont {Ortiz}}, \bibinfo {author}
  {\bibfnamefont {A.~F.}\ \bibnamefont {Izmaylov}}, \bibinfo {author}
  {\bibfnamefont {J.~L.}\ \bibnamefont {Sonnenberg}}, \bibinfo {author}
  {\bibfnamefont {D.}~\bibnamefont {Williams-Young}}, \bibinfo {author}
  {\bibfnamefont {F.}~\bibnamefont {Ding}}, \bibinfo {author} {\bibfnamefont
  {F.}~\bibnamefont {Lipparini}}, \bibinfo {author} {\bibfnamefont
  {F.}~\bibnamefont {Egidi}}, \bibinfo {author} {\bibfnamefont
  {J.}~\bibnamefont {Goings}}, \bibinfo {author} {\bibfnamefont
  {B.}~\bibnamefont {Peng}}, \bibinfo {author} {\bibfnamefont {A.}~\bibnamefont
  {Petrone}}, \bibinfo {author} {\bibfnamefont {T.}~\bibnamefont {Henderson}},
  \bibinfo {author} {\bibfnamefont {D.}~\bibnamefont {Ranasinghe}}, \bibinfo
  {author} {\bibfnamefont {V.~G.}\ \bibnamefont {Zakrzewski}}, \bibinfo
  {author} {\bibfnamefont {J.}~\bibnamefont {Gao}}, \bibinfo {author}
  {\bibfnamefont {N.}~\bibnamefont {Rega}}, \bibinfo {author} {\bibfnamefont
  {G.}~\bibnamefont {Zheng}}, \bibinfo {author} {\bibfnamefont
  {W.}~\bibnamefont {Liang}}, \bibinfo {author} {\bibfnamefont
  {M.}~\bibnamefont {Hada}}, \bibinfo {author} {\bibfnamefont {M.}~\bibnamefont
  {Ehara}}, \bibinfo {author} {\bibfnamefont {K.}~\bibnamefont {Toyota}},
  \bibinfo {author} {\bibfnamefont {R.}~\bibnamefont {Fukuda}}, \bibinfo
  {author} {\bibfnamefont {J.}~\bibnamefont {Hasegawa}}, \bibinfo {author}
  {\bibfnamefont {M.}~\bibnamefont {Ishida}}, \bibinfo {author} {\bibfnamefont
  {T.}~\bibnamefont {Nakajima}}, \bibinfo {author} {\bibfnamefont
  {Y.}~\bibnamefont {Honda}}, \bibinfo {author} {\bibfnamefont
  {O.}~\bibnamefont {Kitao}}, \bibinfo {author} {\bibfnamefont
  {H.}~\bibnamefont {Nakai}}, \bibinfo {author} {\bibfnamefont
  {T.}~\bibnamefont {Vreven}}, \bibinfo {author} {\bibfnamefont
  {K.}~\bibnamefont {Throssell}}, \bibinfo {author} {\bibfnamefont {J.~A.}\
  \bibnamefont {Montgomery}, \bibfnamefont {{Jr.}}}, \bibinfo {author}
  {\bibfnamefont {J.~E.}\ \bibnamefont {Peralta}}, \bibinfo {author}
  {\bibfnamefont {F.}~\bibnamefont {Ogliaro}}, \bibinfo {author} {\bibfnamefont
  {M.~J.}\ \bibnamefont {Bearpark}}, \bibinfo {author} {\bibfnamefont {J.~J.}\
  \bibnamefont {Heyd}}, \bibinfo {author} {\bibfnamefont {E.~N.}\ \bibnamefont
  {Brothers}}, \bibinfo {author} {\bibfnamefont {K.~N.}\ \bibnamefont {Kudin}},
  \bibinfo {author} {\bibfnamefont {V.~N.}\ \bibnamefont {Staroverov}},
  \bibinfo {author} {\bibfnamefont {T.~A.}\ \bibnamefont {Keith}}, \bibinfo
  {author} {\bibfnamefont {R.}~\bibnamefont {Kobayashi}}, \bibinfo {author}
  {\bibfnamefont {J.}~\bibnamefont {Normand}}, \bibinfo {author} {\bibfnamefont
  {K.}~\bibnamefont {Raghavachari}}, \bibinfo {author} {\bibfnamefont {A.~P.}\
  \bibnamefont {Rendell}}, \bibinfo {author} {\bibfnamefont {J.~C.}\
  \bibnamefont {Burant}}, \bibinfo {author} {\bibfnamefont {S.~S.}\
  \bibnamefont {Iyengar}}, \bibinfo {author} {\bibfnamefont {J.}~\bibnamefont
  {Tomasi}}, \bibinfo {author} {\bibfnamefont {M.}~\bibnamefont {Cossi}},
  \bibinfo {author} {\bibfnamefont {J.~M.}\ \bibnamefont {Millam}}, \bibinfo
  {author} {\bibfnamefont {M.}~\bibnamefont {Klene}}, \bibinfo {author}
  {\bibfnamefont {C.}~\bibnamefont {Adamo}}, \bibinfo {author} {\bibfnamefont
  {R.}~\bibnamefont {Cammi}}, \bibinfo {author} {\bibfnamefont {J.~W.}\
  \bibnamefont {Ochterski}}, \bibinfo {author} {\bibfnamefont {R.~L.}\
  \bibnamefont {Martin}}, \bibinfo {author} {\bibfnamefont {K.}~\bibnamefont
  {Morokuma}}, \bibinfo {author} {\bibfnamefont {O.}~\bibnamefont {Farkas}},
  \bibinfo {author} {\bibfnamefont {J.~B.}\ \bibnamefont {Foresman}}, \ and\
  \bibinfo {author} {\bibfnamefont {D.~J.}\ \bibnamefont {Fox}},\ }\href@noop
  {} {\enquote {\bibinfo {title} {Gaussian~16 {R}evision {A}.03},}\ } (\bibinfo
  {year} {2016}),\ \bibinfo {note} {gaussian Inc. Wallingford CT}\BibitemShut
  {NoStop}%
\bibitem [{\citenamefont {Sok}\ \emph {et~al.}(2011)\citenamefont {Sok},
  \citenamefont {Willow}, \citenamefont {Zahariev},\ and\ \citenamefont
  {Gordon}}]{sok2011solvent}%
  \BibitemOpen
  \bibfield  {author} {\bibinfo {author} {\bibfnamefont {S.}~\bibnamefont
  {Sok}}, \bibinfo {author} {\bibfnamefont {S.~Y.}\ \bibnamefont {Willow}},
  \bibinfo {author} {\bibfnamefont {F.}~\bibnamefont {Zahariev}}, \ and\
  \bibinfo {author} {\bibfnamefont {M.~S.}\ \bibnamefont {Gordon}},\ }\bibfield
   {title} {\enquote {\bibinfo {title} {Solvent-induced shift of the lowest
  singlet $\pi\rightarrow\pi$* charge-transfer excited state of p-nitroaniline
  in water: An application of the tddft/efp1 method},}\ }\href@noop {}
  {\bibfield  {journal} {\bibinfo  {journal} {J. Phys. Chem. A}\ }\textbf
  {\bibinfo {volume} {115}},\ \bibinfo {pages} {9801--9809} (\bibinfo {year}
  {2011})}\BibitemShut {NoStop}%
\bibitem [{\citenamefont {Kosenkov}\ and\ \citenamefont
  {Slipchenko}(2010)}]{kosenkov2010solvent}%
  \BibitemOpen
  \bibfield  {author} {\bibinfo {author} {\bibfnamefont {D.}~\bibnamefont
  {Kosenkov}}\ and\ \bibinfo {author} {\bibfnamefont {L.~V.}\ \bibnamefont
  {Slipchenko}},\ }\bibfield  {title} {\enquote {\bibinfo {title} {Solvent
  effects on the electronic transitions of p-nitroaniline: A qm/efp study},}\
  }\href@noop {} {\bibfield  {journal} {\bibinfo  {journal} {J. Phys. Chem. A}\
  }\textbf {\bibinfo {volume} {115}},\ \bibinfo {pages} {392--401} (\bibinfo
  {year} {2010})}\BibitemShut {NoStop}%
\bibitem [{\citenamefont {Eriksen}\ \emph {et~al.}(2013)\citenamefont
  {Eriksen}, \citenamefont {Sauer}, \citenamefont {Mikkelsen}, \citenamefont
  {Christiansen}, \citenamefont {Jensen},\ and\ \citenamefont
  {Kongsted}}]{eriksen2013failures}%
  \BibitemOpen
  \bibfield  {author} {\bibinfo {author} {\bibfnamefont {J.~J.}\ \bibnamefont
  {Eriksen}}, \bibinfo {author} {\bibfnamefont {S.~P.}\ \bibnamefont {Sauer}},
  \bibinfo {author} {\bibfnamefont {K.~V.}\ \bibnamefont {Mikkelsen}}, \bibinfo
  {author} {\bibfnamefont {O.}~\bibnamefont {Christiansen}}, \bibinfo {author}
  {\bibfnamefont {H.~J.~A.}\ \bibnamefont {Jensen}}, \ and\ \bibinfo {author}
  {\bibfnamefont {J.}~\bibnamefont {Kongsted}},\ }\bibfield  {title} {\enquote
  {\bibinfo {title} {Failures of tddft in describing the lowest intramolecular
  charge-transfer excitation in para-nitroaniline},}\ }\href@noop {} {\bibfield
   {journal} {\bibinfo  {journal} {Mol. Phys.}\ }\textbf {\bibinfo {volume}
  {111}},\ \bibinfo {pages} {1235--1248} (\bibinfo {year} {2013})}\BibitemShut
  {NoStop}%
\bibitem [{\citenamefont {Sneskov}\ \emph {et~al.}(2011)\citenamefont
  {Sneskov}, \citenamefont {Schwabe}, \citenamefont {Christiansen},\ and\
  \citenamefont {Kongsted}}]{sneskov2011scrutinizing}%
  \BibitemOpen
  \bibfield  {author} {\bibinfo {author} {\bibfnamefont {K.}~\bibnamefont
  {Sneskov}}, \bibinfo {author} {\bibfnamefont {T.}~\bibnamefont {Schwabe}},
  \bibinfo {author} {\bibfnamefont {O.}~\bibnamefont {Christiansen}}, \ and\
  \bibinfo {author} {\bibfnamefont {J.}~\bibnamefont {Kongsted}},\ }\bibfield
  {title} {\enquote {\bibinfo {title} {Scrutinizing the effects of polarization
  in qm/mm excited state calculations},}\ }\href@noop {} {\bibfield  {journal}
  {\bibinfo  {journal} {Phys. Chem. Chem. Phys.}\ }\textbf {\bibinfo {volume}
  {13}},\ \bibinfo {pages} {18551--18560} (\bibinfo {year} {2011})}\BibitemShut
  {NoStop}%
\bibitem [{\citenamefont {Olsen}, \citenamefont {Aidas},\ and\ \citenamefont
  {Kongsted}(2010)}]{olsen2010excited}%
  \BibitemOpen
  \bibfield  {author} {\bibinfo {author} {\bibfnamefont {J.~M.}\ \bibnamefont
  {Olsen}}, \bibinfo {author} {\bibfnamefont {K.}~\bibnamefont {Aidas}}, \ and\
  \bibinfo {author} {\bibfnamefont {J.}~\bibnamefont {Kongsted}},\ }\bibfield
  {title} {\enquote {\bibinfo {title} {Excited states in solution through
  polarizable embedding},}\ }\href@noop {} {\bibfield  {journal} {\bibinfo
  {journal} {J. Chem. Theory Comput.}\ }\textbf {\bibinfo {volume} {6}},\
  \bibinfo {pages} {3721--3734} (\bibinfo {year} {2010})}\BibitemShut {NoStop}%
\bibitem [{\citenamefont {Eriksen}\ \emph {et~al.}(2012)\citenamefont
  {Eriksen}, \citenamefont {Sauer}, \citenamefont {Mikkelsen}, \citenamefont
  {Jensen},\ and\ \citenamefont {Kongsted}}]{eriksen2012importance}%
  \BibitemOpen
  \bibfield  {author} {\bibinfo {author} {\bibfnamefont {J.~J.}\ \bibnamefont
  {Eriksen}}, \bibinfo {author} {\bibfnamefont {S.~P.}\ \bibnamefont {Sauer}},
  \bibinfo {author} {\bibfnamefont {K.~V.}\ \bibnamefont {Mikkelsen}}, \bibinfo
  {author} {\bibfnamefont {H.~J.~A.}\ \bibnamefont {Jensen}}, \ and\ \bibinfo
  {author} {\bibfnamefont {J.}~\bibnamefont {Kongsted}},\ }\bibfield  {title}
  {\enquote {\bibinfo {title} {On the importance of excited state dynamic
  response electron correlation in polarizable embedding methods},}\
  }\href@noop {} {\bibfield  {journal} {\bibinfo  {journal} {J. Comput. Chem.}\
  }\textbf {\bibinfo {volume} {33}},\ \bibinfo {pages} {2012--2022} (\bibinfo
  {year} {2012})}\BibitemShut {NoStop}%
\bibitem [{\citenamefont {Frutos-Puerto}, \citenamefont {Aguilar},\ and\
  \citenamefont {Fdez.~Galv{'a}n}(2013)}]{frutos2013theoretical}%
  \BibitemOpen
  \bibfield  {author} {\bibinfo {author} {\bibfnamefont {S.}~\bibnamefont
  {Frutos-Puerto}}, \bibinfo {author} {\bibfnamefont {M.~A.}\ \bibnamefont
  {Aguilar}}, \ and\ \bibinfo {author} {\bibfnamefont {I.}~\bibnamefont
  {Fdez.~Galv{'a}n}},\ }\bibfield  {title} {\enquote {\bibinfo {title}
  {Theoretical study of the preferential solvation effect on the solvatochromic
  shifts of para-nitroaniline},}\ }\href@noop {} {\bibfield  {journal}
  {\bibinfo  {journal} {J. Phys. Chem. B}\ }\textbf {\bibinfo {volume} {117}},\
  \bibinfo {pages} {2466--2474} (\bibinfo {year} {2013})}\BibitemShut {NoStop}%
\bibitem [{\citenamefont {DeFusco}\ \emph {et~al.}(2011)\citenamefont
  {DeFusco}, \citenamefont {Minezawa}, \citenamefont {Slipchenko},
  \citenamefont {Zahariev},\ and\ \citenamefont
  {Gordon}}]{defusco2011modeling}%
  \BibitemOpen
  \bibfield  {author} {\bibinfo {author} {\bibfnamefont {A.}~\bibnamefont
  {DeFusco}}, \bibinfo {author} {\bibfnamefont {N.}~\bibnamefont {Minezawa}},
  \bibinfo {author} {\bibfnamefont {L.~V.}\ \bibnamefont {Slipchenko}},
  \bibinfo {author} {\bibfnamefont {F.}~\bibnamefont {Zahariev}}, \ and\
  \bibinfo {author} {\bibfnamefont {M.~S.}\ \bibnamefont {Gordon}},\ }\bibfield
   {title} {\enquote {\bibinfo {title} {Modeling solvent effects on electronic
  excited states},}\ }\href@noop {} {\bibfield  {journal} {\bibinfo  {journal}
  {J. Phys. Chem. Lett.}\ }\textbf {\bibinfo {volume} {2}},\ \bibinfo {pages}
  {2184--2192} (\bibinfo {year} {2011})}\BibitemShut {NoStop}%
\bibitem [{\citenamefont {Mennucci}(2002)}]{mennucci2002hydrogen}%
  \BibitemOpen
  \bibfield  {author} {\bibinfo {author} {\bibfnamefont {B.}~\bibnamefont
  {Mennucci}},\ }\bibfield  {title} {\enquote {\bibinfo {title} {Hydrogen bond
  versus polar effects: An ab initio analysis on n$\rightarrow\pi$* absorption
  spectra and n nuclear shieldings of diazines in solution},}\ }\href@noop {}
  {\bibfield  {journal} {\bibinfo  {journal} {J. Am. Chem. Soc.}\ }\textbf
  {\bibinfo {volume} {124}},\ \bibinfo {pages} {1506--1515} (\bibinfo {year}
  {2002})}\BibitemShut {NoStop}%
\bibitem [{\citenamefont {Pagliai}\ \emph {et~al.}(2017)\citenamefont
  {Pagliai}, \citenamefont {Mancini}, \citenamefont {Carnimeo}, \citenamefont
  {De~Mitri},\ and\ \citenamefont {Barone}}]{pagliai2017electronic}%
  \BibitemOpen
  \bibfield  {author} {\bibinfo {author} {\bibfnamefont {M.}~\bibnamefont
  {Pagliai}}, \bibinfo {author} {\bibfnamefont {G.}~\bibnamefont {Mancini}},
  \bibinfo {author} {\bibfnamefont {I.}~\bibnamefont {Carnimeo}}, \bibinfo
  {author} {\bibfnamefont {N.}~\bibnamefont {De~Mitri}}, \ and\ \bibinfo
  {author} {\bibfnamefont {V.}~\bibnamefont {Barone}},\ }\bibfield  {title}
  {\enquote {\bibinfo {title} {Electronic absorption spectra of pyridine and
  nicotine in aqueous solution with a combined molecular dynamics and
  polarizable qm/mm approach},}\ }\href@noop {} {\bibfield  {journal} {\bibinfo
   {journal} {J. Comput. Chem.}\ }\textbf {\bibinfo {volume} {38}},\ \bibinfo
  {pages} {319--335} (\bibinfo {year} {2017})}\BibitemShut {NoStop}%
\bibitem [{\citenamefont {Biczysko}\ \emph {et~al.}(2012)\citenamefont
  {Biczysko}, \citenamefont {Bloino}, \citenamefont {Brancato}, \citenamefont
  {Cacelli}, \citenamefont {Cappelli}, \citenamefont {Ferretti}, \citenamefont
  {Lami}, \citenamefont {Monti}, \citenamefont {Pedone}, \citenamefont
  {Prampolini}, \citenamefont {Puzzarini}, \citenamefont {Santoro},
  \citenamefont {Trani},\ and\ \citenamefont
  {Villani}}]{biczysko2012integrated}%
  \BibitemOpen
  \bibfield  {author} {\bibinfo {author} {\bibfnamefont {M.}~\bibnamefont
  {Biczysko}}, \bibinfo {author} {\bibfnamefont {J.}~\bibnamefont {Bloino}},
  \bibinfo {author} {\bibfnamefont {G.}~\bibnamefont {Brancato}}, \bibinfo
  {author} {\bibfnamefont {I.}~\bibnamefont {Cacelli}}, \bibinfo {author}
  {\bibfnamefont {C.}~\bibnamefont {Cappelli}}, \bibinfo {author}
  {\bibfnamefont {A.}~\bibnamefont {Ferretti}}, \bibinfo {author}
  {\bibfnamefont {A.}~\bibnamefont {Lami}}, \bibinfo {author} {\bibfnamefont
  {S.}~\bibnamefont {Monti}}, \bibinfo {author} {\bibfnamefont
  {A.}~\bibnamefont {Pedone}}, \bibinfo {author} {\bibfnamefont
  {G.}~\bibnamefont {Prampolini}}, \bibinfo {author} {\bibfnamefont
  {C.}~\bibnamefont {Puzzarini}}, \bibinfo {author} {\bibfnamefont
  {F.}~\bibnamefont {Santoro}}, \bibinfo {author} {\bibfnamefont
  {F.}~\bibnamefont {Trani}}, \ and\ \bibinfo {author} {\bibfnamefont
  {G.}~\bibnamefont {Villani}},\ }\bibfield  {title} {\enquote {\bibinfo
  {title} {Integrated computational approaches for spectroscopic studies of
  molecular systems in the gas phase and in solution: pyrimidine as a test
  case},}\ }\href@noop {} {\bibfield  {journal} {\bibinfo  {journal} {Theor.
  Chem. Acc.}\ }\textbf {\bibinfo {volume} {131}},\ \bibinfo {pages} {1201}
  (\bibinfo {year} {2012})}\BibitemShut {NoStop}%
\bibitem [{\citenamefont {Cossi}\ and\ \citenamefont
  {Barone}(2001)}]{cossi2001time}%
  \BibitemOpen
  \bibfield  {author} {\bibinfo {author} {\bibfnamefont {M.}~\bibnamefont
  {Cossi}}\ and\ \bibinfo {author} {\bibfnamefont {V.}~\bibnamefont {Barone}},\
  }\bibfield  {title} {\enquote {\bibinfo {title} {Time-dependent density
  functional theory for molecules in liquid solutions},}\ }\href@noop {}
  {\bibfield  {journal} {\bibinfo  {journal} {J. Chem. Phys.}\ }\textbf
  {\bibinfo {volume} {115}},\ \bibinfo {pages} {4708--4717} (\bibinfo {year}
  {2001})}\BibitemShut {NoStop}%
\bibitem [{\citenamefont {Mason}(1959)}]{mason1959electronic}%
  \BibitemOpen
  \bibfield  {author} {\bibinfo {author} {\bibfnamefont {S.}~\bibnamefont
  {Mason}},\ }\bibfield  {title} {\enquote {\bibinfo {title} {The electronic
  spectra of n-heteroaromatic systems. part i. the n$\rightarrow\pi$
  transitions of monocyclic azines},}\ }\href@noop {} {\bibfield  {journal}
  {\bibinfo  {journal} {J. Chem. Soc.}\ ,\ \bibinfo {pages} {1240--1246}}
  (\bibinfo {year} {1959})}\BibitemShut {NoStop}%
\bibitem [{\citenamefont {Millefiori}\ \emph {et~al.}(1977)\citenamefont
  {Millefiori}, \citenamefont {Favini}, \citenamefont {Millefiori},\ and\
  \citenamefont {Grasso}}]{millefiori1977electronic}%
  \BibitemOpen
  \bibfield  {author} {\bibinfo {author} {\bibfnamefont {S.}~\bibnamefont
  {Millefiori}}, \bibinfo {author} {\bibfnamefont {G.}~\bibnamefont {Favini}},
  \bibinfo {author} {\bibfnamefont {A.}~\bibnamefont {Millefiori}}, \ and\
  \bibinfo {author} {\bibfnamefont {D.}~\bibnamefont {Grasso}},\ }\bibfield
  {title} {\enquote {\bibinfo {title} {Electronic spectra and structure of
  nitroanilines},}\ }\href@noop {} {\bibfield  {journal} {\bibinfo  {journal}
  {Spectrochim. Acta A}\ }\textbf {\bibinfo {volume} {33}},\ \bibinfo {pages}
  {21--27} (\bibinfo {year} {1977})}\BibitemShut {NoStop}%
\bibitem [{\citenamefont {Cai}\ and\ \citenamefont
  {Reimers}(2000)}]{cai2000low}%
  \BibitemOpen
  \bibfield  {author} {\bibinfo {author} {\bibfnamefont {Z.-L.}\ \bibnamefont
  {Cai}}\ and\ \bibinfo {author} {\bibfnamefont {J.~R.}\ \bibnamefont
  {Reimers}},\ }\bibfield  {title} {\enquote {\bibinfo {title} {The low-lying
  excited states of pyridine},}\ }\href@noop {} {\bibfield  {journal} {\bibinfo
   {journal} {J. Phys. Chem. A}\ }\textbf {\bibinfo {volume} {104}},\ \bibinfo
  {pages} {8389--8408} (\bibinfo {year} {2000})}\BibitemShut {NoStop}%
\bibitem [{\citenamefont {Schreiber}\ \emph {et~al.}(2008)\citenamefont
  {Schreiber}, \citenamefont {Silva-Junior}, \citenamefont {Sauer},\ and\
  \citenamefont {Thiel}}]{schreiber2008benchmarks}%
  \BibitemOpen
  \bibfield  {author} {\bibinfo {author} {\bibfnamefont {M.}~\bibnamefont
  {Schreiber}}, \bibinfo {author} {\bibfnamefont {M.~R.}\ \bibnamefont
  {Silva-Junior}}, \bibinfo {author} {\bibfnamefont {S.~P.}\ \bibnamefont
  {Sauer}}, \ and\ \bibinfo {author} {\bibfnamefont {W.}~\bibnamefont
  {Thiel}},\ }\bibfield  {title} {\enquote {\bibinfo {title} {Benchmarks for
  electronically excited states: Caspt2, cc2, ccsd, and cc3},}\ }\href@noop {}
  {\bibfield  {journal} {\bibinfo  {journal} {J. Chem. Phys.}\ }\textbf
  {\bibinfo {volume} {128}},\ \bibinfo {pages} {134110} (\bibinfo {year}
  {2008})}\BibitemShut {NoStop}%
\bibitem [{\citenamefont {da~Silva}\ \emph {et~al.}(2010)\citenamefont
  {da~Silva}, \citenamefont {Almeida}, \citenamefont {Martins}, \citenamefont
  {Milosavljevi{\'c}}, \citenamefont {Marinkovi{\'c}}, \citenamefont
  {Hoffmann}, \citenamefont {Mason}, \citenamefont {Nunes}, \citenamefont
  {Garcia},\ and\ \citenamefont {Limao-Vieira}}]{da2010electronic}%
  \BibitemOpen
  \bibfield  {author} {\bibinfo {author} {\bibfnamefont {F.~F.}\ \bibnamefont
  {da~Silva}}, \bibinfo {author} {\bibfnamefont {D.}~\bibnamefont {Almeida}},
  \bibinfo {author} {\bibfnamefont {G.}~\bibnamefont {Martins}}, \bibinfo
  {author} {\bibfnamefont {A.}~\bibnamefont {Milosavljevi{\'c}}}, \bibinfo
  {author} {\bibfnamefont {B.}~\bibnamefont {Marinkovi{\'c}}}, \bibinfo
  {author} {\bibfnamefont {S.~V.}\ \bibnamefont {Hoffmann}}, \bibinfo {author}
  {\bibfnamefont {N.}~\bibnamefont {Mason}}, \bibinfo {author} {\bibfnamefont
  {Y.}~\bibnamefont {Nunes}}, \bibinfo {author} {\bibfnamefont
  {G.}~\bibnamefont {Garcia}}, \ and\ \bibinfo {author} {\bibfnamefont
  {P.}~\bibnamefont {Limao-Vieira}},\ }\bibfield  {title} {\enquote {\bibinfo
  {title} {The electronic states of pyrimidine studied by vuv photoabsorption
  and electron energy-loss spectroscopy},}\ }\href@noop {} {\bibfield
  {journal} {\bibinfo  {journal} {Phys. Chem. Chem. Phys.}\ }\textbf {\bibinfo
  {volume} {12}},\ \bibinfo {pages} {6717--6731} (\bibinfo {year}
  {2010})}\BibitemShut {NoStop}%
\bibitem [{\citenamefont {Prabhumirashi}\ and\ \citenamefont
  {Kunte}(1986)}]{prabhumirashi1986solvent}%
  \BibitemOpen
  \bibfield  {author} {\bibinfo {author} {\bibfnamefont {L.}~\bibnamefont
  {Prabhumirashi}}\ and\ \bibinfo {author} {\bibfnamefont {S.}~\bibnamefont
  {Kunte}},\ }\bibfield  {title} {\enquote {\bibinfo {title} {Solvent effects
  on electronic absorption spectra of nitrochlorobenzenes, nitrophenols and
  nitroanilines--i. studies in nonpolar solvents},}\ }\href@noop {} {\bibfield
  {journal} {\bibinfo  {journal} {Spectrochim. Acta A}\ }\textbf {\bibinfo
  {volume} {42}},\ \bibinfo {pages} {435--441} (\bibinfo {year}
  {1986})}\BibitemShut {NoStop}%
\bibitem [{\citenamefont {Kovalenko}\ \emph {et~al.}(2000)\citenamefont
  {Kovalenko}, \citenamefont {Schanz}, \citenamefont {Farztdinov},
  \citenamefont {Hennig},\ and\ \citenamefont
  {Ernsting}}]{kovalenko2000femtosecond}%
  \BibitemOpen
  \bibfield  {author} {\bibinfo {author} {\bibfnamefont {S.}~\bibnamefont
  {Kovalenko}}, \bibinfo {author} {\bibfnamefont {R.}~\bibnamefont {Schanz}},
  \bibinfo {author} {\bibfnamefont {V.}~\bibnamefont {Farztdinov}}, \bibinfo
  {author} {\bibfnamefont {H.}~\bibnamefont {Hennig}}, \ and\ \bibinfo {author}
  {\bibfnamefont {N.}~\bibnamefont {Ernsting}},\ }\bibfield  {title} {\enquote
  {\bibinfo {title} {Femtosecond relaxation of photoexcited para-nitroaniline:
  solvation, charge transfer, internal conversion and cooling},}\ }\href@noop
  {} {\bibfield  {journal} {\bibinfo  {journal} {Chem. Phys. Lett.}\ }\textbf
  {\bibinfo {volume} {323}},\ \bibinfo {pages} {312--322} (\bibinfo {year}
  {2000})}\BibitemShut {NoStop}%
\bibitem [{\citenamefont {de~Almeida}\ \emph {et~al.}(2001)\citenamefont
  {de~Almeida}, \citenamefont {Coutinho}, \citenamefont {De~Almeida},
  \citenamefont {Rocha},\ and\ \citenamefont {Canuto}}]{de2001monte}%
  \BibitemOpen
  \bibfield  {author} {\bibinfo {author} {\bibfnamefont {K.~J.}\ \bibnamefont
  {de~Almeida}}, \bibinfo {author} {\bibfnamefont {K.}~\bibnamefont
  {Coutinho}}, \bibinfo {author} {\bibfnamefont {W.~B.}\ \bibnamefont
  {De~Almeida}}, \bibinfo {author} {\bibfnamefont {W.~R.}\ \bibnamefont
  {Rocha}}, \ and\ \bibinfo {author} {\bibfnamefont {S.}~\bibnamefont
  {Canuto}},\ }\bibfield  {title} {\enquote {\bibinfo {title} {A monte
  carlo--quantum mechanical study of the solvatochromism of pyrimidine in water
  and in carbon tetrachloride},}\ }\href@noop {} {\bibfield  {journal}
  {\bibinfo  {journal} {Phys. Chem. Chem. Phys.}\ }\textbf {\bibinfo {volume}
  {3}},\ \bibinfo {pages} {1583--1587} (\bibinfo {year} {2001})}\BibitemShut
  {NoStop}%
\bibitem [{\citenamefont {Silva-Junior}\ \emph {et~al.}(2010)\citenamefont
  {Silva-Junior}, \citenamefont {Schreiber}, \citenamefont {Sauer},\ and\
  \citenamefont {Thiel}}]{silva2010benchmarks}%
  \BibitemOpen
  \bibfield  {author} {\bibinfo {author} {\bibfnamefont {M.~R.}\ \bibnamefont
  {Silva-Junior}}, \bibinfo {author} {\bibfnamefont {M.}~\bibnamefont
  {Schreiber}}, \bibinfo {author} {\bibfnamefont {S.~P.}\ \bibnamefont
  {Sauer}}, \ and\ \bibinfo {author} {\bibfnamefont {W.}~\bibnamefont
  {Thiel}},\ }\bibfield  {title} {\enquote {\bibinfo {title} {Benchmarks of
  electronically excited states: Basis set effects on caspt2 results},}\
  }\href@noop {} {\bibfield  {journal} {\bibinfo  {journal} {J. Chem. Phys.}\
  }\textbf {\bibinfo {volume} {133}},\ \bibinfo {pages} {174318} (\bibinfo
  {year} {2010})}\BibitemShut {NoStop}%
\bibitem [{\citenamefont {Brehm}\ and\ \citenamefont
  {Kirchner}(2011)}]{brehm2011travis}%
  \BibitemOpen
  \bibfield  {author} {\bibinfo {author} {\bibfnamefont {M.}~\bibnamefont
  {Brehm}}\ and\ \bibinfo {author} {\bibfnamefont {B.}~\bibnamefont
  {Kirchner}},\ }\bibfield  {title} {\enquote {\bibinfo {title} {Travis - a
  free analyzer and visualizer for monte carlo and molecular dynamics
  trajectories},}\ }\href {\doibase 10.1021/ci200217w} {\bibfield  {journal}
  {\bibinfo  {journal} {J. Chem. Inf. Model.}\ }\textbf {\bibinfo {volume}
  {51}},\ \bibinfo {pages} {2007--2023} (\bibinfo {year} {2011})},\ \bibinfo
  {note} {pMID: 21761915},\ \Eprint
  {http://arxiv.org/abs/http://dx.doi.org/10.1021/ci200217w}
  {http://dx.doi.org/10.1021/ci200217w} \BibitemShut {NoStop}%
\bibitem [{\citenamefont {Perkampus}(1996)}]{perkampus1996uv}%
  \BibitemOpen
  \bibfield  {author} {\bibinfo {author} {\bibfnamefont {H.-H.}\ \bibnamefont
  {Perkampus}, \bibfnamefont {Ed.}},\ }\href@noop {} {\emph {\bibinfo {title}
  {UV-VIS atlas of organic compounds}}}\ (\bibinfo  {publisher} {Wiley,
  Weinheim},\ \bibinfo {year} {1996})\BibitemShut {NoStop}%
\bibitem [{\citenamefont {Loco}\ \emph {et~al.}(2018)\citenamefont {Loco},
  \citenamefont {Buda}, \citenamefont {Lugtenburg},\ and\ \citenamefont
  {Mennucci}}]{loco2018dynamic}%
  \BibitemOpen
  \bibfield  {author} {\bibinfo {author} {\bibfnamefont {D.}~\bibnamefont
  {Loco}}, \bibinfo {author} {\bibfnamefont {F.}~\bibnamefont {Buda}}, \bibinfo
  {author} {\bibfnamefont {J.}~\bibnamefont {Lugtenburg}}, \ and\ \bibinfo
  {author} {\bibfnamefont {B.}~\bibnamefont {Mennucci}},\ }\bibfield  {title}
  {\enquote {\bibinfo {title} {The dynamic origin of color tuning in proteins
  revealed by a carotenoid pigment},}\ }\href@noop {} {\bibfield  {journal}
  {\bibinfo  {journal} {J. Phys. Chem. Lett.}\ }\textbf {\bibinfo {volume}
  {9}},\ \bibinfo {pages} {2404--2410} (\bibinfo {year} {2018})}\BibitemShut
  {NoStop}%
\bibitem [{\citenamefont {Egidi}\ \emph {et~al.}(2018)\citenamefont {Egidi},
  \citenamefont {Lo~Gerfo}, \citenamefont {Macchiagodena},\ and\ \citenamefont
  {Cappelli}}]{egidi2018nature}%
  \BibitemOpen
  \bibfield  {author} {\bibinfo {author} {\bibfnamefont {F.}~\bibnamefont
  {Egidi}}, \bibinfo {author} {\bibfnamefont {G.}~\bibnamefont {Lo~Gerfo}},
  \bibinfo {author} {\bibfnamefont {M.}~\bibnamefont {Macchiagodena}}, \ and\
  \bibinfo {author} {\bibfnamefont {C.}~\bibnamefont {Cappelli}},\ }\bibfield
  {title} {\enquote {\bibinfo {title} {On the nature of charge-transfer
  excitations for molecules in aqueous solution: a polarizable qm/mm study},}\
  }\href@noop {} {\bibfield  {journal} {\bibinfo  {journal} {Theor. Chem.
  Acc.}\ }\textbf {\bibinfo {volume} {137}},\ \bibinfo {pages} {82} (\bibinfo
  {year} {2018})}\BibitemShut {NoStop}%
\bibitem [{\citenamefont {Maschietto}\ \emph {et~al.}(2018)\citenamefont
  {Maschietto}, \citenamefont {Campetella}, \citenamefont {Frisch},
  \citenamefont {Scalmani}, \citenamefont {Adamo},\ and\ \citenamefont
  {Ciofini}}]{maschietto2018charge}%
  \BibitemOpen
  \bibfield  {author} {\bibinfo {author} {\bibfnamefont {F.}~\bibnamefont
  {Maschietto}}, \bibinfo {author} {\bibfnamefont {M.}~\bibnamefont
  {Campetella}}, \bibinfo {author} {\bibfnamefont {M.~J.}\ \bibnamefont
  {Frisch}}, \bibinfo {author} {\bibfnamefont {G.}~\bibnamefont {Scalmani}},
  \bibinfo {author} {\bibfnamefont {C.}~\bibnamefont {Adamo}}, \ and\ \bibinfo
  {author} {\bibfnamefont {I.}~\bibnamefont {Ciofini}},\ }\bibfield  {title}
  {\enquote {\bibinfo {title} {How are the charge transfer descriptors affected
  by the quality of the underpinning electronic density?}}\ }\href@noop {}
  {\bibfield  {journal} {\bibinfo  {journal} {J. Comput. Chem.}\ }\textbf
  {\bibinfo {volume} {39}},\ \bibinfo {pages} {735--742} (\bibinfo {year}
  {2018})}\BibitemShut {NoStop}%
\bibitem [{\citenamefont {Santoro}\ and\ \citenamefont
  {Jacquemin}(2016)}]{santoro2016going}%
  \BibitemOpen
  \bibfield  {author} {\bibinfo {author} {\bibfnamefont {F.}~\bibnamefont
  {Santoro}}\ and\ \bibinfo {author} {\bibfnamefont {D.}~\bibnamefont
  {Jacquemin}},\ }\bibfield  {title} {\enquote {\bibinfo {title} {Going beyond
  the vertical approximation with time-dependent density functional theory},}\
  }\href@noop {} {\bibfield  {journal} {\bibinfo  {journal} {WIREs Comput. Mol.
  Sci.}\ }\textbf {\bibinfo {volume} {6}},\ \bibinfo {pages} {460--486}
  (\bibinfo {year} {2016})}\BibitemShut {NoStop}%
\end{thebibliography}%

\end{document}